\documentclass[twocolumn,trackchanges,floatfix]{aastex7}
\usepackage[utf8]{inputenc}
\usepackage{xcolor}
\usepackage{graphicx}
\usepackage{amsmath}
\usepackage{array}
\usepackage{comment}
\usepackage{gensymb}
\usepackage{caption}

\begin{document}

\title{On the Uniqueness and Causal Relationship of Precursor Activity to Solar Energetic Events: I. Transient Brightenings -- Introduction and Overview}

\author[0000-0001-5661-9759]{Karin Dissauer}
\affiliation{NorthWest Research Associates \\
3380 Mitchell Lane, Boulder, CO 80301, USA}
\email{dissauer@nwra.com}

\author[0000-0003-3571-8728]{Graham Barnes}
\affiliation{NorthWest Research Associates \\
3380 Mitchell Lane, Boulder, CO 80301, USA}
\email{graham@nwra.com}

\author[0000-0003-0026-931X]{KD Leka}
\affiliation{NorthWest Research Associates \\
3380 Mitchell Lane, Boulder, CO 80301, USA}
\affiliation{Institute for Space-Earth Environmental Research, Nagoya University, \\
Furo-cho Chikusa-ku, Nagoya, Aichi 464-8601, JAPAN}
\email{leka@nwra.com}

\author[0000-0002-7709-723X]{Eric L. Wagner}
\affiliation{NorthWest Research Associates \\
3380 Mitchell Lane, Boulder, CO 80301, USA}
\email{wagneric@nwra.com}

\correspondingauthor{Karin Dissauer}
\email{dissauer@nwra.com}
\begin{abstract}
The physical role played by small-scale activity that occurs before the sudden onset of solar energetic events (SEEs, i.e., solar flares and coronal mass ejections) remains in question, in particular as related to SEE initiation and early evolution.  It is still unclear whether such precursor activity, often interpreted as plasma heating, particle acceleration, or early filament activation, is indicative of a pre-event phase or simply on-going background activity.
In this series, we statistically investigate the uniqueness and causal connection between precursors and SEEs using paired activity-quiet epochs.
This first paper specifically introduces transient brightenings (TBs) and presents analysis regimes to study their role as precursors, including imaging of the solar atmosphere, magnetic field, and topology analysis. Applying these methods qualitatively to three cases, we find that prior to solar flares, TBs 1) tend to occur in one large cluster close to the future flare ribbon location and below the separatrix surface of a coronal magnetic null point, 2) are co-spatial with reconnection signatures in the lower solar atmosphere, such as bald patches and null point fan traces and 3) cluster in the vicinity of strong-gradient polarity inversion lines and regions of increased excess magnetic energy density.
TBs are also observed during quiet epochs of the same active regions, but they appear in smaller clusters not following a clear spatial pattern, although sometimes associated with short, spatially-intermittent bald patches and fan traces, but predominantly away from strong gradient polarity inversion lines in areas with little excess energy density.
 
\end{abstract}
\section{Introduction}
Solar energetic events, i.e., solar flares and coronal mass ejections (CMEs), are the result of free magnetic energy stored in the solar corona suddenly being released via magnetic reconnection. The physical mechanisms that trigger and initiate these energetic events still remain elusive, as most trigger mechanisms cannot be observed directly \citep[][]{Green:2018}.  Various phenomena are frequently observed in coronal and chromospheric data in the minutes to hours before solar flares and CMEs \citep[e.g.][]{Tappin:1991, Farnik:1996, Farnik:1998, Farnik:2003, Joshi:2011, Chifor:2007, Bamba:2014, Hernandez-Perez:2019b, Wang:2017}, that are interpreted as ``precursors''.
These phenomena include localized, small-scale, short-lived transient brightenings in multiple wavelengths including the optical, (extreme-) ultraviolet and X-rays \citep[e.g.,][]{Chifor:2007, Joshi:2011, Hernandez-Perez:2019a,Hudson:2020}, strong blue-shifts in H$\alpha$ \citep[e.g.,][]{Cho:2016}, and pre-event extreme-ultraviolet (EUV) coronal dimmings \citep[e.g.,][]{Qiu:2007, Zhang:2017, Zhu:2024}.

The physical role of precursors is not clear.
Precursors in the form of small-scale transient brightenings (TBs) may be responsible for destabilizing the overlying magnetic field in accordance with the breakout scenario \citep[e.g.][]{Sterling:2001, Sterling:2004}, or they may signify reconnection deep in the sheared core magnetic field region as required from the tether-cutting mechanism \citep[e.g.][]{Chifor:2007, Hernandez-Perez:2019a, He:2023}. TBs may signal emerging flux or magnetic flux cancellation \citep[e.g.][]{Sterling:2005,Bamba:2014} that act as catalyst for the main reconnection event \citep{Kusano_etal_2012,IshiguroKusano2017, Bamba:2017}. Pre-event activity in the form of Doppler shifts seem to correspond to the activation and rise of filaments, while pre-event coronal dimmings (PCDs) may indicate the expansion and stretching of overlying loops during a slow rise phase before an impulsive eruption \citep[e.g.][]{Qiu:2017, Zhang:2017}. Hot onsets, recently detected in GOES/XRS total irradiance lightcurves in soft X-rays are characterized by a gradual and almost linear growth in emission measure at isothermal temperatures in the range of 10--16~MK prior to the flare’s impulsive phase, as determined by the start of the hard X-ray emission \citep[][]{Hudson:2021, Battaglia:2023, daSilva:2023}. Observations indicate that soft X-ray emission occurs within the footpoint and low-lying loop regions and not in the corona and is therefore hypothesized to be physically distinct from the main flaring process.

Alternatively, the observed transient phenomena may simply reflect the general dynamics and restructuring of the chromosphere and corona, with no causal link to subsequent energetic events. For example, UV bursts -- small ($\leq$  2''), intense brightenings in active regions -- are not directly associated with flares and typically show weak or no signals in AIA coronal imaging channels \citep{Peter:2014, Young:2018}. They are thought to be small-scale reconnection events that occur from the photosphere to the upper chromosphere, often in magnetic environments such as emerging active regions, bald patches, sunspot moats, and light bridges \citep{Young:2018}.

So what makes small-scale pre-event activity actual event ``precursors''?  Are physical properties of these phenomena any different from random activity that occurs without any subsequent solar energetic event?

This series investigates, for the first time, the uniqueness and causal relationships of precursor signatures -- such as transient brightenings and pre-event coronal dimmings -- to solar energetic events to better understand the triggers for energy release and magnetic reconnection. Our work builds on \cite{Leka:2023}, which identified the kurtosis of running-difference images in AIA channels (sensitive to hot flaring plasma, the transition region, and chromospheric plasma at 94, 131, and 304 \AA) as key parameters for distinguishing flare-active from flare-quiet epochs. The kurtosis in general provides sensitivity to the far wings of a distribution, while running-difference images track variation in intensity on a 72s cadence, suggesting that short-lived, small-scale brightening and dimming events are more frequent before solar flares than during flare-quiet periods. The spatial association between these precursors and the magnetic and plasma environments could not be fully captured  in that study due to its reliance on FOV-integrated quantities for parametrization.

Therefore, this series, focuses on statistically quantifying differences in the spatial, temporal, and physical characteristics of event-precursor signatures against the distributions of similar phenomena that occur during activity-quiet epochs, thus solidifying any statistical association and identifying unique physical signatures. In addition, we investigate the relationships between precursor phenomena and the related event's observational signatures of magnetic reconnection, eruption dynamics, and energy release, thus solidifying a causal relationship.

In this first paper of the series we introduce transient brightenings and the different regimes that are combined to study them. Observational imaging diagnostics, including cumulative TB masks, heat maps, and differential emission measure analysis, are combined with magnetic field diagnostics. The latter includes parameters describing the magnetic environment relevant to TBs and/or the subsequent flare, including bald patches in the observed field. Additionally, coronal magnetic field modeling and the reconstruction of the magnetic skeleton are used, highlighting features like separatrix surfaces associated with coronal magnetic null points and between open and closed magnetic flux.
We apply these methods qualitatively to three different case study events (see Table~\ref{tab:events}) for demonstration purposes; a full quantitative statistical analysis is presented in a future part of the series.
\section{Definitions}\label{sec:defs}
\paragraph{Precursor activity} is defined as small-scale \textit{transient brightenings} or coronal dimming activity (\textit{pre-event coronal dimmings}) prior to solar flares and coronal mass ejections. We do not consider smaller flares, i.e., C-class, prior to larger flares, i.e., M-, and X-class, to be precursors \citep[][]{Gyenge:2016}. C-class flares are classified as events, while activity below the C-class threshold is not considered an event and may therefore be analyzed within \textit{analysis epochs}.
\paragraph{Transient brightenings} are small-scale, short-lived (but without a defined lifetime) emission enhancements in SDO/AIA 1600\AA, interpreted as observational signatures of small-scale reconnection events. TBs will be detected based on a global thresholding algorithm (Sec.~\ref{sec:detection}) similar to flare ribbon detection, since we hypothesize they form due to the same physical process but on a different spatial scale. 
A cumulative binary detection mask that contains at least one contiguous group of non-zero pixels is defined as a TB event.
\paragraph{Pre-event coronal dimmings} are temporary decreases in emission at EUV wavelengths that are hypothesized to indicate early filament activation or 
the rise of overlying fields that enable a subsequent CME.  A future paper in the series introduces pre-event coronal dimmings, and presents an overview of their relationship to the main eruption as well as to the magnetic topological skeleton, their magnetic and plasma environment.
\paragraph{Analysis epochs} To investigate the uniqueness of precursor activity, TBs will be detected prior to the GOES soft X-ray start of the flare event (\textit{active epochs}) and also during specifically selected quiet times of the same active region (\textit{quiet epochs}). We define an active epoch as the pre-flare phase, specifically the period preceding a C1.0 or larger solar flare listed in the GOES flare catalog, during which no other flaring events above the C1.0 level occur.
For the purpose of transient brightenings, a \textit{quiet epoch} is thus defined as a time period during which no flares are detected above C1.0 level in the GOES flare catalog\footnote{\url{ftp://ftp.swpc.noaa.gov/pub/warehouse/}} during an \textit{analysis interval}. Quiet epochs are selected from the same active regions as active epochs, both prior to and after the active epoch of interest (if possible), to serve as control data and also allow intercomparison between quiet epochs.
Our definition of the pre-flare phase is in contrast to the pre-flash phase \citep[][]{Benz:1983}, or the time period prior to the hard X-ray onset, also referred to as pre-flare phase when, usually, soft X-ray activity is already detected \citep[e.g.,][]{Battaglia:2009, Hudson:2021}.
\paragraph{Analysis intervals} are time intervals of approximately two hours, assigned for either an active or quiet epoch, over which the analysis is conducted. For active epochs these intervals start two hours prior to the GOES flare start time and end four minutes before it. The same interval lengths are used for quiet epochs to optimize comparison between the epochs. 
The four-minute gap between the end of the analysis interval and the flare/quiet epoch start time corresponds to the same significance criterion\footnote{The start time of a solar flare observed by GOES/XRS is defined as follows: ``The begin time of an X-ray event is defined as the first minute, in a sequence of 4 minutes, of steep monotonic increase in 0.1--0.8~nm flux.''} used to define the start time in the 1--8\AA~GOES/XRS flux. A gradual increase in the flux before the flare start time is often observed and therefore the start time can be ambiguous. We try to resolve this ambiguity by introducing this additional lag time, so TB signatures are not mistaken as early flare ribbons. The closest-in-time valid quiet epoch is selected for each active epoch, provided it is within 24 hours.
The selected events under study (see Table~\ref{tab:events}) are part of a large statistical sample, for which all events are selected to be on-disk from the SDO perspective within
$45\degree$ around the central meridian, and part of the \texttt{RibbonDB} database \citep[][]{Kazachenko:2017}.
\section{Observations}\label{sec:data}
\subsection{Data and Pre-Processing}
In order to investigate precursor activity we primarily use data from the Solar Dynamics Observatory \citep[SDO;][]{Pesnell:2012}, specifically from the Atmospheric Imaging Assembly \citep[AIA;][]{Lemen:2012} and the Helioseismic and Magnetic Imager \citep[HMI;][]{Scherrer:2012, Hoeksema:2014}.
AIA acquires coronal images of the Sun in nine different extreme ultraviolet and ultraviolet (UV) passbands with a
native cadence of 12s and a spatial resolution of $1.5^{\prime\prime}$ sampled at $0.6^{\prime\prime}$. 
The full range of channels (94, 131, 171, 193, 211, 304, and 335 \AA), plus the 1600 and 1700 \AA~UV channels, provide data sensitive to temperatures between $\approx$ 5,000~K -- 10~MK, {\it i.e.,} from the photosphere, chromosphere and transition region up through the corona. 
HMI provides vector magnetic field data with a default cadence of 720s and a spatial resolution of $1.0^{\prime\prime}$ sampled at $0.5^{\prime\prime}$.

For all magnetic field analysis in this series, we use the vector field data from HMI, specifically the {\tt hmi.B\_720s} series, with the {\tt hmi.Mharp\_720s} series providing auxilliary information (see below).  The ``random'' option is chosen for the disambiguation of weak-signal areas, and the pipeline product
is otherwise used with no further downsampling or modifications. 
We use sub-area regions-of-interest (ROIs) data cubes of three hours (two hours of analysis time plus one hour to characterize the subsequent flare event or quiet state) down-sampled time-series of AIA data based on the ``HMI Active Region Patches'' \citep[HARPs;][]{Hoeksema:2014}, called ``AIA Active Region Patches'' \citep[AARPs;][]{Dissauer:2023}. 
HARP metadata ({\tt hmi.Mharp\_720s} series) are used to extract these ROI-defined images at all wavelengths of interest, with down-sampling to 72s cadence for the
shorter-wavelength data to match the UV-image cadence.
AARP image-plane spatial dimensions are kept constant over the target time range, and cover the HARP bounding boxes for straightforward cross-source analysis. We note that compared to the AARP definition, the AIA extracted field-of-views have been reduced to match the HARP box dimension. 
The first image of each time series is a reference against which subsequent extracted images are tracked and corrected for differential rotation. 
We detect transient brightenings primarily in 1600\AA~AIA data and characterize their plasma properties in the corona performing Differential Emission Measure (DEM) analysis using the code of \cite{Cheung:2015}. We use co-temporal AIA data of 94, 131, 171, 193, 211, and 335\AA~channels, limit the temperature range of the inversions to $\log T=[5.7-7.7]$ with a binsize of $\Delta \log T=0.1$.
Standard Solarsoft IDL and Python software is used to analyze the data.
\subsection{Events Overview}
\begin{table*}[htp!]
\centering
\begin{tabular}{l p{1.2cm} p{1.1cm} p{1.7cm} p{1.0cm} p{2.4cm} p{2.4cm} p{2.4cm}}
\tableline
\tableline
Date & NOAA AR No. & HARP No. & Flare Start time & Flare Class & Active epoch  & Pre-event quiet  epoch & Post-event quiet epoch \\ 
\tableline
16-Oct-2010 & 11112 & 211 &  19:07~UT & M2.9 & \mbox{17:07--19:03~UT} (17:48~UT) & \mbox{13:00--14:56~UT} (13:48~UT) &  \mbox{21:12--23:08~UT} (22:00~UT)\\
12-Nov-2010 & 11123 & 245 & 07:59~UT & C1.5 & \mbox{05:59--07:55~UT} (06:36~UT) & none\footnote{Due to enhanced flaring activity, no such epoch could be identified.} & \mbox{15:48--17:44~UT} (16:36~UT) \\
06-Sep-2011 & 11283 & 833 & 22:12~UT & X2.1 & \mbox{20:12--22:08~UT} (21:00~UT)& \mbox{14:12--16:08~UT}; \mbox{16:12--18:08~UT} (\mbox{17:00~UT}); \mbox{18:12--20:08~UT*} & 07-Sep-2011 \mbox{00:24--02:20~UT} (01:12~UT) \\
\noalign{\smallskip}
\tableline
\tableline
\end{tabular}
\caption{Event overview. The date, NOAA active region and HMI HARP number, flare start time, GOES/XRS Soft X-ray peak (``Flare Class''), and the analysis intervals for the active, and the selected pre-event and post-event quiet epochs are listed. The times indicated in bracket refer to the time instant the PFSS model is reconstructed for each epoch, respectively. *An extended period of non-flaring activity prior to the flare is observed, therefore a sequential
investigation of TBs during three consecutive pre-event quiet epochs leading up to the active epoch is performed. Pre-event quiet epoch \#2, i.e., 16:12--18:08~UT is used for magnetic skeleton, magnetic and plasma environment investigations.}
\label{tab:events}
\end{table*}
We focus here on three flare events that have distinctly different behavior.  The events, dates, host regions, active epoch and
selected pre-event and post-event quiet epochs are listed in Table~\ref{tab:events}. 
The M2.9 flare event in NOAA AR~11112 on October 16, 2010 is classified as a circular ribbon flare and is associated with an EUV late-phase emission episode \citep[][]{Woods:2011,Chen:2020}.
The C1.5 flare event in NOAA AR~11123 on November 12, 2010 is randomly selected from the \texttt{RibbonDB} database to serve as an example of a smaller flare (below M- and X-class).
The X2.1 flare event that occurred in NOAA AR~11283 on September 6, 2011 is well-studied from both an observational and modelling point of view \citep[][]{Jiang:2013, Jiang:2014, Janvier:2016,Jiang:2018,Prasad:2020}. In addition, an extended period of non-flaring activity prior to this flare is observed, therefore a sequential investigation
of TBs during three consecutive pre-event quiet epochs leading up to the active epoch is performed (see Table~\ref{tab:events} for details). The active epoch of November 12, 2010 is used throughout the methodology section to illustrate the application of the different methods (c.f., Figures~\ref{fig:detection_tbs}-\ref{fig:efree_schrijver_tbs}).
\section{Methodology}\label{sec:method}
\subsection{Transient Brightening Detection}\label{sec:detection}
\begin{figure*}
    \centering
    \includegraphics[width=0.9\textwidth, clip, trim = 0mm 0mm 2mm 2mm]{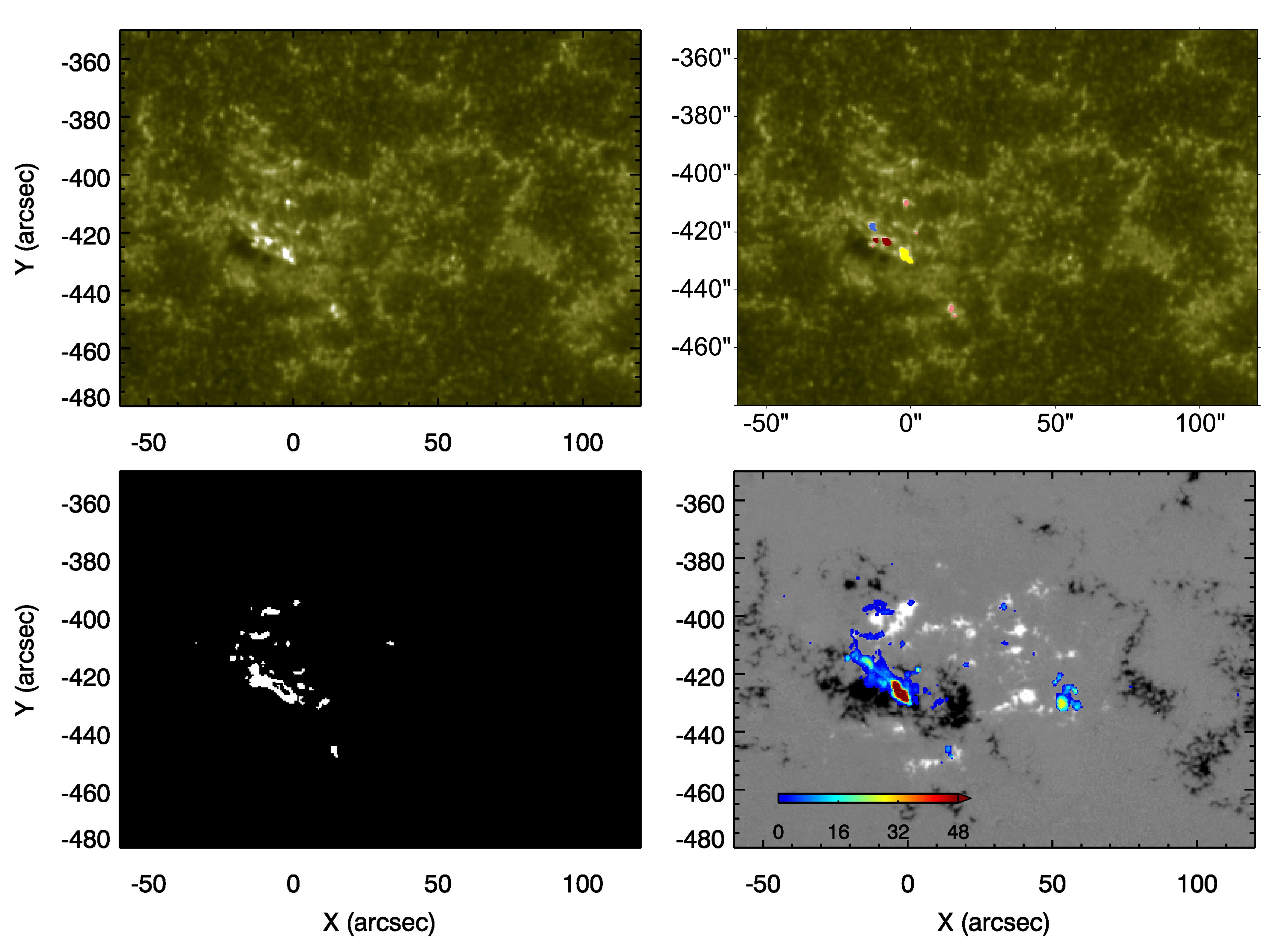}
    \caption{Example of the detection of transient brightenings in NOAA AR~11123. Top row: SDO/AIA 1600\AA~images at 06:41~UT during the pre-flare phase of a C1.5 flare on November 12, 2010 showing small-scale brightening activity (left) and the detected TB pixels identified as four individual clusters to illustrate HDBSCAN for this time step (right). Bottom row: Cumulative TB mask up to this time step starting at 05:59~UT (left) and TB heat map for the full analysis interval (i.e., 05:59--07:55~UT) overlaid on an image of the radial component of the magnetic field. The colorbar shows how often each individual pixel was detected as TB during the analysis interval. An animation of this figure is available. It shows original SDO/AIA 1600\AA~observations together with the detection of TBs over the full analysis interval of two hours and the formation of the corresponding cumulative TB mask. The animation does not include the TB heat map panel of the figure.}
    \label{fig:detection_tbs}
\end{figure*}
Transient brightenings are detected based on a global thresholding algorithm in 1600\AA~originally developed for the analysis of flare ribbons \citep[e.g.,][]{Qiu:2002, Qiu:2004, Qiu:2007, Kazachenko:2012, Kazachenko:2017}.
Flare ribbons are interpreted to represent the footpoints of newly formed loops as a result of magnetic reconnection. 
Detecting TBs using the same algorithm is a first step to target the detection of the same physical phenomena, namely magnetic reconnection, only during a different phase (i.e., prior to a solar flare or during a quiet epoch).
For each image $k$ at time step $t_k$, transient brightenings are identified as pixels $p$ with intensity values higher than a pre-defined cutoff intensity $I_{c}$, which is a multiple $M$ of the median intensity of $k$ over the AARP field-of-view.  
$M\approx6$ was empirically found to detect TBs as they are smaller and shorter lived compared to main-flare ribbons, which presents the lower limit of values used for flare ribbon detections \citep[][]{Kazachenko:2017}. The same value of $M$ is used for all events and their associated epochs.
The instantaneous brightening mask $R_{k}(p)$ at each time step $t_{k}$ is defined to have a value of 1 at pixels identified as brightening and 0 elsewhere \citep[e.g.,][]{Kazachenko:2017, Dissauer:2018a}.

A heat map $H_{0:n}(p)$ is used to study TB activity over analysis intervals \citep[data are available at][]{tb_data}.
It is based on the pixel-by-pixel sum of instantaneous brightening masks $R_{k}(p)$ at each time step $t_{k}$, which contain all TB pixels identified at time step $t_k$, 
\textit{i.e.,} $H_{0:n}(p)= R_{n}(p) + R_{n-1}(p) + \ldots + R_{0}(p)$.  
Heat maps represent cumulative maps, \textit{i.e.,} they include information over the entire analysis interval, and have the advantage of containing weighted information on multiple brightenings for individual pixels and their brightening duration. Heat maps resolve ``hot spot'' areas of enhanced activity. TBs that occur multiple times at the
same location or are longer-lived have a higher value within the mask\footnote{We note that the difference between TBs being above the detection threshold for multiple, successive frames (long-duration brightening) versus exceeding the threshold during multiple different times (multiple short-duration bursts) during the time range under investigation is not resolved with this method.}.
Figure~\ref{fig:detection_tbs} illustrates the basic detection of transient brightenings during the active epoch on November 12, 2010.

\subsection{Spatiotemporal Clustering}\label{sec:clustering}
To identify underlying patterns and objectively quantify hot spot areas in the heat maps of transient brightenings, we apply Hierarchical Density-Based Spatial Clustering of Applications with Noise \citep[HDBSCAN;][]{Campello:2013, Campello:2015}. HDBSCAN is a density-based clustering algorithm: given a set of points in space, it groups together points that are closely packed together (points with many nearby neighbors), marking points as outliers that occur in isolation or are located in very sparse regions (i.e., whose nearest neighbors are far away). HDBSCAN calculates a density-based clustering hierarchy, which enables the recognition of clusters of various shapes and sizes by taking into account the stability of cluster assignments across different levels of the hierarchy; it identifies stable clusters as those with consistent memberships at multiple levels, ensuring robustness in cluster formation. HDBSCAN has two key parameters, the number of \textit{min\_samples}-neighbors used to calculate the distance between each point in the sample and its nearest neighbors to empirically estimate density and the minimum number of points required to form a cluster (\textit{min\_cluster\_size}).
For our application we choose $min\_samples=5$ and $min\_cluster\_size=10$. In addition, we impose a lower limit on the sparsity of a cluster by defining the cluster neighborhood ($\epsilon$) to be 1.5~Mm (see \footnote{https://hdbscan.readthedocs.io/en/latest/dbscan\_from\_hdbscan.html} for details), which corresponds to approximately 2'', the typical length scale of UV bursts \citep{Young:2018}. The top right panel of Fig.~\ref{fig:detection_tbs} shows three TB clusters and noise identified by HDBSCAN for one time instant during the active epoch on November 12, 2010.
\subsection{Plasma Context}\label{sec:plasma}
\begin{figure*}
\centering
\includegraphics[width=1.0\textwidth, clip, trim = 5mm 3mm 0mm 0mm]{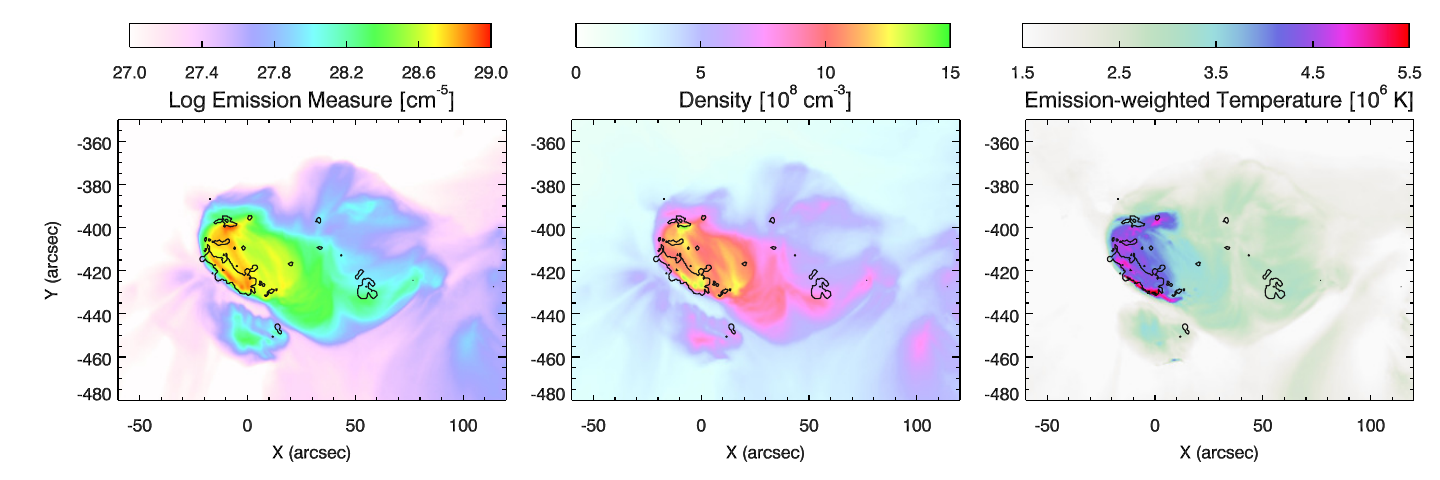}
\caption{DEM-constructed average emission measure (left), density (middle) and emission-weighted temperature (right) maps for NOAA AR~11123 during the active epoch on November 12, 2010. The black contours represent the detected transient brightenings over the same analysis intervals.}
\label{fig:dem}
\end{figure*}
Differential Emission Measure diagnostics \citep{Cheung:2015} are employed here to study the properties of, and the coronal response to, the transient brightenings detected in the upper photosphere/transition region and their surroundings \citep{dem_data}. Following the notation of~\cite{Cheung:2015,Saqri:2020}, narrow-band EUV observations, e.g., from SDO/AIA can
be inversely related to the physical properties of optically thin coronal plasma as an integral over temperature space:
\begin{equation}
y_{i}=\int_{0}^{\infty} K_{i}(T)\,{D\!E\!M}(T)\,dT\,,
\end{equation}
where $y_{i}$ is the exposure time-normalized pixel value in the $i$th AIA channel (in units of DN s$^{-1}$ pixel$^{-1}$) and $K_{i}(T)$ is the temperature response function (in units of DN cm$^{5}$ s$^{-1}$ pixel$^{-1}$).
For optically-thin emission, the temperature distribution of the plasma along the column height $h$ is described by the
differential emission measure (in units cm$^{-5}$ K$^{-1}$)
\begin{equation}
D\!E\!M(T)=n(T)^{2}\,\frac{dh}{dT} \, ,
\end{equation}
where $n$ is the number density dependent on the temperature $T$ along $h$.
The integral of $D\!E\!M(T)$ over a finite temperature range represents the total emission measure $E\!M$ of the plasma (in units cm$^{-5}$)
\begin{equation}
E\!M=\int_{T} D\!E\!M(T)\,dT \, .
\end{equation}
Temperature and density are calculated as DEM-weighted average temperature $T$ and density $n$  \citep{Cheng:2012, Vanninathan:2015}:
\begin{equation}
\bar{T}=\frac{\int_{T} D\!E\!M(T)\, T\, dT}{E\!M} \,,
\end{equation}
\begin{equation}
\bar{n}=\sqrt{\frac{E\!M}{h}}. 
\end{equation}
The DEM of the full field-of-view of active regions, on a pixel-by-pixel basis, is reconstructed, this means that $h$ will change, depending on the structures present. Therefore, we use a pixel-by-pixel estimate of $h$ based on the hydrostatic scale height of coronal plasma \citep[e.g.,][]{Aschwanden:2004}:
\begin{equation}
  h =\frac{2 k_{B}\,\bar{T}}{\mu\, m_{H}\,g} \,,
\end{equation}
where we assume a fully-ionized solar plasma with $\mu=0.64$ and $m_{H}= 1.67\times10^{-27}$~kg, and use the DEM-weighted average temperature $\bar{T}$ in each individual pixel to calculate $h$.
We note that this approach to estimate \textit{h} may not represent its ``true'' value for each pixel, as loops with varying temperatures could appear along the line of sight and different loop geometries apply for each pixel (pixels covering the top of loops versus their footpoints). However, we believe that the scale height might serve as a more accurate estimate of \textit{h} than using a constant value across the entire field of view given the variation in loops along the line of sight. While the hydrostatic scale height represents a reasonable estimate of $h$ at the loop top, it is an overestimation for the loop footpoints, leading to an underestimation of density for the given EM.

Plasma property maps are calculated at each time step during the different analysis epochs, and average maps over each analysis interval are used for comparison with TBs. 
Figure~\ref{fig:dem} shows the mean emission measure (left), density (middle) and emission-measure weighted temperature map (right) during the active epoch on November 12, 2010. The black contours represent the cumulative map of TBs detected during this epoch.
The TBs generally lie in areas of enhanced emission measure, density, and temperature.
\subsection{Magnetic topological skeleton}\label{sec:topology}
TBs might be chromospheric signatures of reconnection occurring higher in the Sun's atmosphere.
Reconnection typically occurs at null points, separators, and quasi-separators, locations at which the connectivity of the magnetic field is rapidly or discontinuously changing \citep{Pontin:2022}.
Particles accelerated during reconnection near these locations then propagate along separatrix or quasi-separatrix layers, resulting in heating when they reach the chromosphere.
To understand the causes of the TBs, we examine their location in the context of the magnetic skeleton \citep{Longcope:2005}, specifically the intersection of separatrix surfaces with the photosphere \citep{skeleton_data}.

To model the coronal magnetic field, we use a high-resolution Potential Field Source Surface \citep[PFSS;][]{Altschuler:1969,Schatten:1969} model.
Our implementation uses the SHTools \citep[][]{Wieczorek:2018} package to compute spherical harmonics, which is accurate up to maximum degree $\approx 2800$, and as a boundary condition uses a full-disk image of the radial component of the HMI vector magnetogram $B_{r}$, combined with an assumption of antisymmetry in the far hemisphere.\footnote{The PFSS code will be made available on gitlab upon acceptance of the manuscript.}
This allows small scale photospheric magnetic features to be resolved that may be associated with TBs, but is not suitable for global modelling of the corona.
We note that more sophisticated coronal models could be used, e.g., Nonlinear Force-Free Field (NLFFF) models \citep{Wiegelmann2021}, but it was shown by \citet{Regnier:2012} that the presence of a magnetic null point in a potential field is relatively robust to the introduction of moderate currents, thus we anticipate that elements of the magnetic skeleton found in the PFSS model are likely to also be present in non-potential models.
Further, for our planned statistical analysis it is currently still unrealistic to validate this many NLFFF extrapolations following the recommendations of \cite{DeRosa:2015}.

We also determine the location of bald patches in the observed field that are a signature of dipped field lines in the corona \citep{Seehafer:1986,Titov:1993}. Currents flowing in the corona are likely to result in dipped field lines.
By using the combination of the skeleton of the potential field and the observed bald patches, we can determine most of the locations that could be expected to exhibit signs of reconnection in the corona. 

The magnetic skeleton consisting of coronal null points and their topological features as well as open field footpoints is reconstructed at one time step, i.e., in the middle of each analysis epoch (see Table~\ref{tab:events} for details). Bald patches are identified during all available HMI vector magnetograms of the analysis intervals and evaluated as cumulative maps.
\subsubsection{Coronal null points and associated topological structures}
Magnetic null points are locations where the magnetic field vanishes, $\bf{B}=\bf{0}$.
In the vicinity of a null point, there are two field lines directed toward (away from) the null, called spine field lines, and a set of field lines directed away from (toward) the null that lie in a surface, referred to as the fan surface, which separate domains of different connectivity \citep{Longcope:2005}.
The intersection of two fan surfaces is a separator field line, which is a particularly favorable location for reconnection.
These structures comprise the elements of the magnetic skeleton associated with the null.
The fan traces (intersection with the lower boundary of the null separatrix surfaces) are the most likely place to find a chromospheric signature of null point or separator reconnection. 
\begin{figure}
\centering
\includegraphics[width=0.5\textwidth, clip, trim = 9mm 2mm 19mm 11mm]{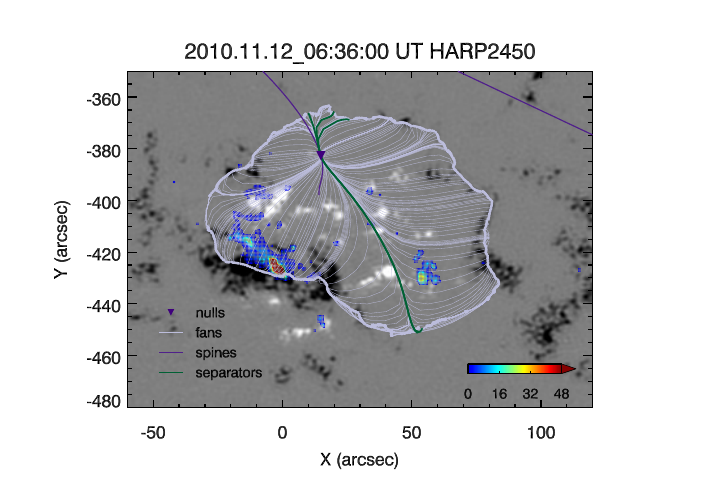}
\caption{Transient brightenings (in rainbow color, see Fig.~\ref{fig:detection_tbs} for details) identified during the active epoch of NOAA AR~11123 on November 12, 2010 overlaid on the radial component of the magnetic field $B_{r}$ from the PFSS model at 06:36~UT using spherical harmonics up to degree 2800. TB hot spot areas are marked in red, areas of lower TB activity are shown in blue. The building blocks of the magnetic skeleton are shown: a coronal null point (purple), field lines in its separatrix surface (lilac), separators (green) and two spine field lines (purple).}
\label{fig:topology_nulls_fans}
\end{figure}

To check if pre-event activity occurs at locations favorable for signatures of reconnection, we locate null points in the PFSS model using the approach of \citet{Haynes:2007}, then determine the other elements of the magnetic skeleton based on the approach of \citet{Haynes:2010}.
Figure~\ref{fig:topology_nulls_fans} illustrates the magnetic skeleton associated with a single null point within NOAA AR\,11123 on November 12, 2010 together with the detected transient brightenings in color. The majority of TBs during this epoch occur within the null point dome. 

Determining the magnetic topological skeleton for a specific AR is a rather complex task, as numerous coronal null points, especially low-lying ones associated with quiet Sun magnetic fields are present everywhere on the Sun. Identifying the relevant topological features within our active-region-sized FOVs is therefore key. The following approach is used: First, null points are located in the full coronal volume above the AR, and their spine field lines are traced. Next, the remaining elements of the magnetic skeleton (fan surfaces and separators) are determined for null points that have at least one spine end in the vicinity of a TB. Separators from these nulls are used to determine a second set of null points for which elements of the skeleton are also determined. 
\subsubsection{Open field footpoints}
\begin{figure}
\centering
\includegraphics[width=0.5\textwidth, clip, trim = 9mm 2mm 19mm 11mm]{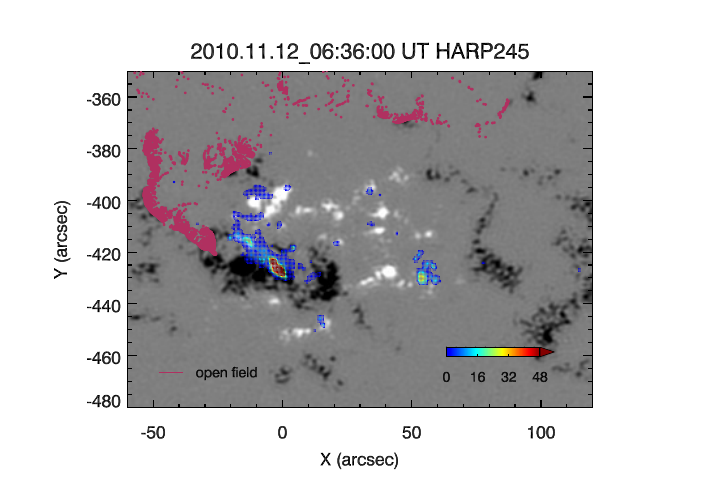}
\caption{Same as  Figure~\ref{fig:topology_nulls_fans} but showing open field footpoints (rose-pink) as the topological feature of interest.}
\label{fig:topology_open_field}
\end{figure}
The boundary between open and closed field is a separatrix surface, and thus another possible location for signatures of reconnection.
To determine the footpoints of open magnetic field, we initiate field lines on a regular longitude-sin(latitude) grid covering the entire source surface, and trace them down to their intersection with the lower boundary.  
Figure~\ref{fig:topology_open_field} shows open field footpoints of the PFSS extrapolation of AR 11123 at the center of the active epoch on November 12, 2010.  
The high spherical harmonic degree used is the reason for the patchiness in the areas of open field as small spatial scale opposite polarity features are resolved and result in small areas of closed field within otherwise open areas.
The boundaries of the open flux footpoints do not show an obvious correspondence with the TBs.
\subsubsection{Bald Patches}\label{sec:BPs}
\begin{figure}
\centering
\includegraphics[width=0.49\textwidth, clip, trim = 9mm 2mm 19mm 11mm]{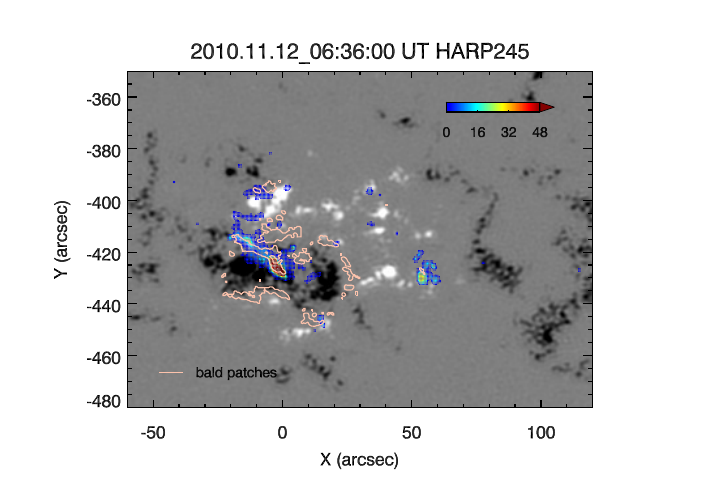}
\caption{Same as Figure~\ref{fig:topology_nulls_fans} but showing bald patches (light pink contours) as the topological feature of interest.}
\label{fig:topology_bald_patches}
\end{figure}
Bald patches are segments of a polarity inversion line (PIL) where the horizontal field is directed from the negative to the positive polarity of the radial field \citep[e.g.,][]{Seehafer:1986,Titov:1993,Titov:1999}: 
\begin{equation}
{\bf B}_h \cdot {\bf \nabla}_h B_r \vert_{B_r = 0} > 0.
\end{equation}
Such regions indicate the presence of dips in the overlying magnetic field lines.
The field line connectivity is discontinuous across a bald patch PIL, hence the bald patch separatrix surface, threaded by field lines that originate in the bald patch, is a potential site of reconnection and thus an interesting topological feature to consider for TBs.
By looking for bald patches in the observed field, we circumvent the need to model the coronal field. 
However, we are unable to determine the footpoints of the field lines in the bald patch separatrix surface which lie away from the PIL.

The existence of a bald patch is consistent with the presence of a twisted magnetic flux rope in the corona above.
However, bald patches are not unique to flux ropes - even potential fields can contain bald patches and flux ropes may have very small bald patch at the photosphere.
Here, we mostly focus on bald patches simply as a potential site for reconnection, without trying to determine whether a flux rope is present.

We use IDL's \textit{contour} procedure to return points along the PIL, then check at each point whether the direction of the horizontal field satisfies the bald patch criterion.
We restrict our analysis to points at which the confidence in the disambiguation of the magnetic field is large\footnote{\textsc{conf\_disambig}=90 in the HMI series {\tt hmi.B\_720s}.}, and hence the inferred direction of the horizontal field is more reliable.
Figure~\ref{fig:topology_bald_patches} shows the locations of magnetic bald patches as light pink contours on top of a colored TB heat map for November 12, 2010 illustrating promising spatial agreement between the two.

\subsection{The Broader Magnetic Landscape}\label{sec:mag}
\begin{figure}
\centering
\includegraphics[width=0.5\textwidth, clip, trim = 5mm 4mm 19mm 11mm]{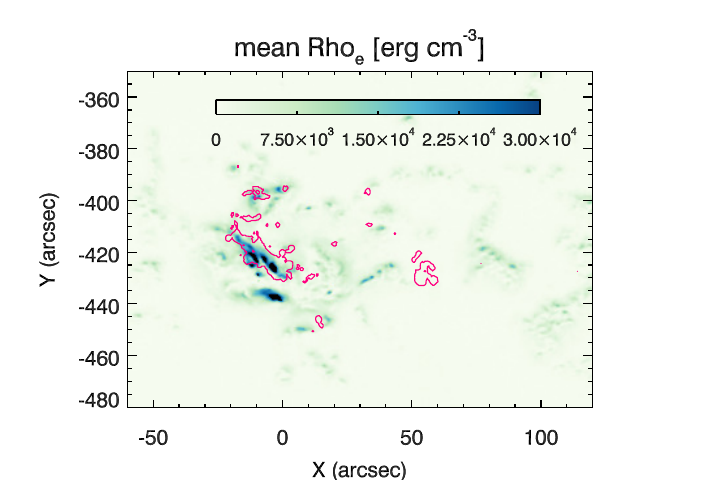}
\includegraphics[width=0.5\textwidth, clip, trim = 5mm 0mm 19mm 11mm]{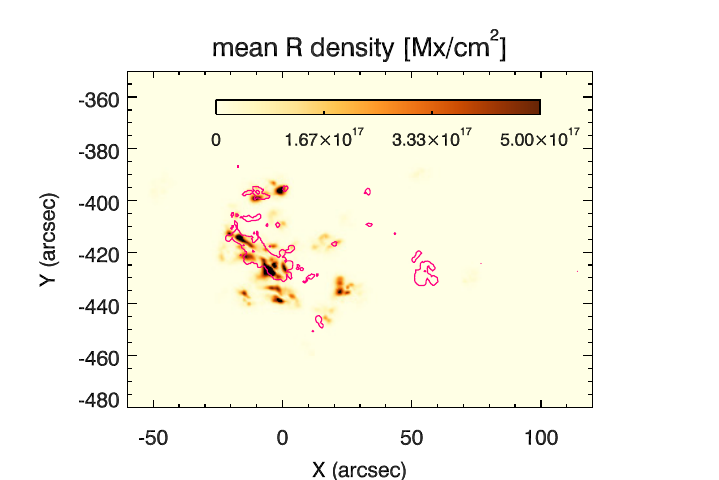}
\caption{Average excess magnetic energy density $\rho_{e}$ (top) and average magnetic flux in strong-gradient PIL areas $r$ (bottom) for AR 11123 on November 12, 2010 during the active epoch. The pink contours outline all transient brightenings detected over the same textbf{analysis} interval.}
\label{fig:efree_schrijver_tbs}
\end{figure}
Complementing the coronal magnetic topological skeleton, we place the TBs into the broader magnetic landscape through the analysis of the vector components of the photospheric magnetic field, as provided by HMI. Here we focus on two (of many, c.f. \citet{params, Leka:2018, Georgoulis:2021}) 
ways of characterizing the locally stored excess magnetic energy and propensity for magnetic reconnection \citep{mag_data}.
We evaluate the $R$ parameter \citep{Schrijver2007}, a measure to identify strong-field, high-gradient polarity inversion lines which are proxies of (near-) photospheric electrical currents. $R$ is defined as total unsigned magnetic flux within $\sim15$~Mm of strong-field, high-gradient PILs; following \citet{Leka:2018}, we compute $R$ using physical distances rather than a static number of pixels as was done originally.  A Gaussian-weighted mask of distance from the strong-gradient PILs is multiplied by the magnetic flux at each pixel (``$p$'') $\phi_{p}$ to define a density (the spatial integral of which is $R$) against which we can compare the spatial association of $r(p)$ with the TBs.  

The photospheric excess magnetic energy density $\rho_{e}$ \citep{params} is defined as the difference between the observed ${\bf B}^{o}$ and the potential fields ${\bf B}^{p}$ on a pixel-by-pixel basis
\begin{equation}
    \rho_{e}(p)=({\bf B}^{p}(p)-{\bf B}^{o}(p))^{2}/8\pi \;,
\end{equation}
and therefore represents an estimate of the energy available for solar energetic events. The total excess energy $E_{e}=\sum \rho_{e}\,dA$ computed over the active region, consistently ranks as one of the best single-performing parameters in discriminating flare-imminent from flare-quiet active regions, indicating that active regions with substantial magnetic ``free energy'' are more likely to produce major solar flares \citep{dfa3,Welsch_etal_2009,BobraCouvidat2015,Florios_etal_2018}.

Both parameters provide complementary information to the topological features introduced in Sec.~\ref{sec:topology} which are based on the evaluation of a potential field. For example, neither bald patches nor fan traces of coronal null points are required to be associated with strong-gradient PILs, as is the $R$ parameter.  The excess magnetic energy density is a measure with respect to the potential field, therefore it is not available by the information from the skeleton itself.

We calculate average maps of both parameters using all available HMI vector magnetograms over the analysis intervals of each epoch.
Figure~\ref{fig:efree_schrijver_tbs} shows an example of the excess energy density $\rho_{e}(p)$ (top) and the spatial map of flux that would contribute to $R$, or the $r(p)$ ``density'' (bottom), of NOAA active region 11123 on November 12, 2010, during the active epoch. The pink contours outline the detected TBs during the same analysis interval. Some TB clusters are co-spatial with the high values of $\rho_{e}(p)$ and $r(p)$, indicating the availability of energy in magnetic configurations known to be associated with magnetic reconnection.
\section{Results}\label{sec:results}
In the following, qualitative comparisons between transient brightenings, their location within the magnetic skeleton, and magnetic and plasma environments are shown for three events of interest including their matched quiet epochs (Table~\ref{tab:events}). 
This serves as an overview of regimes that are considered; detailed quantitative comparisons for a large statistical sample will be presented in a future paper of the series.
\subsection{October 16, 2010}
\begin{figure*}
    \centering
    \includegraphics[width=1.0\linewidth]{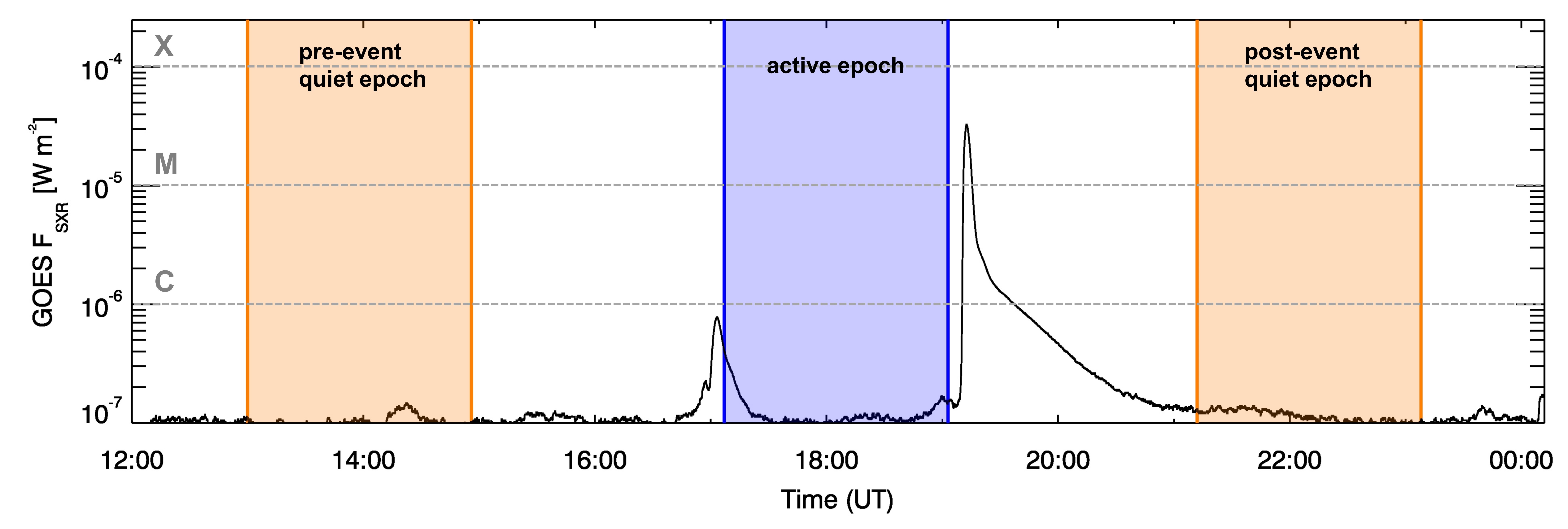}
    \caption{Overview of the analysis epochs for the October 16, 2010 event. GOES/XRS soft X-ray 1--8\AA~time evolution during the time period of interest. The colored areas indicate the active epoch (blue; during the pre-flare phase of the associated M2.9 flare), as well as pre- and post-event quiet epochs (orange). We note that flares below C-class are not considered as events.}
    \label{fig:goes_epochs_20101016}
\end{figure*}

\begin{figure*}
\centering{
\includegraphics[width=0.329\textwidth,clip, trim = 0mm 2mm 24mm 14mm]{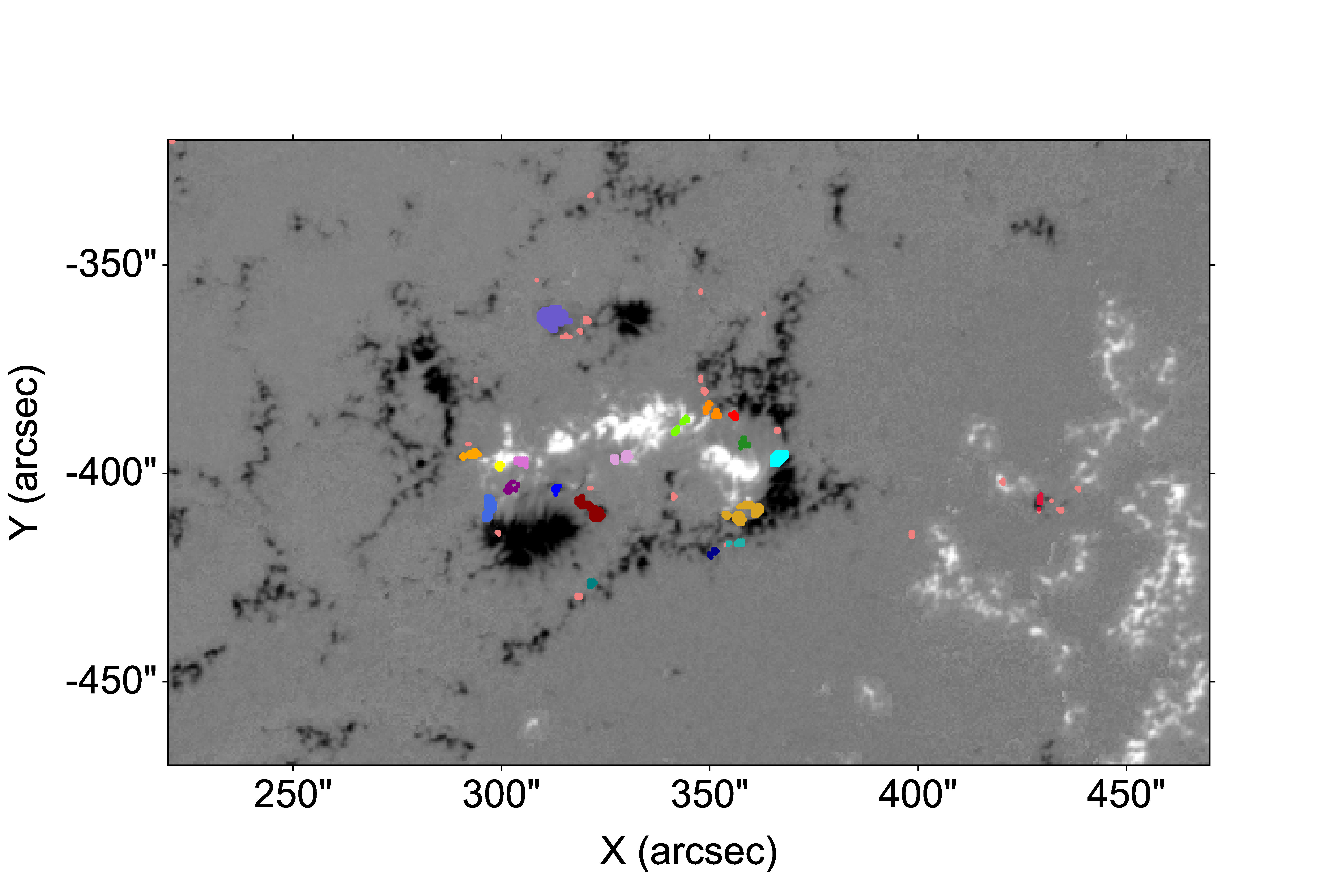}
\includegraphics[width=0.329\textwidth,clip, trim = 0mm 2mm 24mm 14mm]{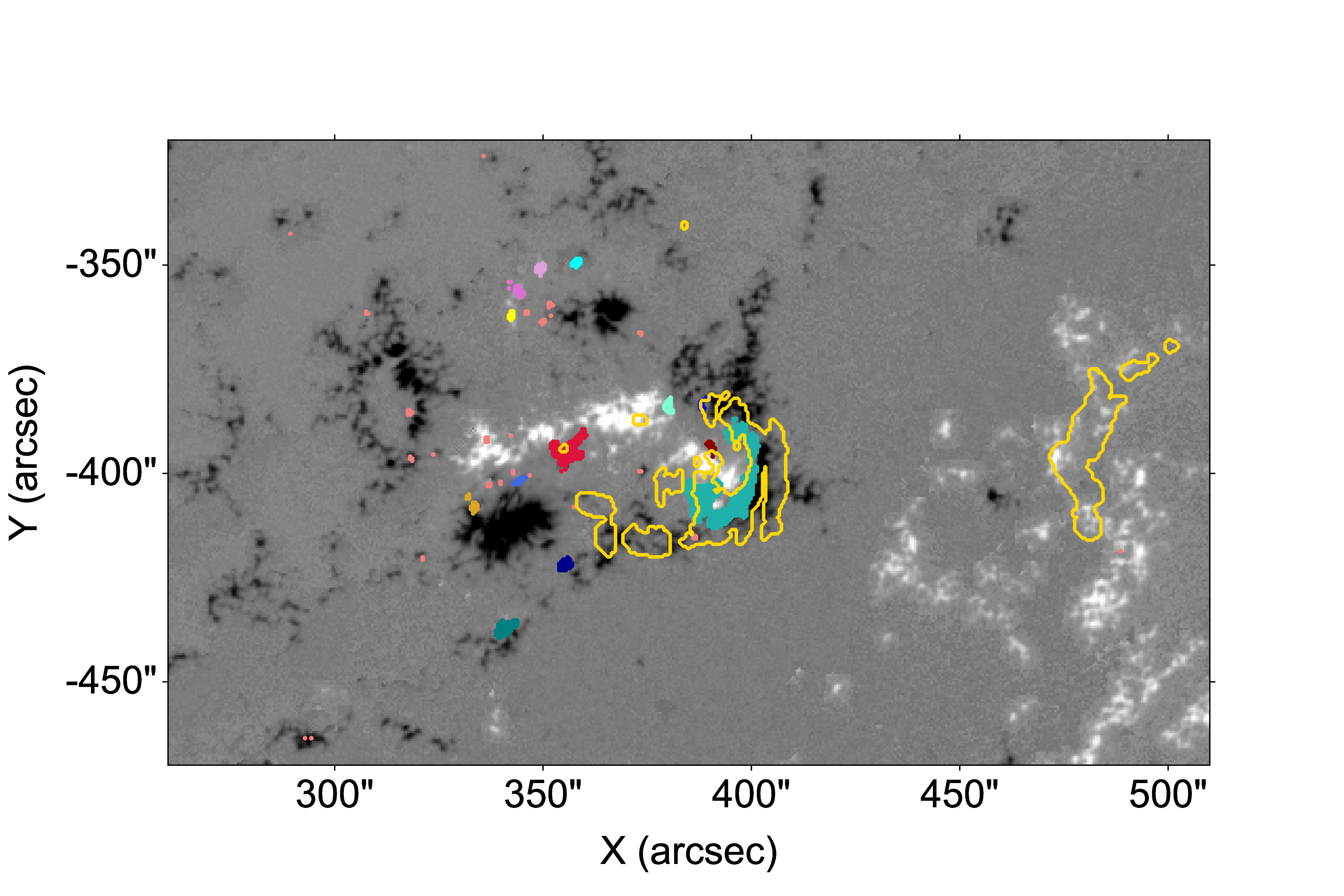}
\includegraphics[width=0.329\textwidth,clip, trim = 0mm 2mm 24mm 14mm]{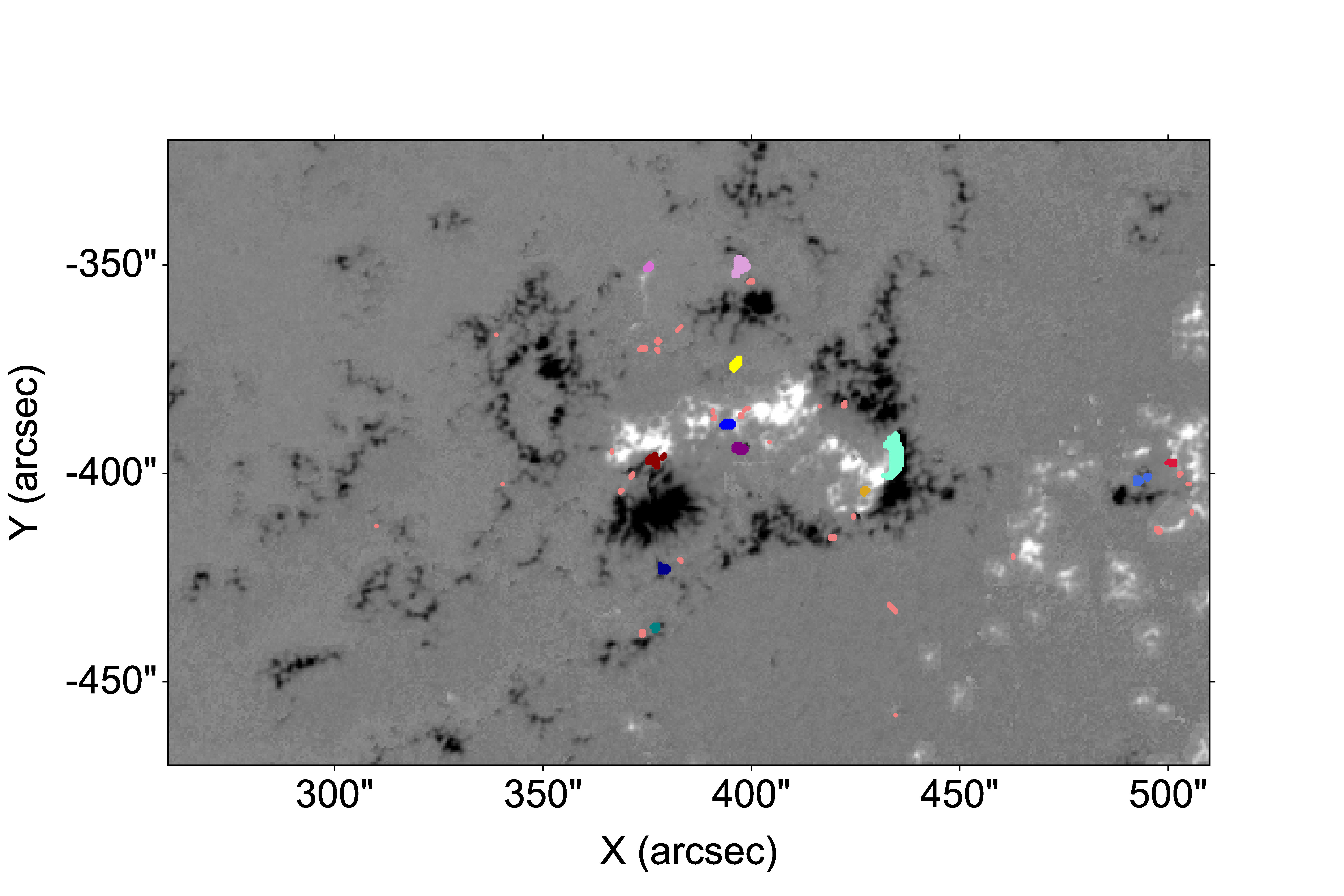}}
\centerline{
\includegraphics[width=0.33\textwidth,clip, trim = 4mm 8mm 15mm 14mm]{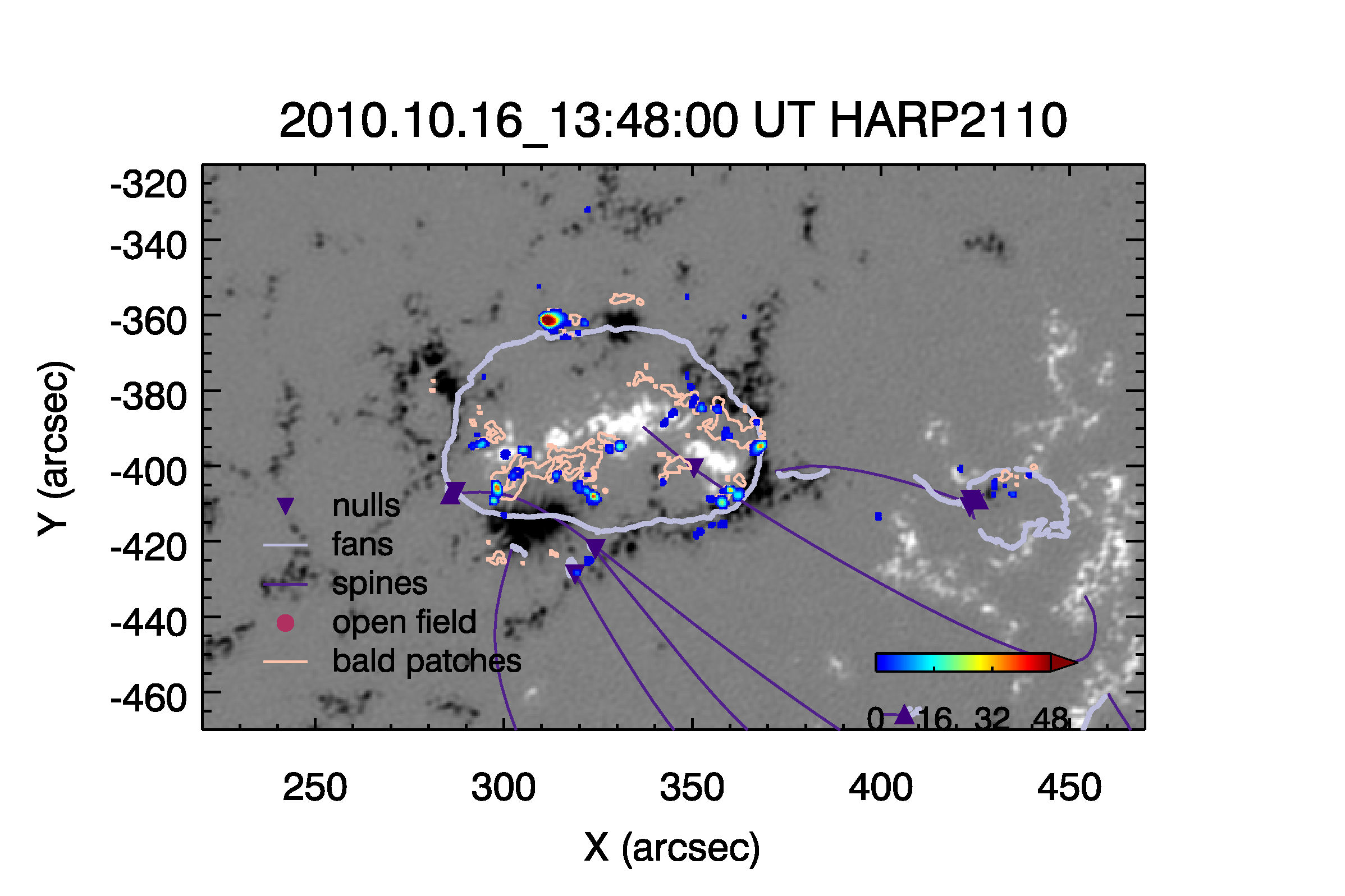}
\includegraphics[width=0.33\textwidth,clip, trim = 4mm 8mm 15mm 14mm]{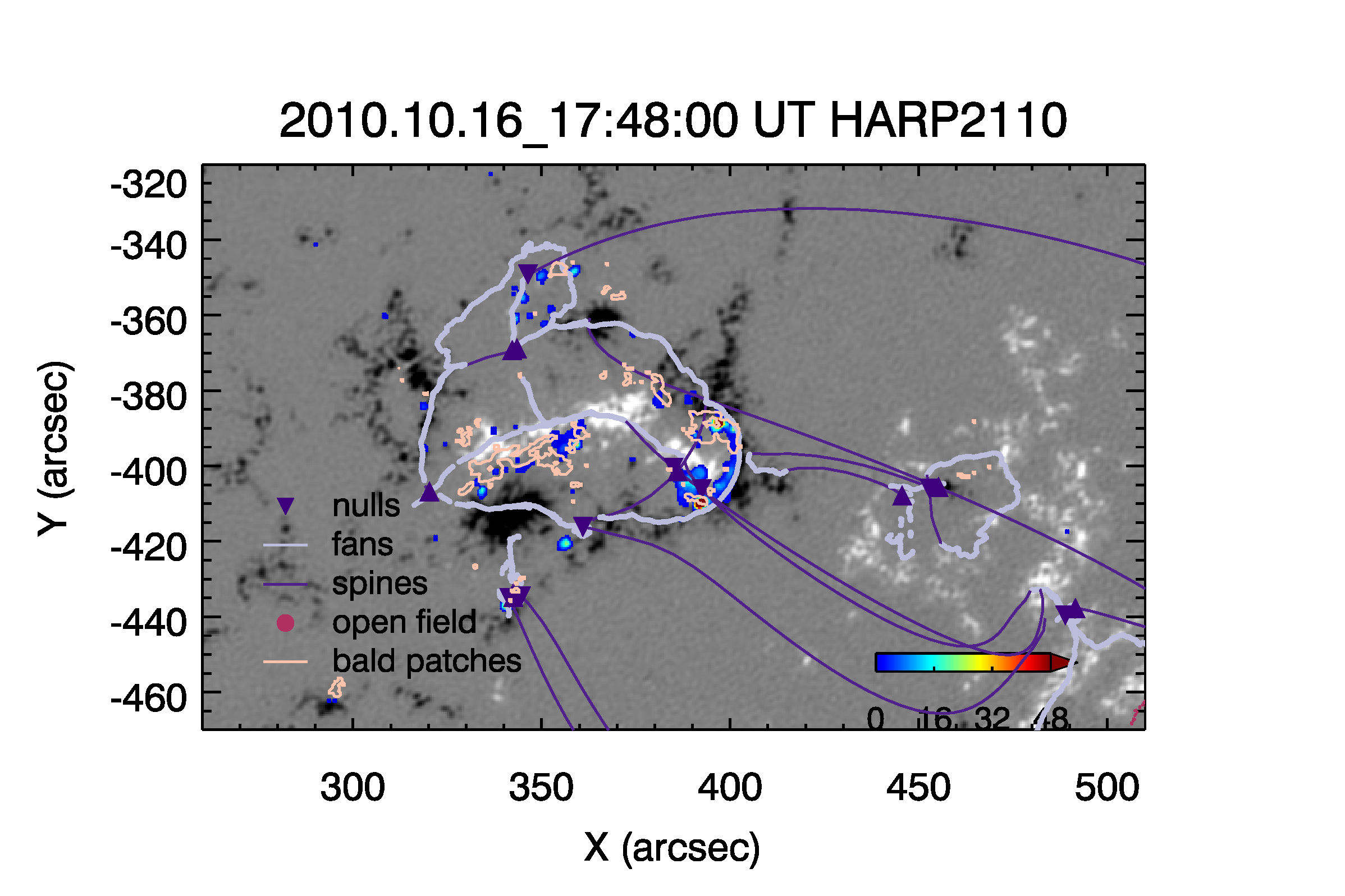}
\includegraphics[width=0.33\textwidth,clip, trim = 4mm 8mm 15mm 14mm]{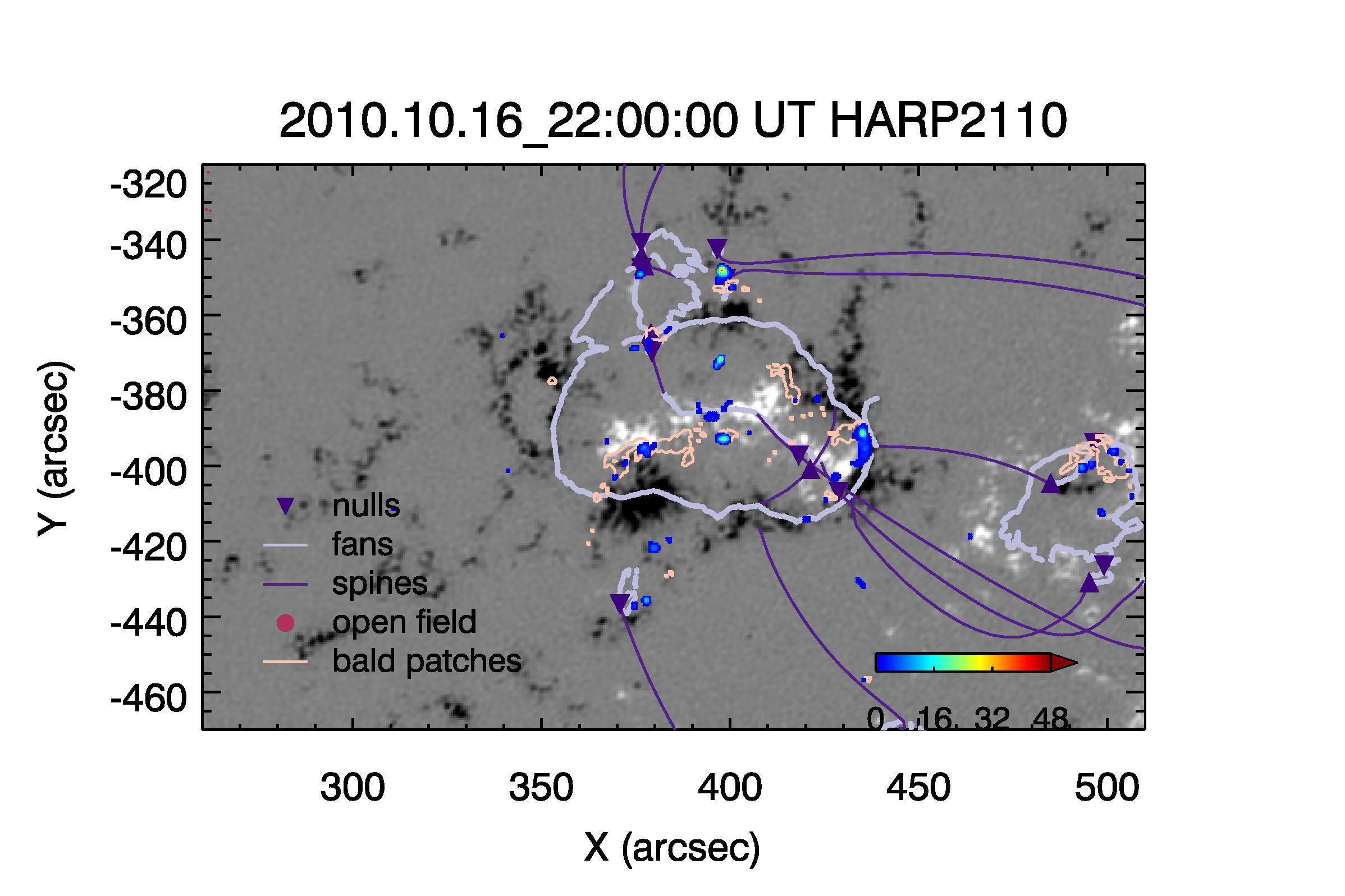}}
\centerline{
\includegraphics[width=0.33\textwidth,clip, trim = 4mm 8mm 15mm 14mm]{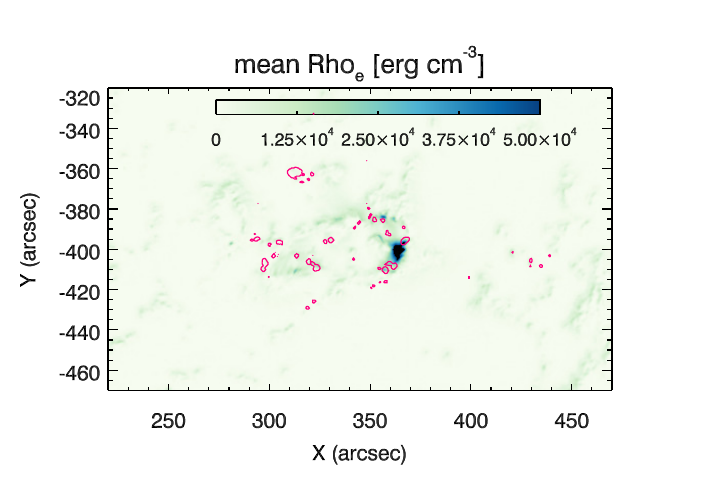}
\includegraphics[width=0.33\textwidth,clip, trim = 4mm 8mm 15mm 14mm]{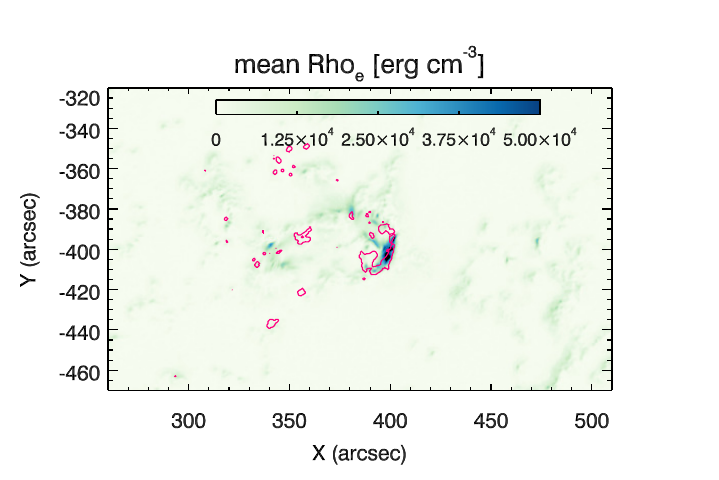}
\includegraphics[width=0.33\textwidth,clip, trim = 4mm 8mm 15mm 14mm]{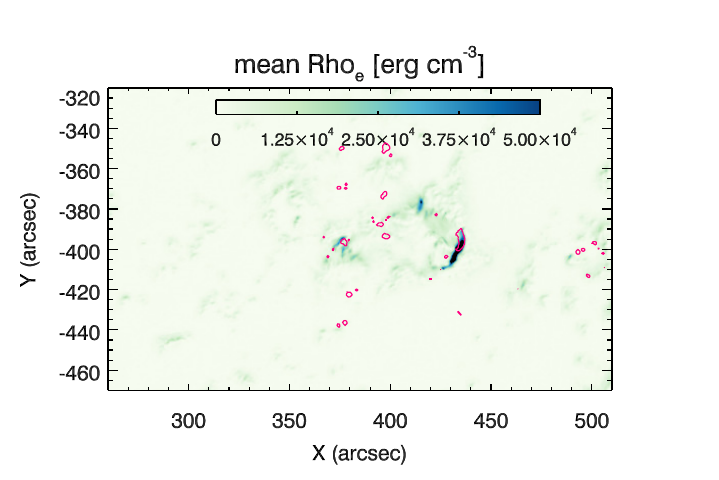}}
\centerline{
\includegraphics[width=0.33\textwidth,clip, trim = 4mm 3mm 15mm 14mm]{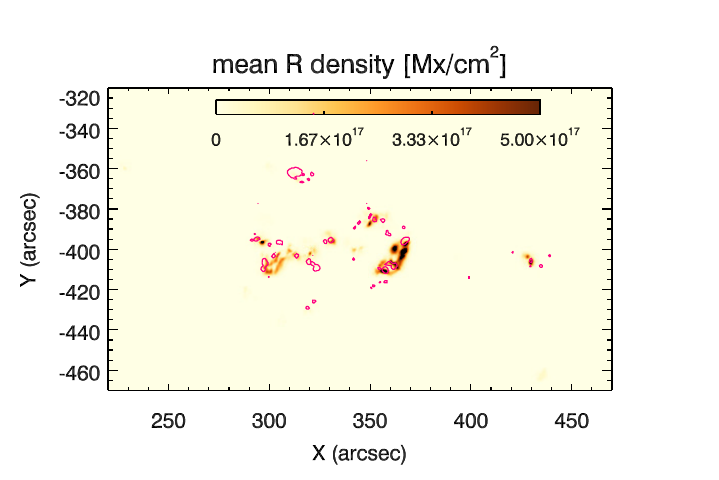}
\includegraphics[width=0.33\textwidth,clip, trim = 4mm 3mm 15mm 14mm]{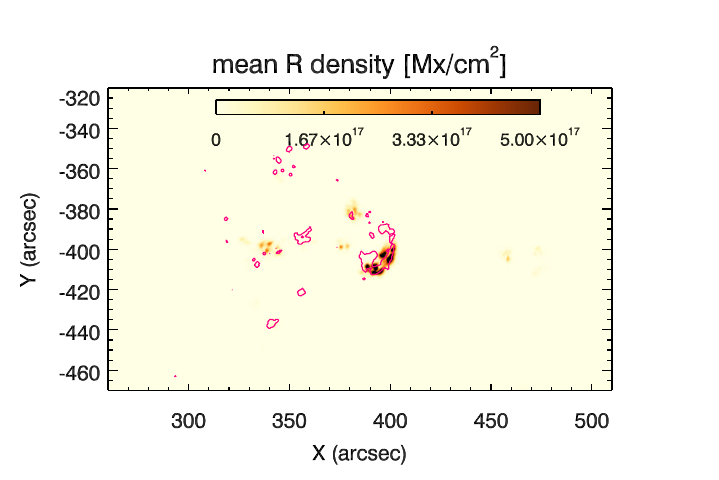}
\includegraphics[width=0.33\textwidth,clip, trim = 4mm 3mm 15mm 14mm]{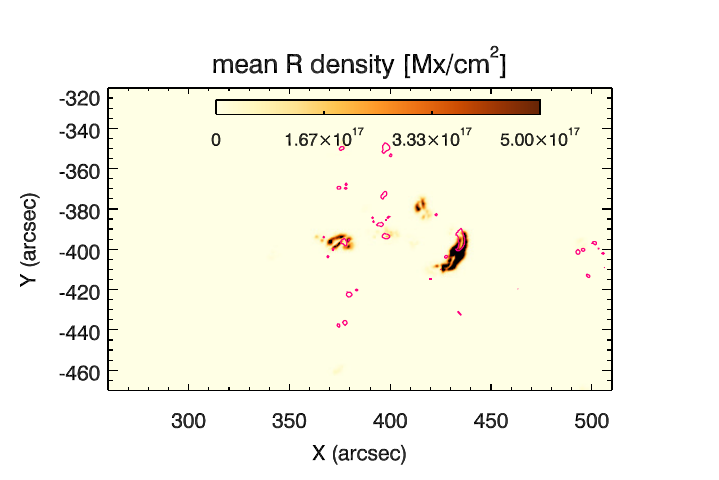}}
\vspace{-0.7cm}
\caption{Transient brightenings together with the magnetic skeleton of the active region and the magnetic environment for the active epoch in AR 11112 on October 16, 2010 prior to a M2.9 flare (middle column; 17:07--19:03~UT) and two matched quiet epochs, one prior to the event (left column, pre-event quiet, i.e., 13:00--14:56~UT) and one after the event (right column, post-event quiet, i.e., 21:12--22:08~UT). \textbf{First row:} DBSCAN clustering results of the cumulative TB map during each epoch. Each identified cluster appears in a different color, light rose pixels indicate noise, not corresponding to any cluster. The radial component of the magnetic field $B_{r}$ of the vector magnetogram is plotted in the background for reference. The golden contour in the middle panel outlines the flare ribbons during the rise time of the associated flare. \textbf{Second row:} Transient brightenings (in rainbow color, see Fig.~\ref{fig:detection_tbs} for details) together with elements of the magnetic skeleton of the host active region (see legend). The reconstruction time of each skeleton is listed in Table~\ref{tab:events} for each epoch. The radial component of the magnetic field $B_{r}$ from the PFSS model using spherical harmonics up to degree 2800 is plotted in the background. \textbf{Third row:} Excess magnetic energy density $\rho_{e}(p)$ averaged over the two-hour analysis intervals of the different epochs. The pink contours outline the cumulative TB map during each epoch. \textbf{Fourth row:} Magnetic flux in strong-gradient PIL areas $r(p)$ averaged over the two-hour analysis intervals of the different epochs. The pink contours outline the cumulative TB map during each epoch.}
\label{fig:combined_topology_20101016}
\end{figure*}

\begin{figure*}
    \centerline{
    \includegraphics[width=0.33\textwidth, clip, trim = 4mm 6mm 15mm 14mm]{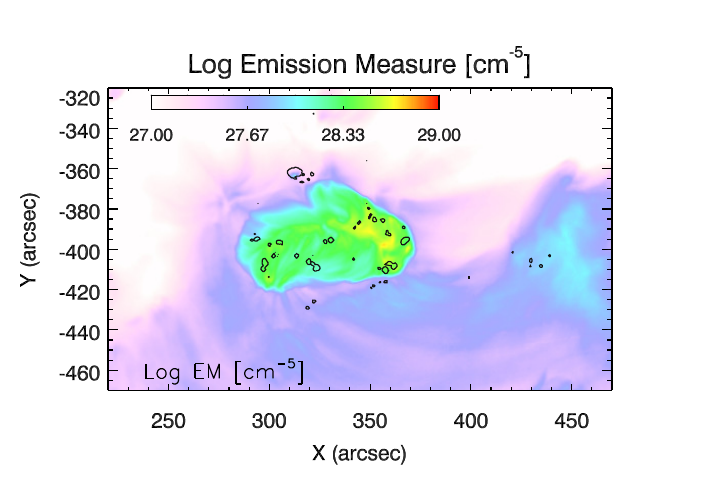}
    \includegraphics[width=0.33\linewidth, clip, trim = 4mm 6mm 15mm 14mm]{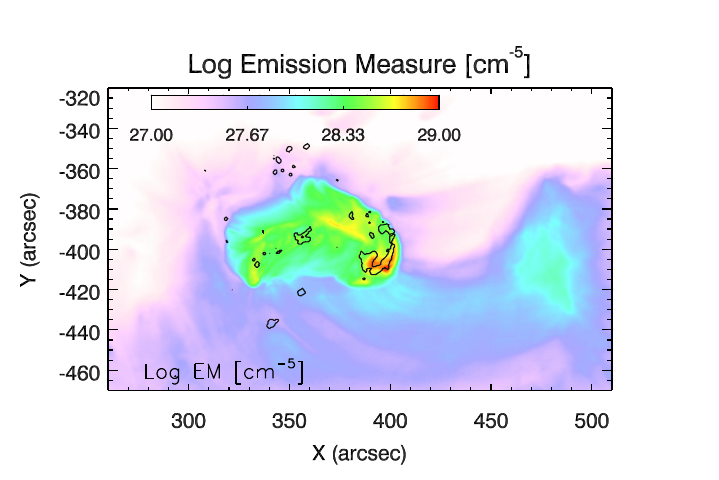}
    \includegraphics[width=0.33\linewidth, clip, trim = 4mm 6mm 15mm 14mm]{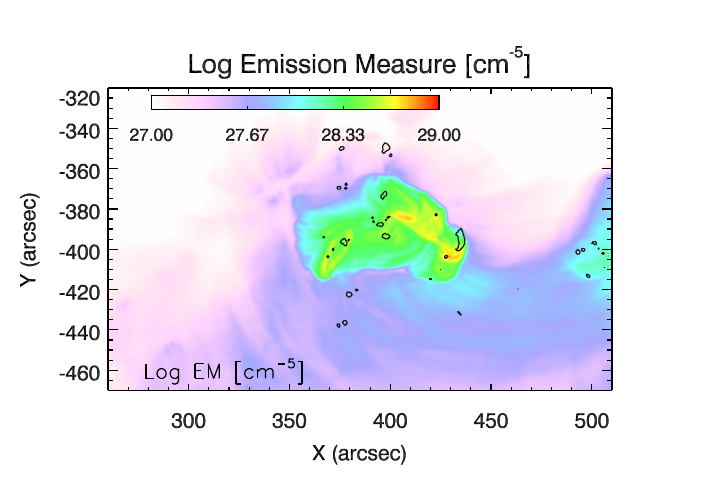}}
    \centerline{
    \includegraphics[width=0.33\textwidth, clip, trim = 4mm 6mm 15mm 14mm]{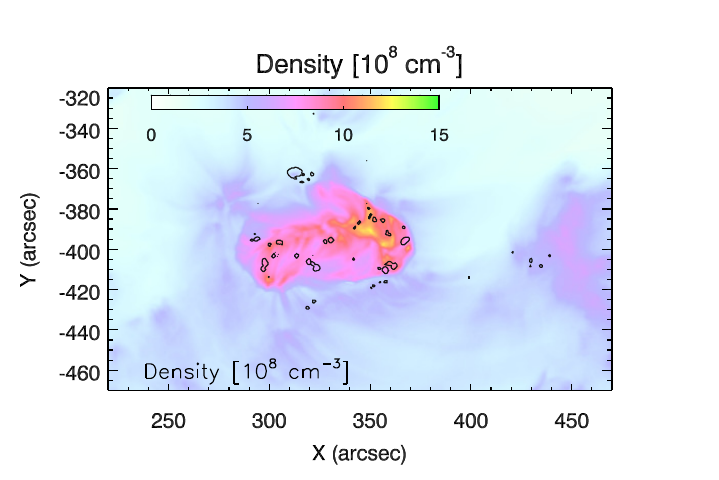}
    \includegraphics[width=0.33\linewidth, clip, trim = 4mm 6mm 15mm 14mm]{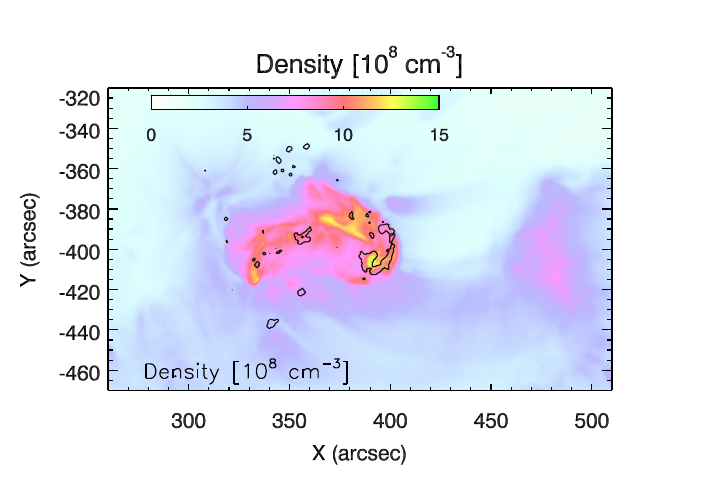}
    \includegraphics[width=0.33\linewidth, clip, trim = 4mm 6mm 15mm 14mm]{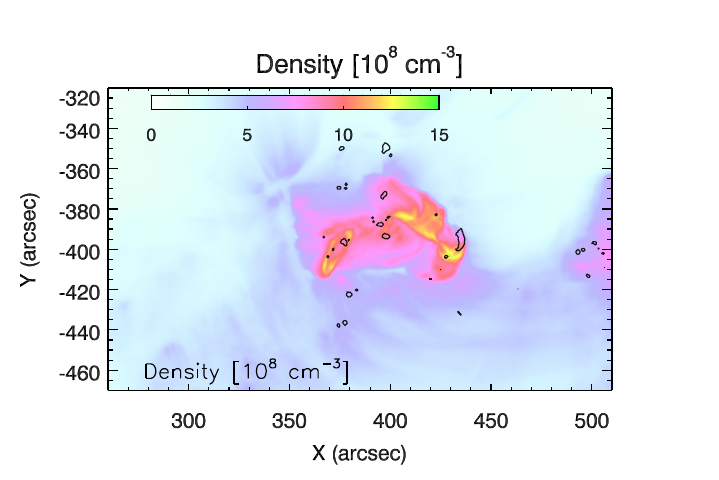}}
        \centerline{
    \includegraphics[width=0.33\textwidth, clip, trim = 4mm 2mm 15mm 14mm]{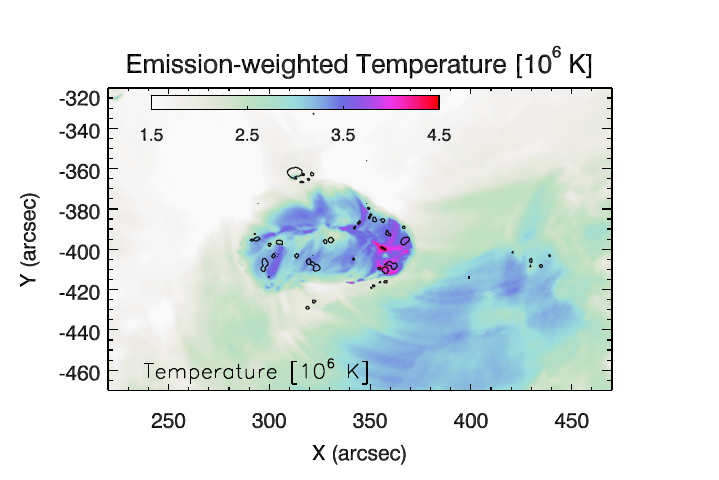}
    \includegraphics[width=0.33\linewidth, clip, trim = 4mm 2mm 15mm 14mm]{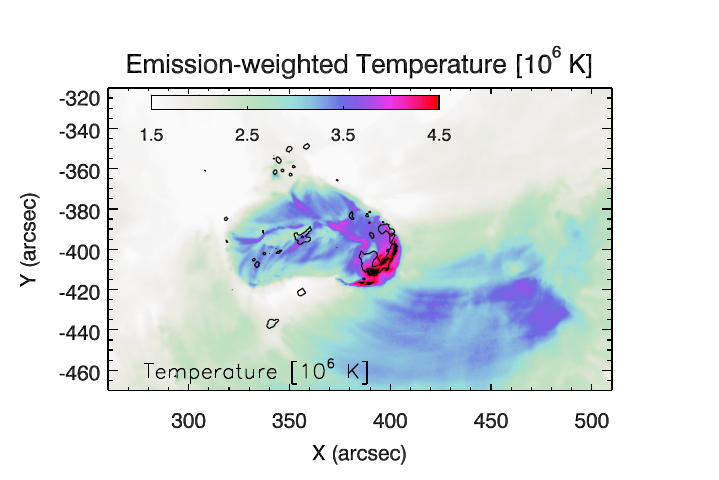}
    \includegraphics[width=0.33\linewidth, clip, trim = 4mm 2mm 15mm 14mm]{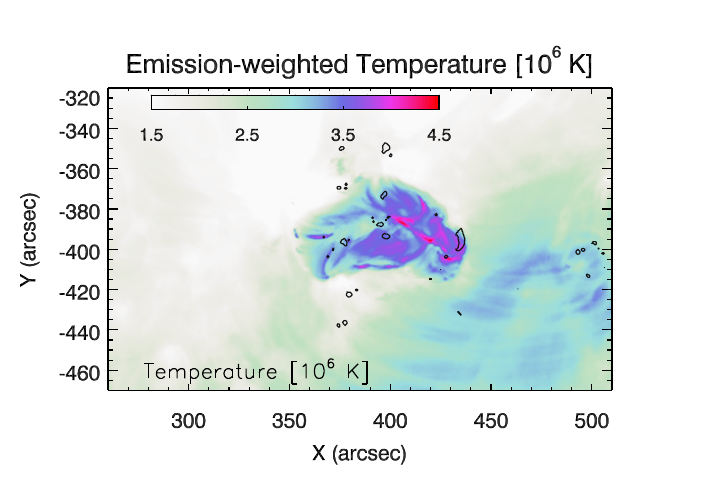}}
    \caption{Transient brightenings together with the plasma environment for the active epoch in AR 11112 on October 16, 2010 prior to a M2.9 flare (middle column; 17:07--19:03~UT) and two matched quiet epochs, same as in Fig.~\ref{fig:combined_topology_20101016} (left column: pre-event quiet epoch, i.e., 13:00--14:56~UT; right column: post-event quiet epoch, i.e., 21:12--22:08~UT). DEM reconstructed emission measure (first row, in log), density (second row) and temperature (third row) averaged over the two-hour analysis intervals of the different epochs. The black contours outline all TBs detected during each epoch, respectively.}
    \label{fig:dem_20101016}
\end{figure*}

The M2.9 flare occurring in NOAA AR 11112 on October 16, 2010 has been classified as circular-ribbon flare, starting at 19:07~UT, and reaching its peak intensity at 19:12~UT \citep[][]{Chen:2020}.
Figure~\ref{fig:goes_epochs_20101016} shows the GOES/XRS soft X-ray flux evolution for an extended time period around this flare to illustrate the selection of the different epochs transient brightenings are studied in. Highlighted in blue is the pre-flare, active epoch, and two time periods prior to (after) the active epoch, i.e., the pre-event (post-event) quiet epochs in orange.

Figure~\ref{fig:combined_topology_20101016} illustrates transient brightening activity together with the topological skeleton and the magnetic environment during the active epoch (middle column) and the quiet epochs (left column: pre-event quiet, right column: post-event quiet).
The first row shows the results of the clustering algorithm DBSCAN applied to the TB heat maps of each epoch. Different colors represent different clusters of TBs identified by the method. 
The golden contour in the middle panel represents the flare ribbons identified during the flare rise time (i.e., the time period from GOES soft X-ray flare start to peak time). In addition to the main semi-circular flare ribbon, a remote ribbon associated with the scattered positive fluxes in the facular region to the West is observed \citep{Chen:2020}. The active epoch is characterized by the appearance of multiple TB clusters of varying sizes, ranging from a few to several tens of arcseconds. One large cluster clearly exceeds the size of the others, and it is
co-spatial with later flare ribbons in the core of the active region. TBs are not observed at the remote flare ribbon. While the quiet epochs show even more TB clusters compared to the active epoch, they are all smaller in size. No systematic differences in the clustering behavior between the two quiet epochs is identified.

The second row of Figure~\ref{fig:combined_topology_20101016} shows the location of transient brightenings as a heat map together with the magnetic topological skeleton, consisting of coronal null points, their spines and fan traces, open field footpoints and bald patch locations (see figure legend).
For all epochs, the majority of TBs occur within or close to the fan traces of the central null domes.
For the active epoch, the largest cluster partially lies within a fan trace, but in close proximity to its border. It contains the only two compact hot spot areas (defined as TBs detected for more than half of the time within the epoch, indicated in red) observed within the FOV. Both hot spot areas are co-spatial with bald patches. Most of the other, smaller TB clusters during this epoch occur within fan traces of nulls and are sometimes co-spatial with bald patches.
During the quiet epochs, the majority of TBs occur as well within or in close proximity to fan traces, some are co-spatial with PILs that exhibit short, spatially intermittent lengths of bald patches, while a few show no correspondence to a topological feature at all; it may be that some TBs that do not correspond to any topological feature are associated with the intersection of a bald patch separatrix surface away from the PIL, which we do not locate (see \S\ref{sec:BPs}).
We note that more, compact TB clusters are observed in quiet epochs compared to the active one but not more hot spot areas. Most of these clusters are co-spatial with bald patches or fan traces, or both but a systematic pattern could not be identified.
The footpoints of spine field lines do not seem to be preferential locations for TBs in any analyzed epoch.
No open field regions of interest are present within the field-of-view of this active region during the investigated epochs.

The third and fourth rows of Figure~\ref{fig:combined_topology_20101016} show the excess magnetic energy density $\rho_{e}(p)$ and the magnetic flux in strong-gradient PIL areas $r(p)$.  
For all the epochs, the majority of TBs do not correspond to high-$\rho_{e}(p)$ values.
However, the largest TB cluster for the active epoch almost fully encloses the high-$\rho_{e}(p)$ area, and we note the spatial agreement of a big cluster in the post-event quiet epoch with the largest values of $\rho_{e}(p)$. 
This is slightly different for $r(p)$, where some small clusters during the quiet epochs match locations of interest of the parameter as well. But also here, the majority of TBs do not spatially co-align with strong-gradient PIL areas. Good spatial agreement with high-$r(p)$ areas is found for the largest TB cluster during the active epoch. Hot spot areas do not seem to occur at preferred locations of these parameters.

Figure~\ref{fig:dem_20101016} shows the plasma environment of AR 11112 during the three analysis epochs in the form of average emission measure (first row), density (second row), and emission-weighted temperature (third row). During the quiet epochs, the majority of TBs occur in regions of enhanced emission measure ($\sim 10^{28}$cm$^{-5}$), but not in the highest ones. Similarly, most TBs occur in areas with unremarkable values of temperature and density, averaging $\sim3$MK and $\sim 7.0 \times 10^{8}$cm$^{-3}$. Smaller clusters associated with low emission measure locations also exist. During the active epoch, the highest emission measure and temperature values over the investigated analysis intervals are observed. These regions are associated with the large cluster of TBs. Parts of this cluster correspond to either hot ($>4.0$MK) or dense regions ($>1.0\times10^{9}$cm$^{-3}$), indicative of plasma accumulation and plasma heating. Again hot spot areas or locations of enhanced activity of transient brightenings are not exceptional locations in terms of emission measure, density or temperature. 
\subsection{November 12, 2010}
\begin{figure*}
    \centering
    \includegraphics[width=1.0\linewidth]{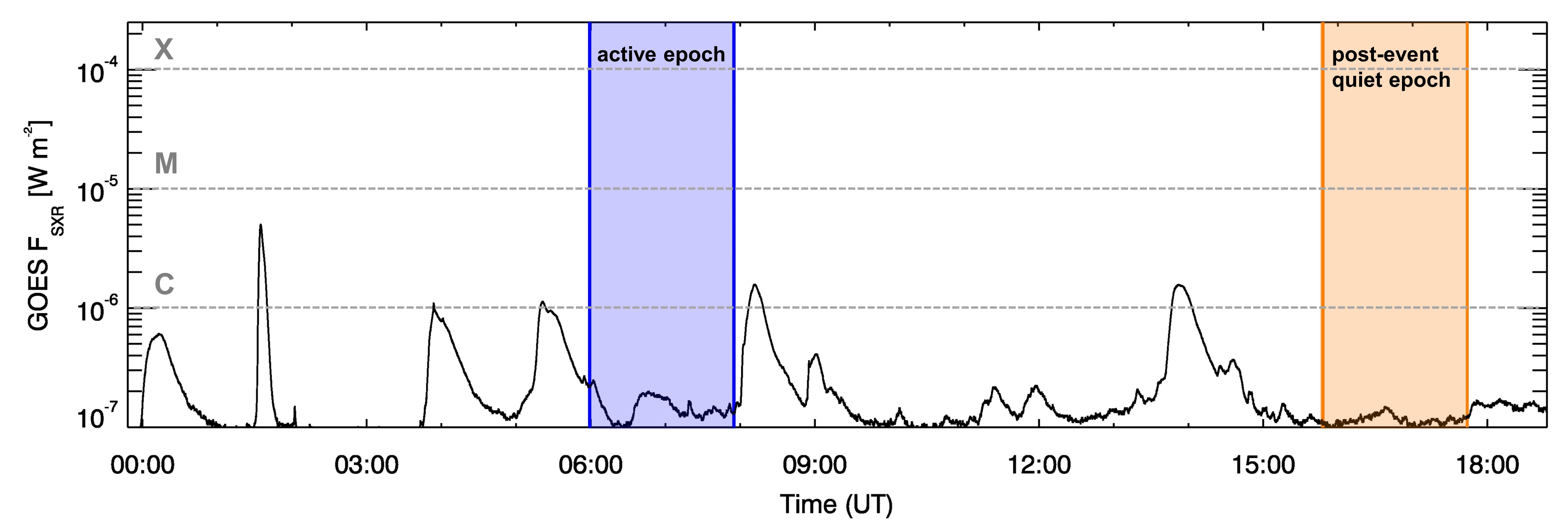}
    \caption{Overview of the analysis epochs for the November 12, 2010 event. GOES/XRS soft X-ray 1--8\AA~time evolution during the time period of interest. The colored areas indicate the active epoch (blue; during the pre-flare phase of the associated C1.5 flare) and the post-event quiet epoch (orange). No valid pre-event quiet epoch could be identified for this event.}
    \label{fig:goes_epochs_20101112}
\end{figure*}

\begin{figure*}
\centerline{
\includegraphics[width=0.33\textwidth, clip, trim = 6mm 2mm 19mm 12mm]{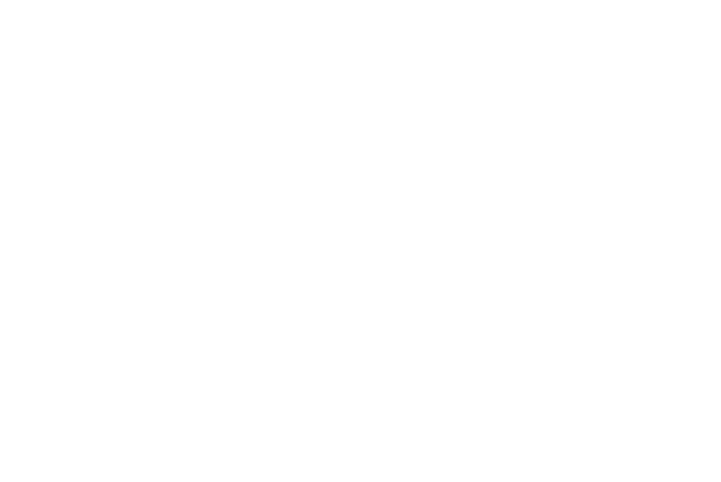}
\includegraphics[width=0.33\textwidth, clip, trim = 6mm 0mm 35mm 23mm]{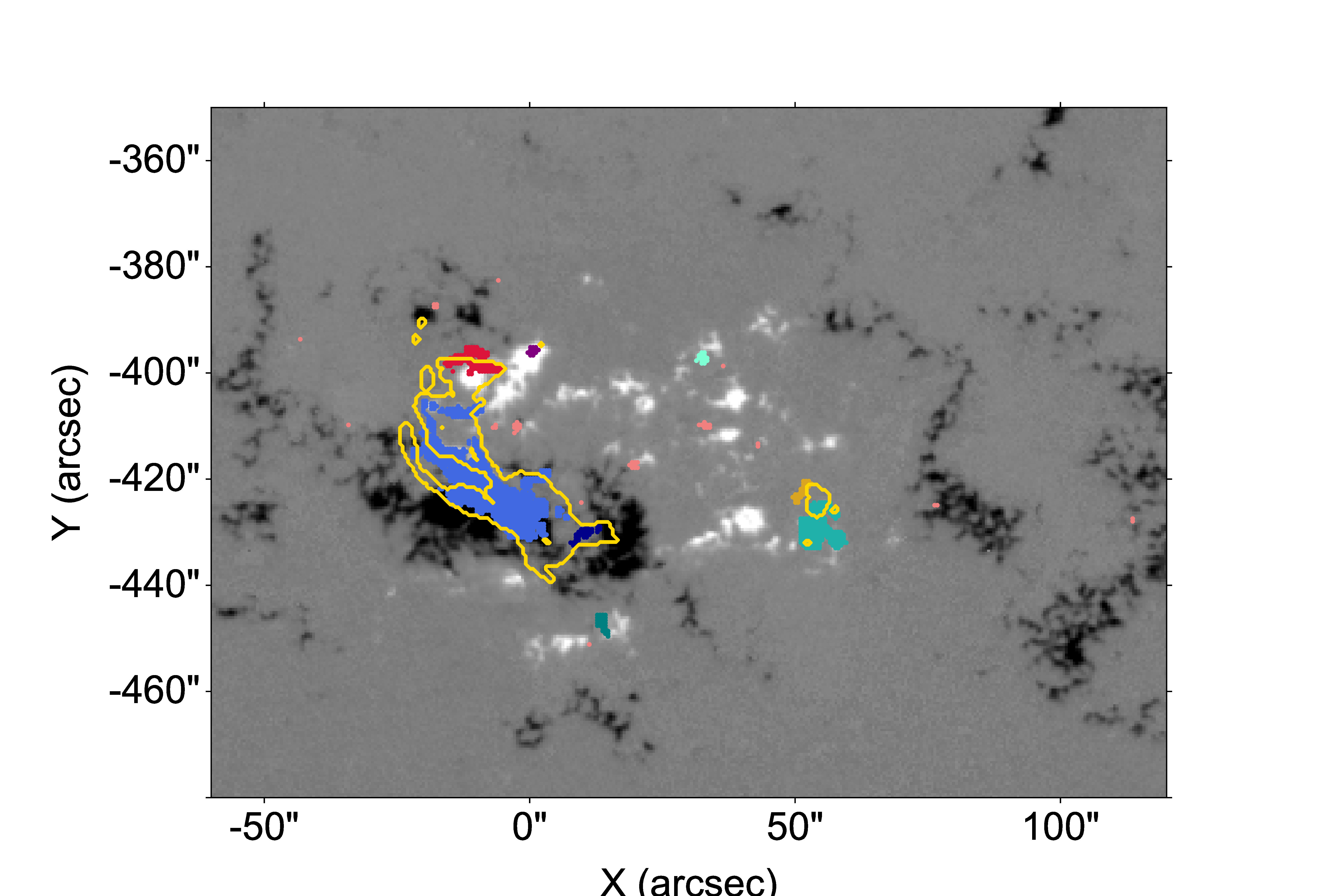}
\includegraphics[width=0.33\textwidth, clip, trim = 6mm 0mm 35mm 23mm]{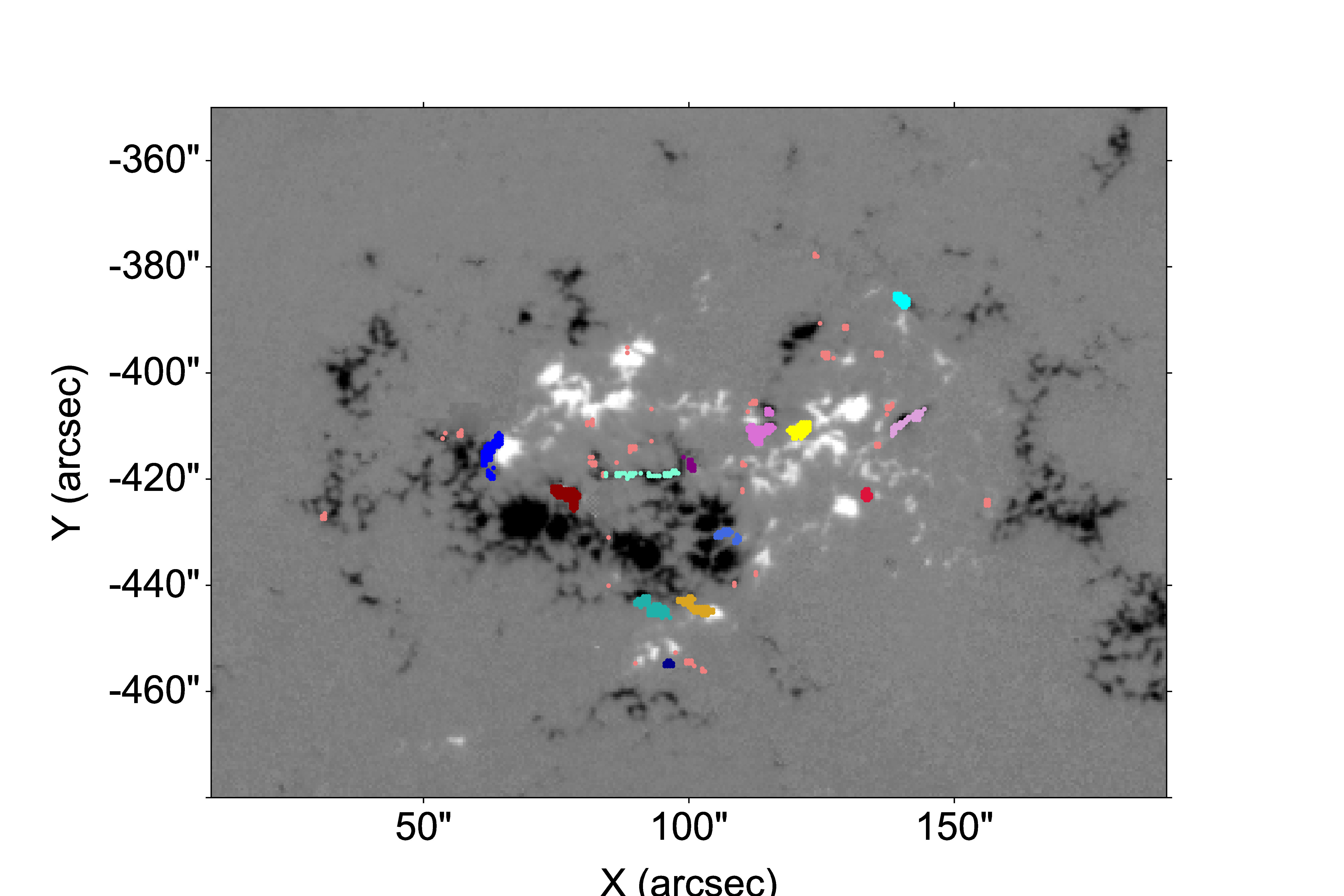}}
\centerline{
\includegraphics[width=0.33\textwidth, clip, trim = 6mm 2mm 19mm 12mm]{20101112_blank_space.pdf}
\includegraphics[width=0.33\textwidth, clip, trim = 6mm 4mm 19mm 11mm]{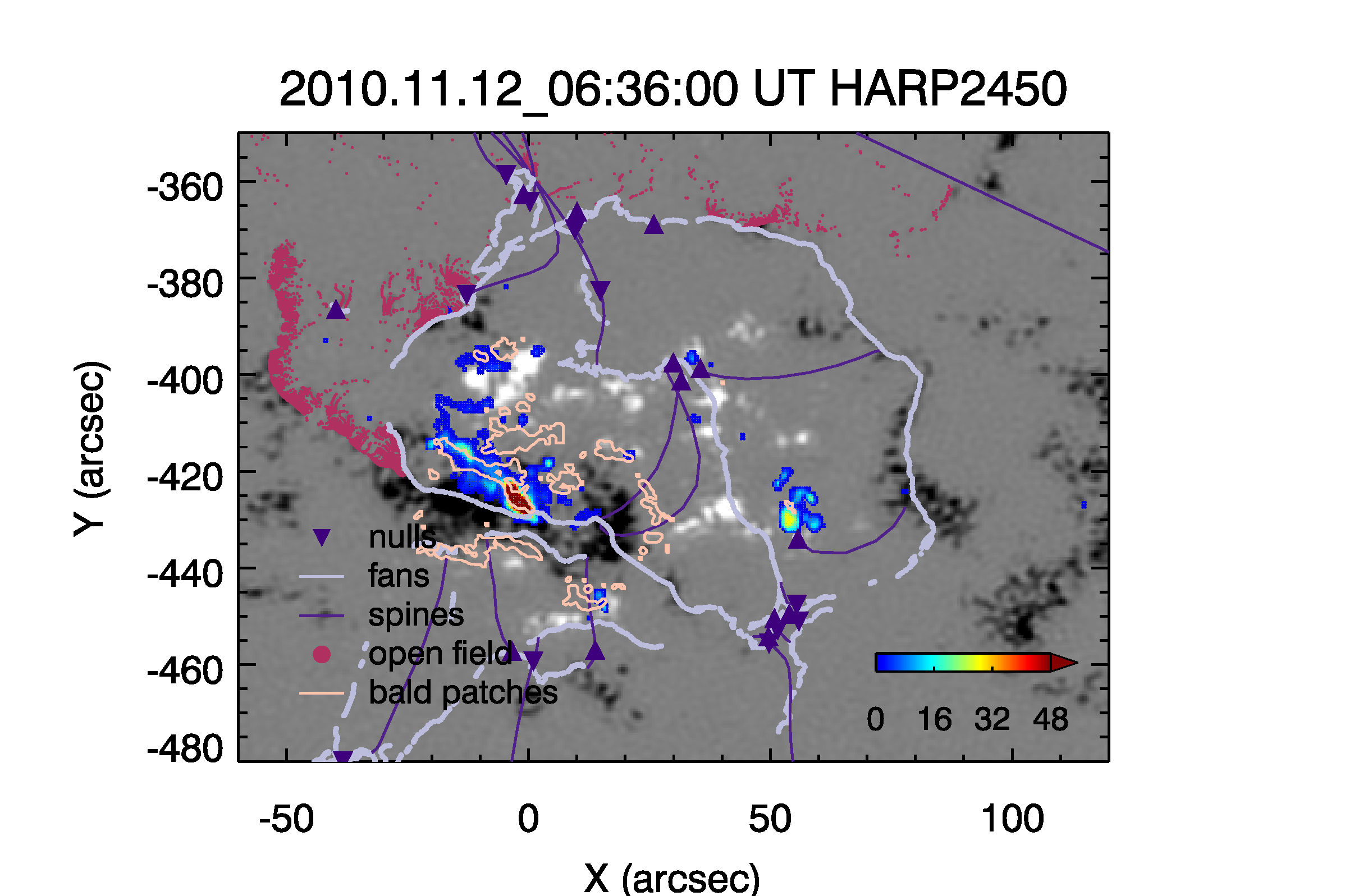}
\includegraphics[width=0.33\textwidth, clip, trim = 6mm 4mm 19mm 11mm]{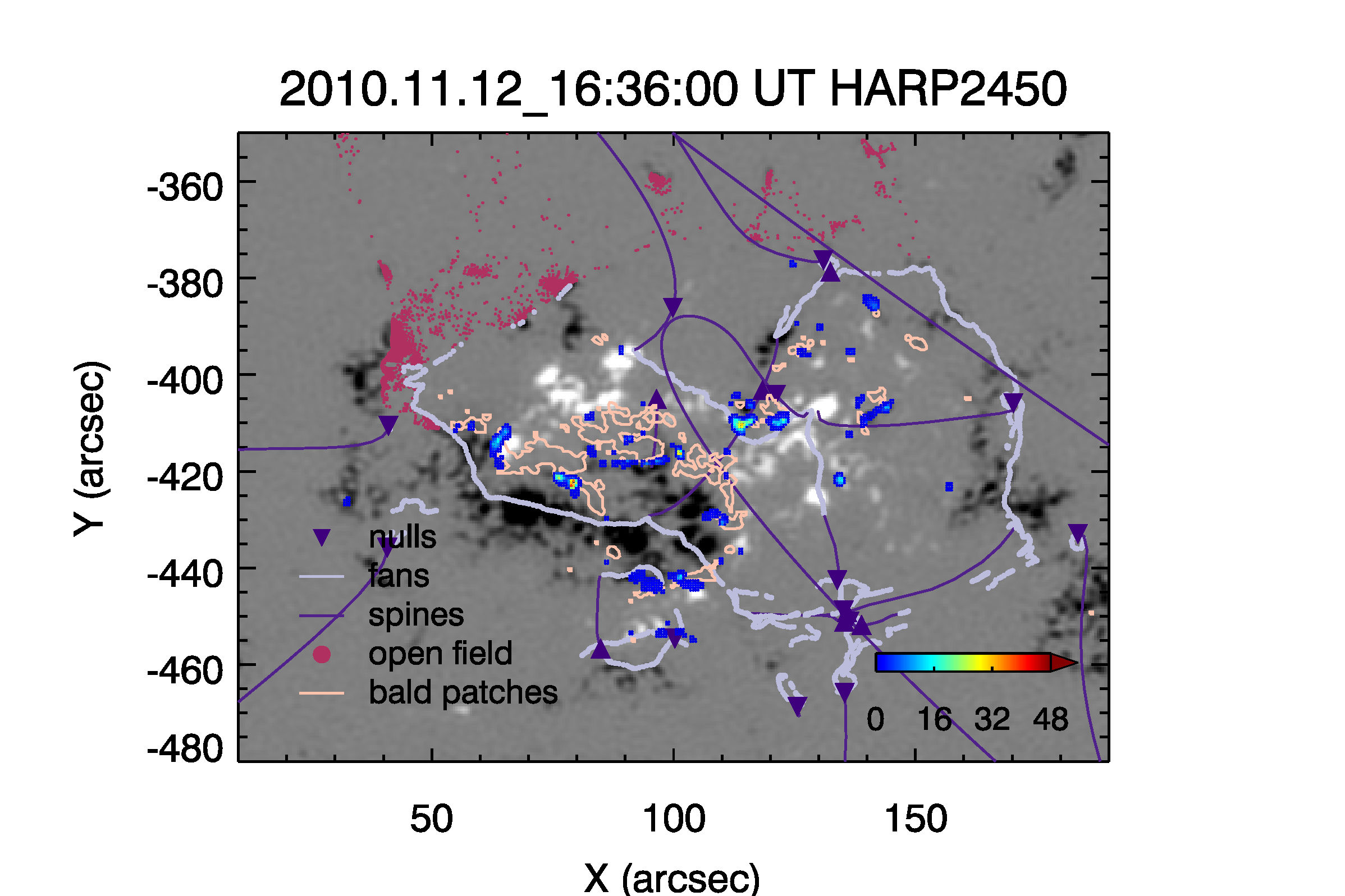}}
\centerline{
\includegraphics[width=0.33\textwidth, clip, trim = 6mm 2mm 19mm 12mm]{20101112_blank_space.pdf}
\includegraphics[width=0.33\linewidth, clip, trim = 6mm 4mm 19mm 12mm]{20101112_0559_245_efree_proxy_heat_map_update.pdf}
\includegraphics[width=0.33\linewidth, clip, trim = 6mm 4mm 19mm 12mm]{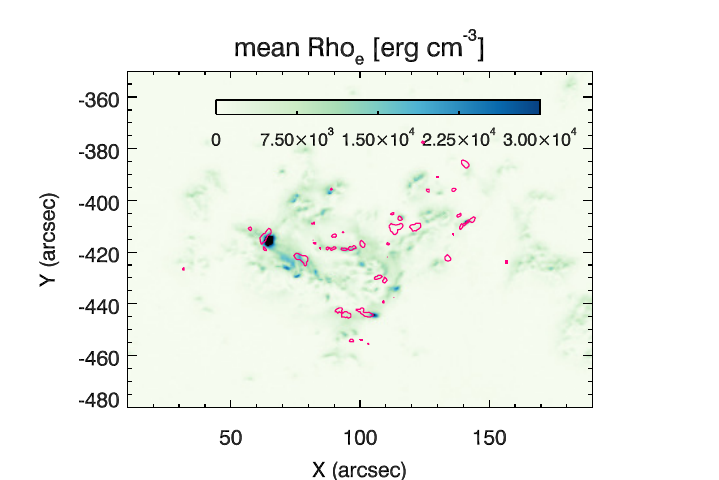}}
\centerline{
\includegraphics[width=0.33\textwidth, clip, trim = 6mm 2mm 19mm 12mm]{20101112_blank_space.pdf}
\includegraphics[width=0.33\linewidth, clip, trim = 6mm 0mm 19mm 12mm]{20101112_0559_245_schrijver_heat_map_update.pdf}
\includegraphics[width=0.33\linewidth, clip, trim = 6mm 0mm 19mm 12mm]{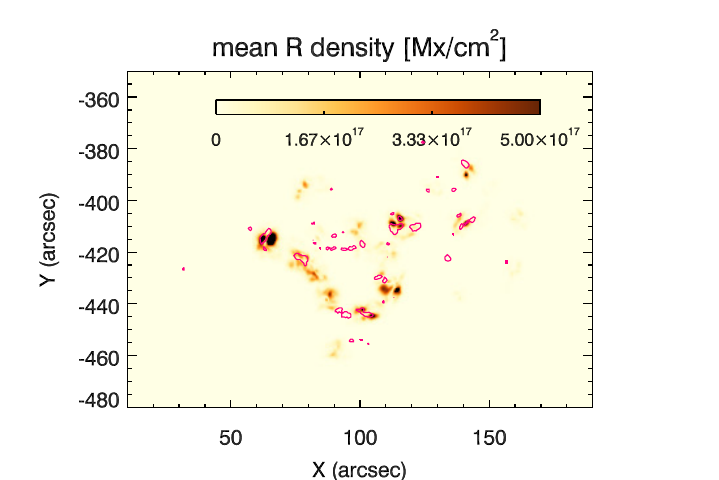}}
\caption{Same as Fig.~\ref{fig:combined_topology_20101016} but for the active epoch in AR 11123 on November 12, 2010 prior to a C1.5 flare (left column; 05:59--07:55~UT) and
one matched quiet epoch after the event (right
column, post-event quiet, i.e., 15:48--17:44~UT). No valid pre-event quiet epoch could be identified for this event.}
\label{fig:combined_topology_20101112}
\end{figure*}

\begin{figure*}
    \centerline{
    \includegraphics[width=0.33\textwidth, clip, trim = 7mm 3mm 19mm 12mm]{20101112_blank_space.pdf}
    \includegraphics[width=0.33\linewidth, clip, trim = 7mm 3mm 19mm 12mm]{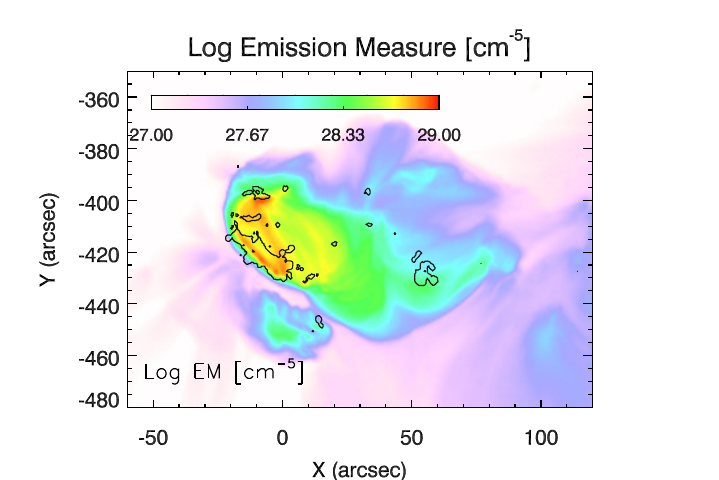}
    \includegraphics[width=0.33\linewidth, clip, trim = 7mm 3mm 19mm 12mm]{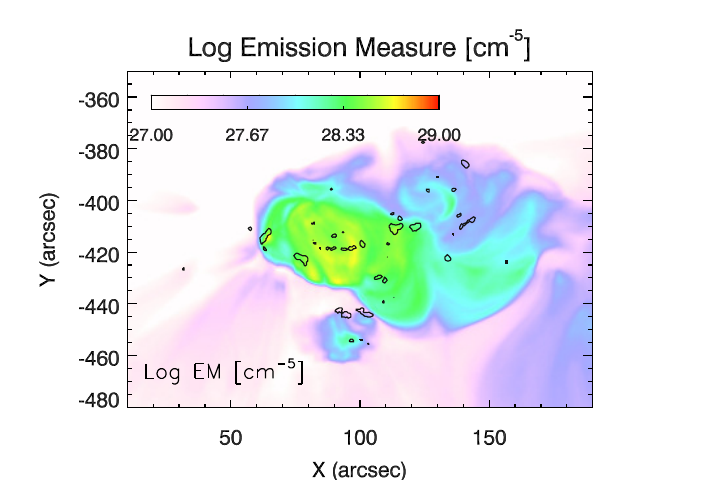}}
    \centerline{
    \includegraphics[width=0.33\textwidth, clip, trim = 7mm 3mm 19mm 12mm]{20101112_blank_space.pdf}
    \includegraphics[width=0.33\linewidth, clip, trim = 7mm 3mm 19mm 12mm]{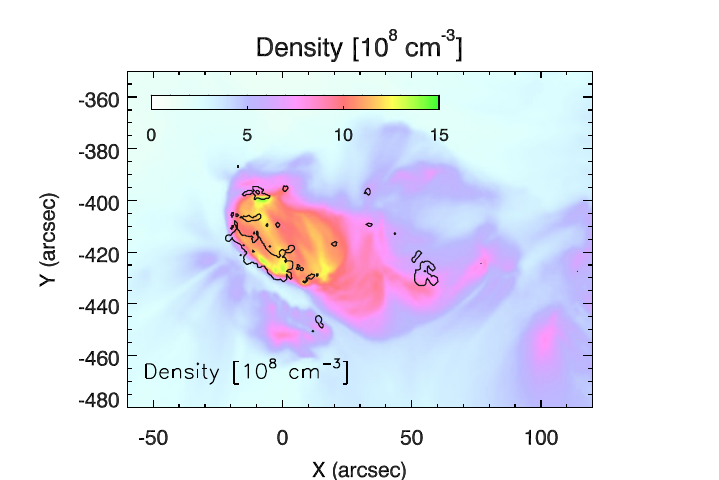}
    \includegraphics[width=0.33\linewidth, clip, trim = 7mm 3mm 19mm 12mm]{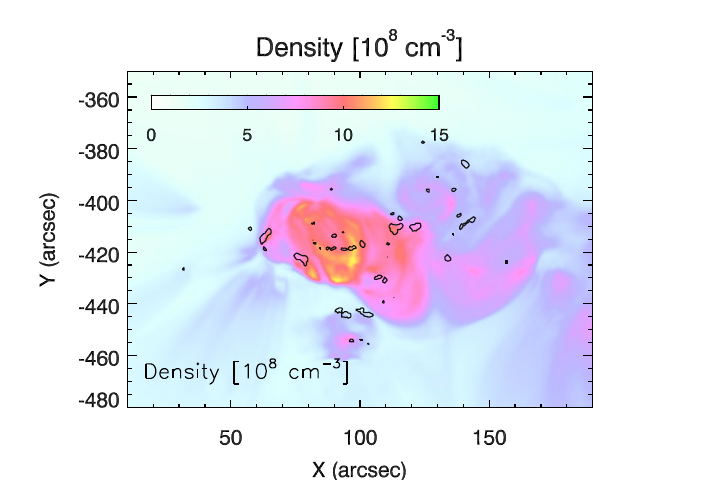}}
        \centerline{
    \includegraphics[width=0.33\textwidth, clip, trim = 7mm 0mm 19mm 12mm]{20101112_blank_space.pdf}
    \includegraphics[width=0.33\linewidth, clip, trim = 7mm 0mm 19mm 12mm]{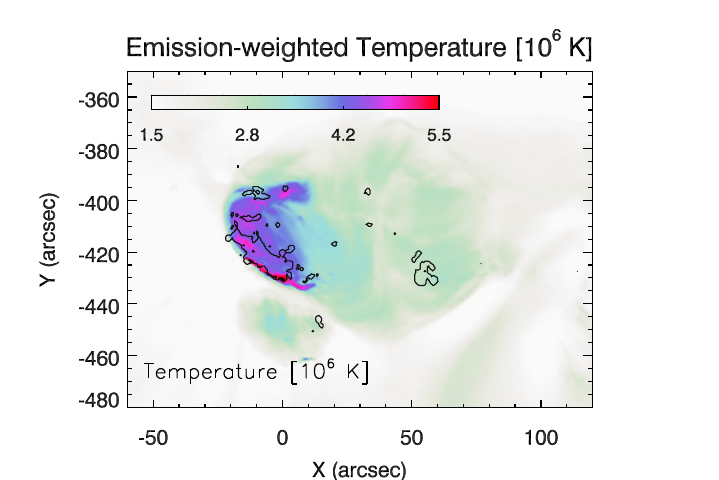}
    \includegraphics[width=0.33\linewidth, clip, trim = 7mm 0mm 19mm 12mm]{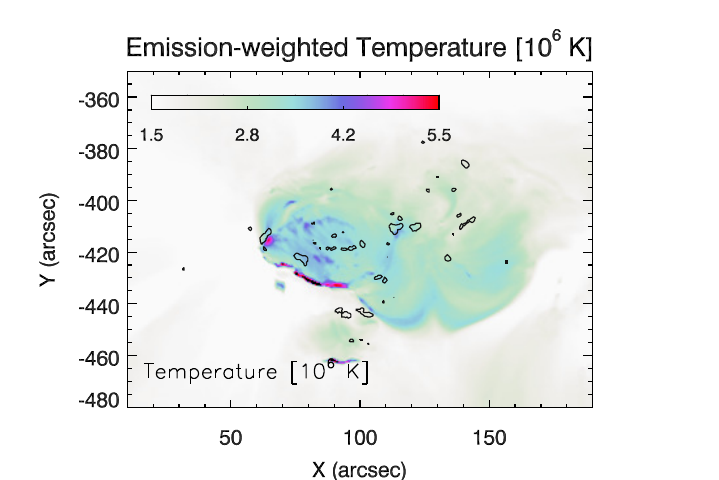}}
    \caption{Same as Fig.~\ref{fig:dem_20101016} but for the active epoch in AR 11123 on November 12, 2010 prior to a C1.5 flare (left column; 05:59--07:55~UT) and one matched quiet epoch after the event (right column, post-event quiet, i.e., 15:48--17:44~UT). No valid pre-event quiet epoch could be identified for this event.}
    \label{fig:dem_20101112}
\end{figure*}
NOAA AR 11123, that newly emerged on November 10, 2010, is part of AR complex HARP 245 that also includes magnetically connected AR 11121 that is at this point already decaying. 
Here, we investigate the C1.5 flare of AR 11123 on November 12, 2010 that starts at 07:59~UT and reaches its maximum at 08:11~UT. 
Since no major precursor activity was detected in AR 11121 during the analysis intervals we only focus on AR 11123.

Figure~\ref{fig:goes_epochs_20101112} shows the time evolution of the GOES/XRS soft X-ray flux in 1--8\AA~around the time of the flare, indicating the analysis intervals for the active and post-event quiet epoch.
Due to enhanced flaring activity prior to the active epoch, no valid pre-event quiet epoch could be selected.

Figure~\ref{fig:combined_topology_20101112} shows transient brightening activity for the active and the post-event quiet epoch in relation to the topology of the active region and its magnetic environment. 
During the active epoch, transient brightening clustering at the future flaring location is observed. The largest cluster identified by DBSCAN (blue) is co-spatial with ribbons during the flare's rise time (Fig.~\ref{fig:combined_topology_20101112}, first row, left). Comparing the TB heat map with the magnetic skeleton reveals that TB hot spots occur in the vicinity of an extended bald patch contour, within a big null point dome footprint (Fig.~\ref{fig:combined_topology_20101112}, second row, left). In terms of the magnetic environment, the majority of TBs cluster around regions of high-value $\rho_{e}(p)$ and $r(p)$ (Fig.~\ref{fig:combined_topology_20101112}, third and fourth row, left).

During the post-event quiet epoch, more individual TB clusters are identified but they are smaller, all of similar size and appear at random locations within the FOV of the active region (Fig.~\ref{fig:combined_topology_20101112}, first row, right). These smaller TB clusters also appear within a similar bigger null point dome structure, close to locations of short and intermittent bald patch segments (Fig.~\ref{fig:combined_topology_20101112}, second row, right). Smaller-scale areas of enhanced activity occur but they are not hot spots, compared to the active epoch. Regarding the magnetic environment, TBs during the post-event quiet epoch show no spatial correspondence with areas of increased excess magnetic energy $\rho_{e}(p)$ (Fig.~\ref{fig:combined_topology_20101112}, third row, right) or flux-enhanced regions of strong-gradient polarity inversion lines (Fig.~\ref{fig:combined_topology_20101112}, fourth row, right).
For both epochs, open field areas are located outside of the big separatrix surface to the North, but no TBs occur at these locations (Fig.~\ref{fig:combined_topology_20101112}, second row).

Figure~\ref{fig:dem_20101112} shows the plasma environment of AR 11123 together with transient brightening locations. The largest cluster identified during the active epoch is co-spatial with the highest emission measure regions, which enclose high density ($>1.0\times10^{9}$cm$^{-3}$) and temperature ($>5$~MK) areas. The elongated high temperature region is partially co-spatial with the fan trace of a coronal null point identified within the skeleton. While a similar, but smaller high-temperature region is observed in the post-event epoch, no TBs are occurring there. The smaller TB clusters appear mostly at enhanced emission measure areas ($\sim 5\times 10^{28}$cm$^{-5}$), but do not match particular regions-of-interest in terms of density and temperature. 

\subsection{September 6, 2011}
\begin{figure*}
   \centering
   \includegraphics[width=1.0\textwidth]{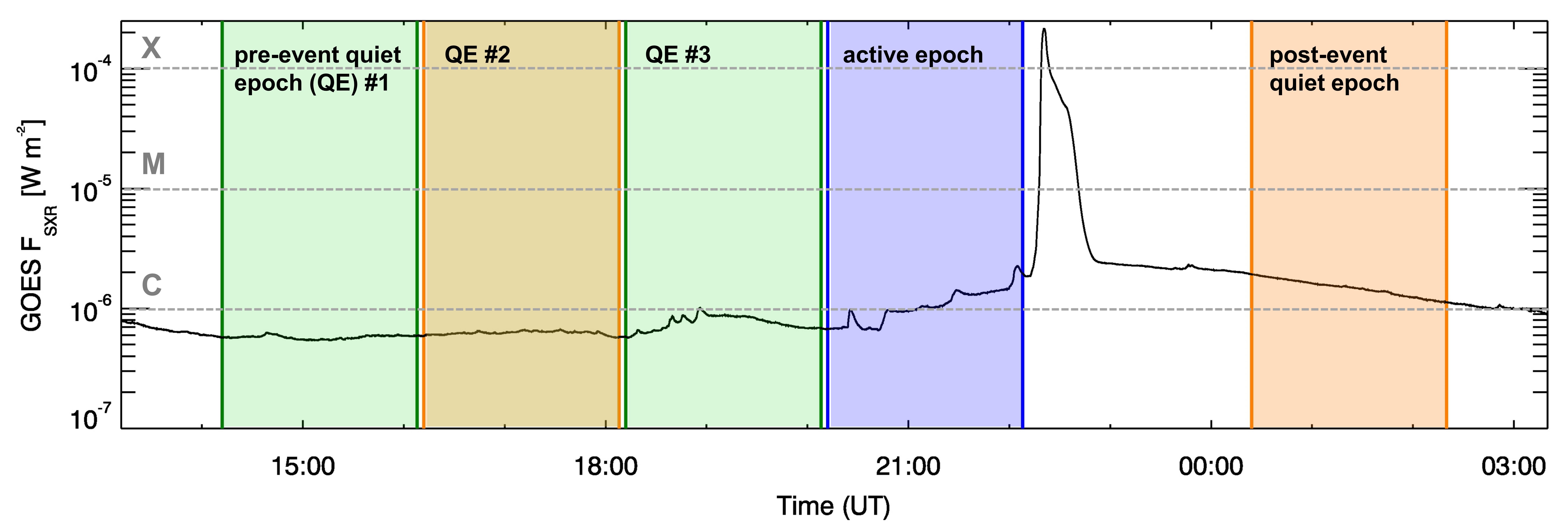}
   \caption{Overview of the analysis epochs for the September 6, 2011 event. GOES/XRS soft X-ray 1--8\AA~time evolution during the time period of interest. The colored areas indicate the active epoch (blue; during the pre-flare phase of the associated X2.1 flare), the consecutive pre-event quiet epochs leading up to it (in green and orange/green, labeled QE \#1-3) and the post-event quiet epoch (orange). For the magnetic skeleton, magnetic and plasma environment analysis QE~\#2, the active epoch and the post-event quiet epoch are used.}
   \label{fig:goes_epochs_20110906}
\end{figure*}

The X2.1 flare event in NOAA active region 11283 on September 6, 2011 starts at 22:12~UT and reaches its peak around 22:20~UT. 
It is the only event being examined that features an extended period of non-flaring activity before the flare itself. This allows for a sequential analysis of TBs during the quiet periods leading up to the active epoch (see Table~\ref{tab:events} for details).
Figure~\ref{fig:goes_epochs_20110906} illustrates the selection of active (blue) and sequential pre-event (green, and orange-green) as well as post-event quiet epochs (orange) for the event under study. 
\begin{figure*}
\centering
\includegraphics[width=1.0\textwidth, clip, trim = 2mm 3mm 2mm 2mm]{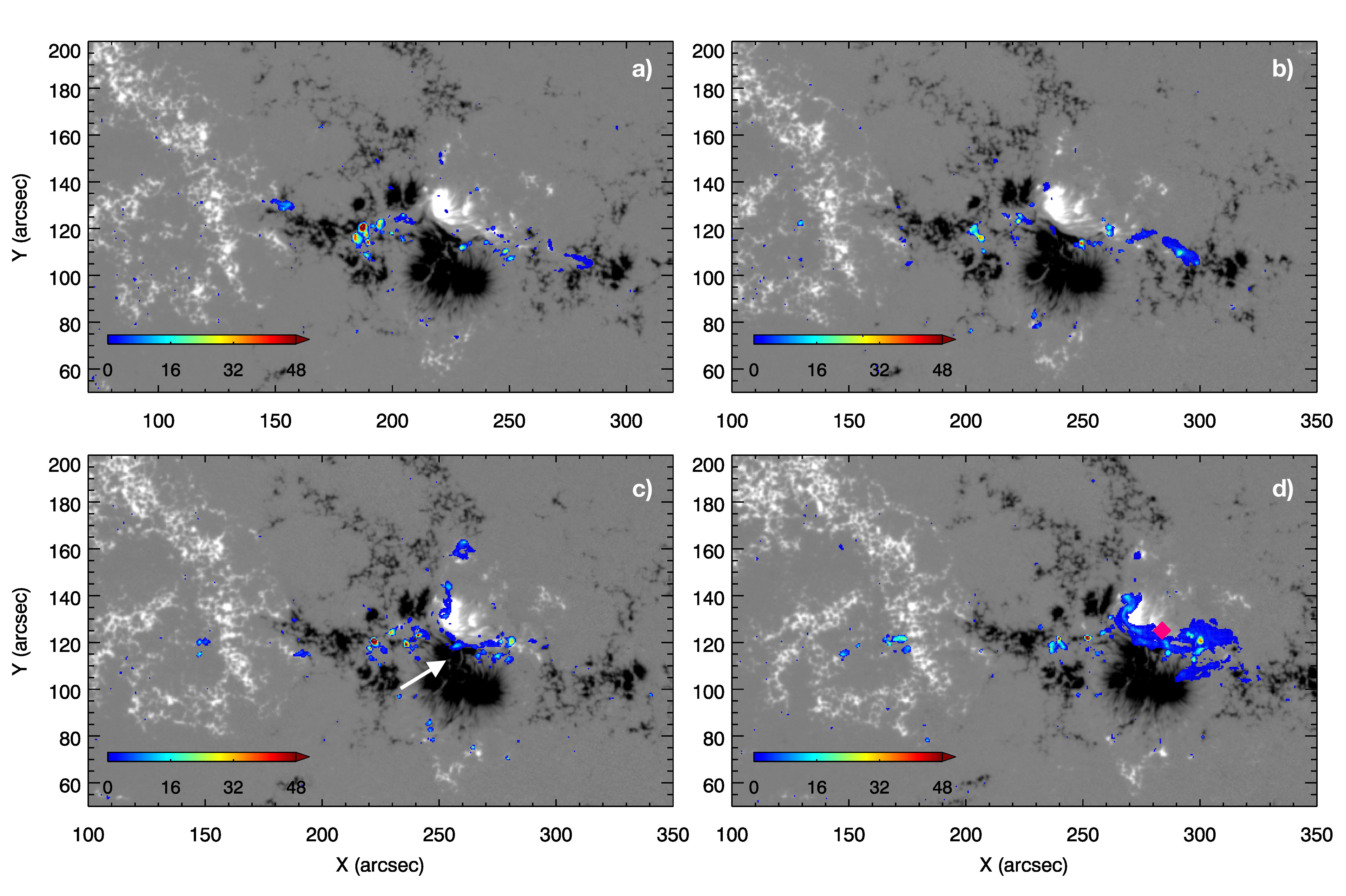}
\caption{Evolution of transient brightenings detected over consecutive quiet and active analysis intervals for AR 11283 on September 6, 2011. The background shows the radial component of the magnetic
field $B_{r}$ of the SDO/HMI vector magnetogram, TB heat maps are shown in rainbow color on top (see Fig.~\ref{fig:detection_tbs} for details). Pre-event quiet epochs at (a) 14:12--16:08~UT (QE\#1), (b)
16:12--18:08~UT (QE\#2), (c) 18:12--20:08~UT (QE\#3) and the active epoch at (d) 20:12--22:08~UT are shown.
A gradual shift and clustering of TBs towards the strong-gradient polarity inversion line
(indicated by the white arrow) can be observed over time. A X2.1 flare is occurring at
22:12~UT (pink diamond) close to the location of the largest TB cluster in the active epoch.}
\label{fig:evolution_clustering}
\end{figure*}
Figure~\ref{fig:evolution_clustering} shows the evolution of TB heat maps over approximately 8~hrs on September 6, 2011, with panels (a-c) showing consecutive pre-event quiet epochs and panel (d) showing the active epoch.  
In the early stages of this time sequence (panels a,b), there is no clear spatial pattern to the TBs, but over time a distinct gradual shift and clustering of TBs towards the strong-field polarity inversion line and the future flare location can be observed (panels c,d).

\begin{figure*}
\centerline{
\includegraphics[width=0.33\textwidth, clip, trim = 2mm 2mm 20mm 27mm]{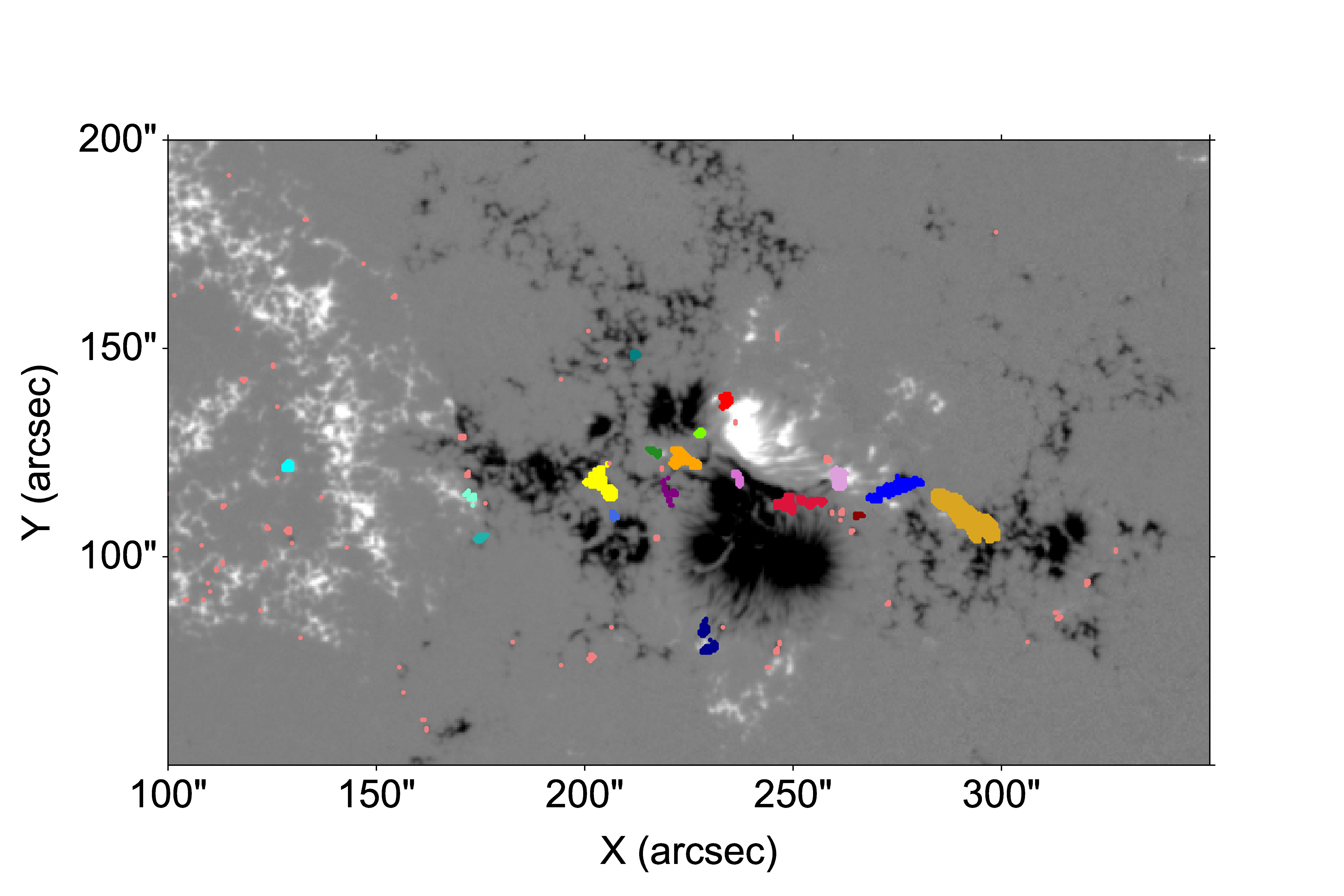}
\includegraphics[width=0.33\textwidth, clip, trim = 2mm 2mm 20mm 27mm]{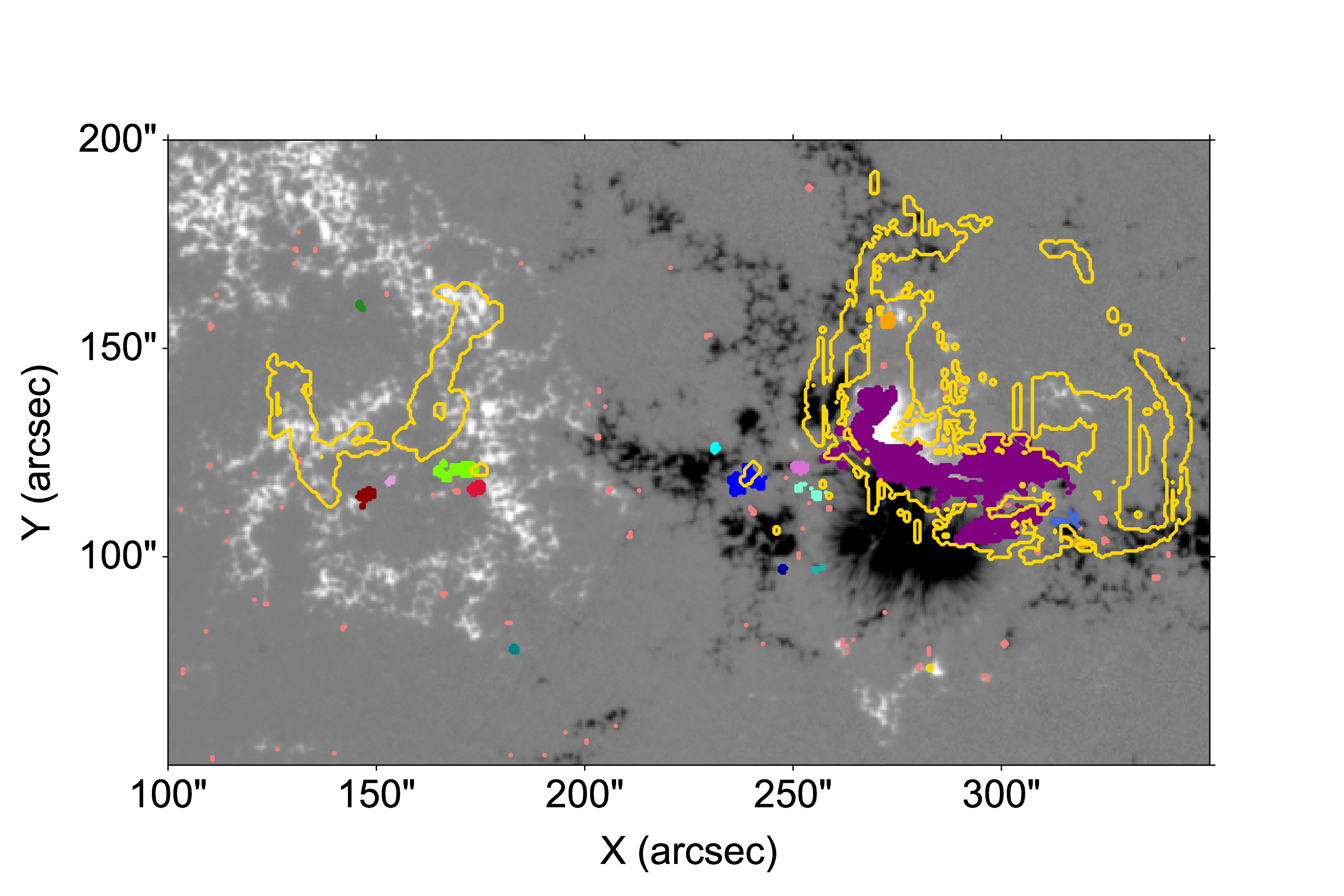}
\includegraphics[width=0.33\textwidth, clip, trim = 2mm 2mm 20mm 27mm]{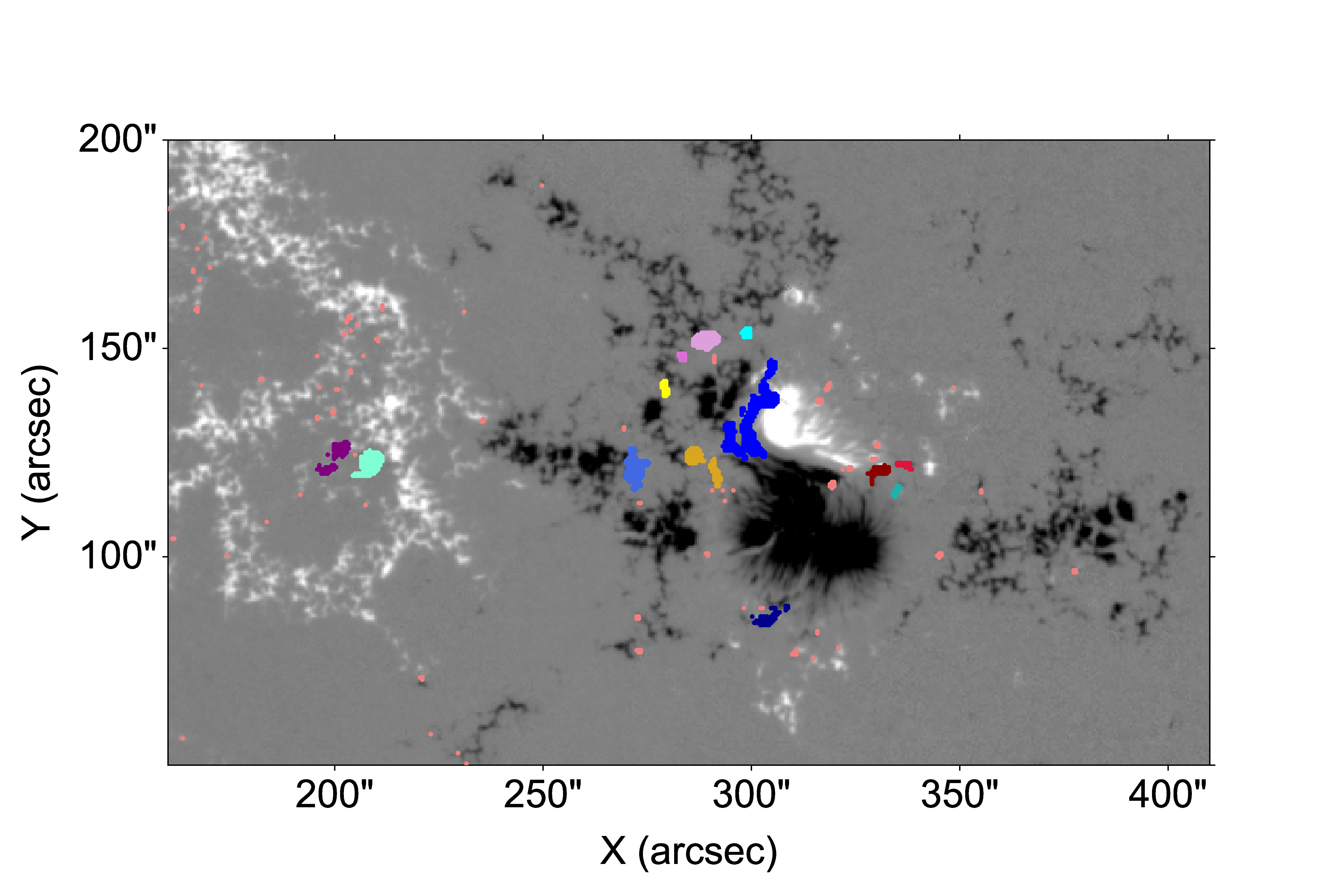}}
\centerline{
\includegraphics[width=0.33\textwidth, clip, trim = 5mm 8mm 15mm 14mm]{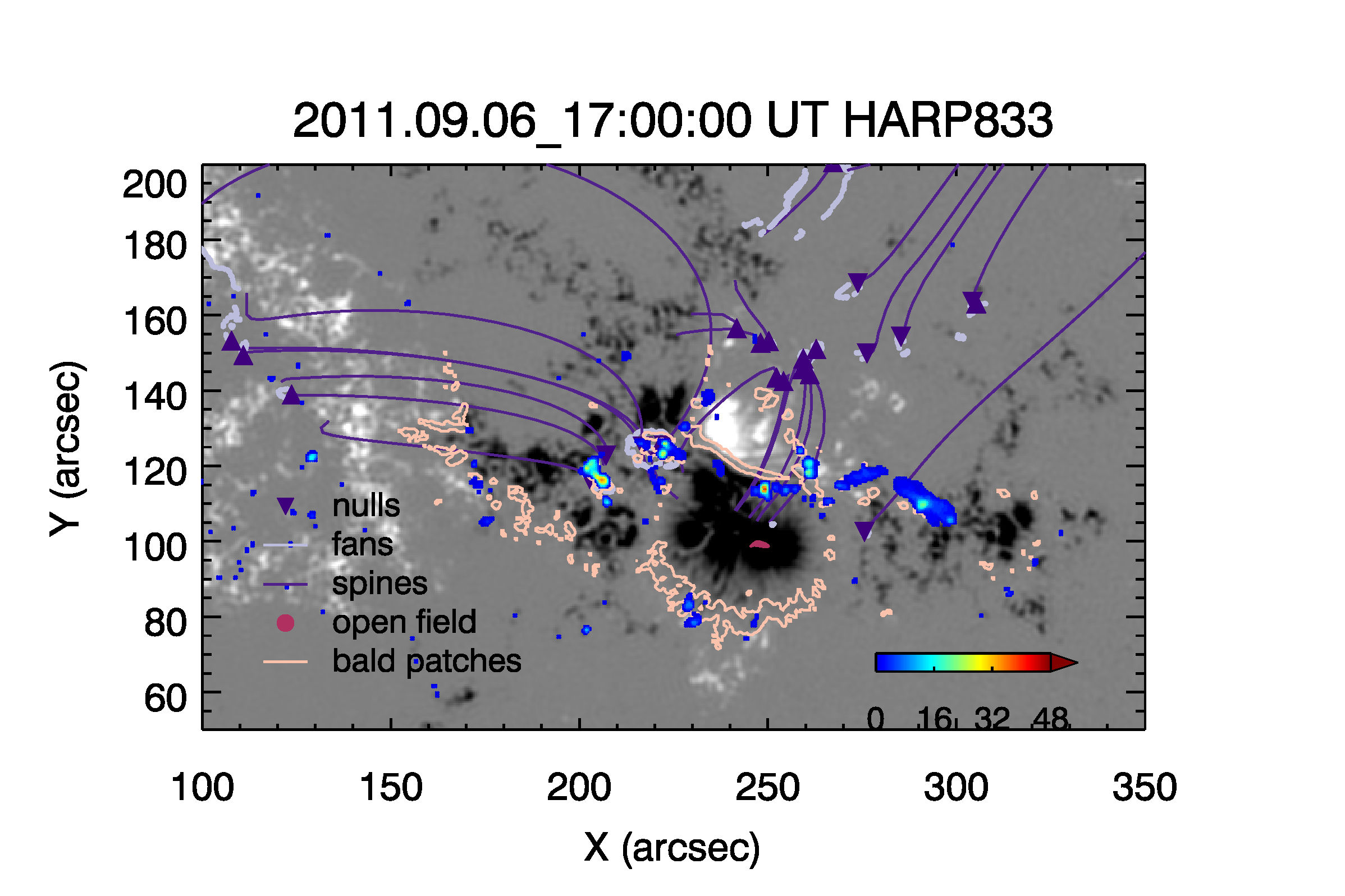}
\includegraphics[width=0.33\textwidth, clip, trim = 5mm 8mm 15mm 14mm]{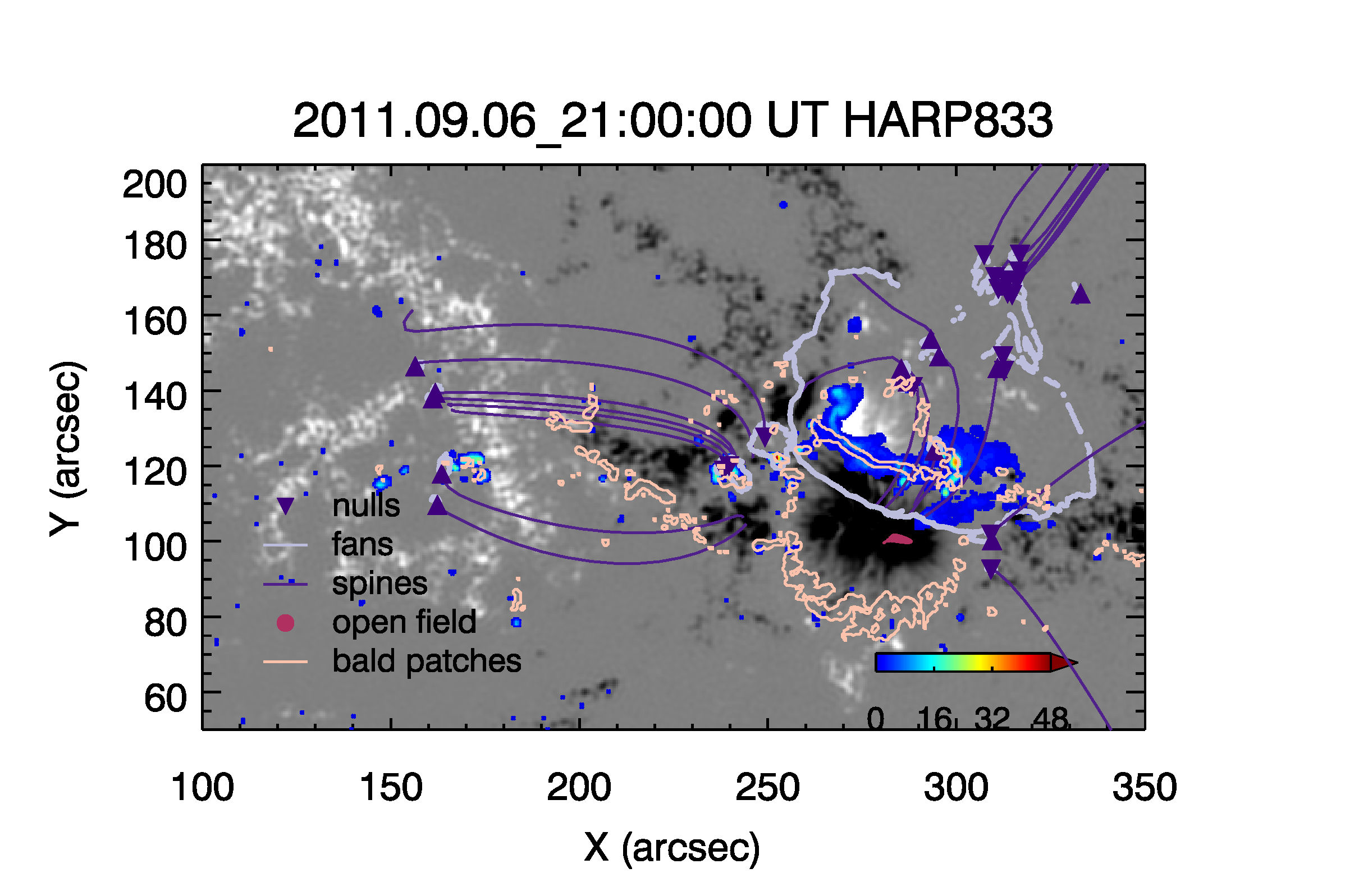}
\includegraphics[width=0.33\textwidth, clip, trim = 5mm 8mm 15mm 14mm]{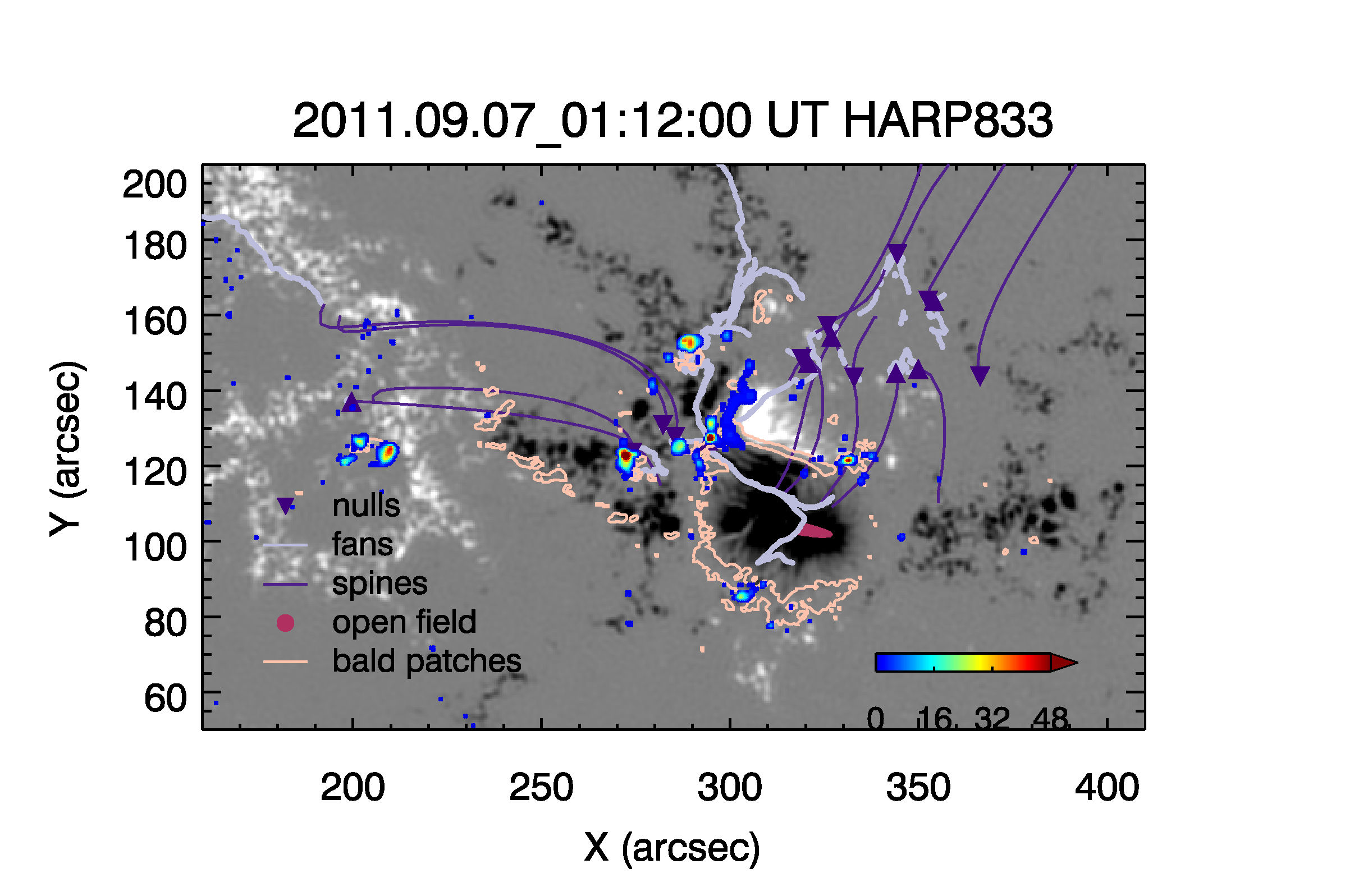}}
\centerline{
\includegraphics[width=0.33\textwidth, clip, trim = 5mm 8mm 15mm 14mm]{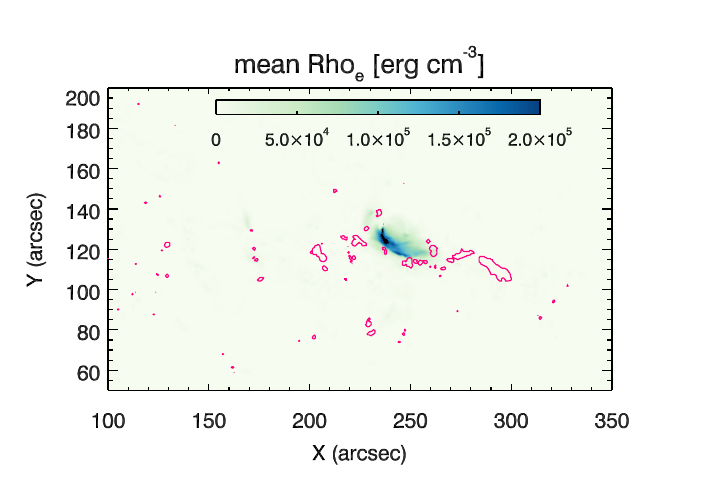}
\includegraphics[width=0.33\textwidth, clip, trim = 5mm 8mm 15mm 14mm]{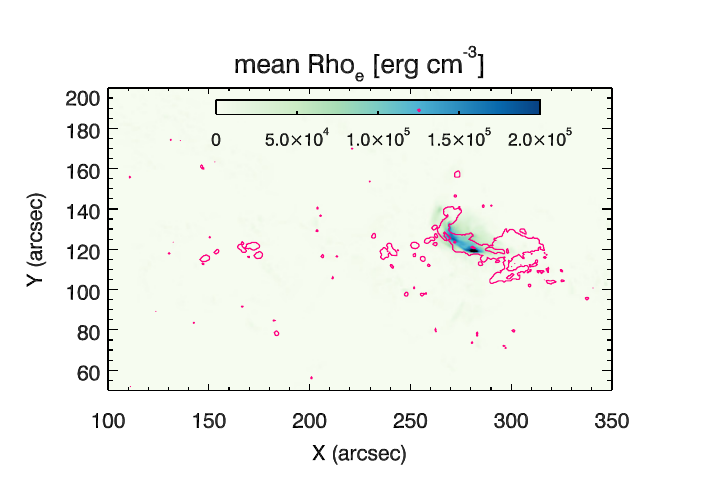}
\includegraphics[width=0.33\textwidth, clip, trim = 5mm 8mm 15mm 14mm]{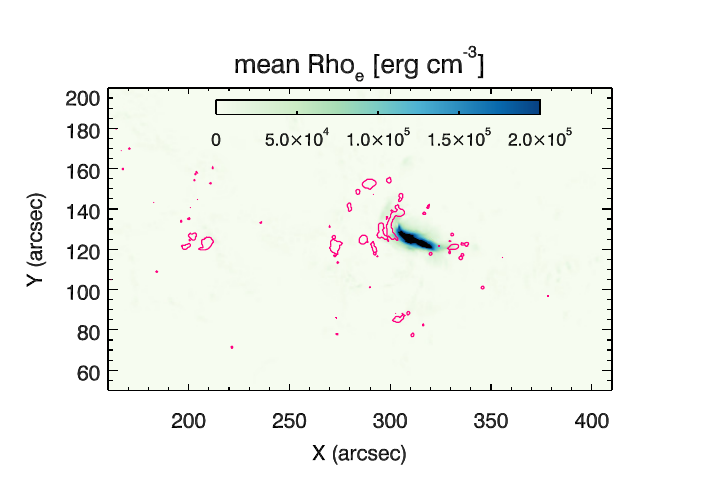}}
\centerline{
\includegraphics[width=0.33\textwidth, clip, trim = 5mm 2mm 15mm 14mm]{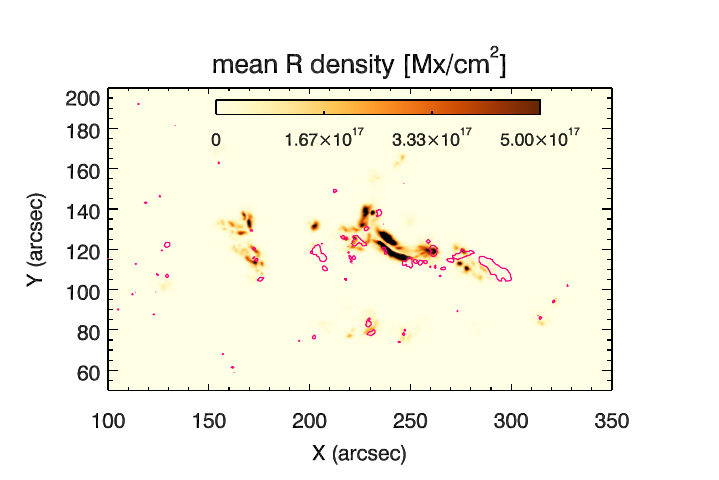}
\includegraphics[width=0.33\textwidth, clip, trim = 5mm 2mm 15mm 14mm]{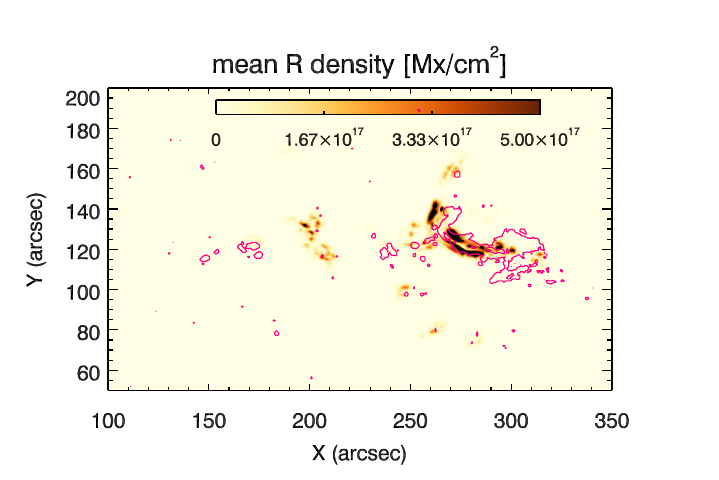}
\includegraphics[width=0.33\textwidth, clip, trim = 5mm 2mm 15mm 14mm]{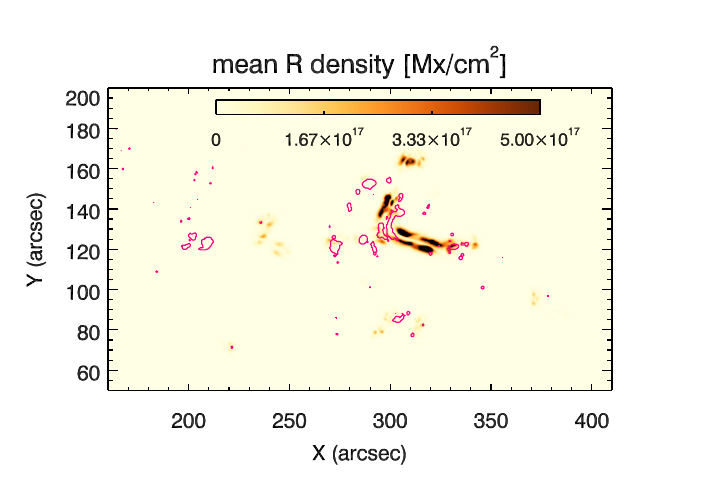}}
\caption{Same as Fig.~\ref{fig:combined_topology_20101016} but for the active epoch in AR 11283 on September 6, 2011 prior to a X2.1 flare (middle column; 20:12--22:08~UT) and two matched quiet epochs, one prior to the event (left
column, pre-event quiet QE\#2, i.e., 16:12--18:08~UT) and one after the event (right column, post-event quiet, i.e., September 7, 2011 00:24--02:20~UT).}
\label{fig:combined_topology_20110906}
\end{figure*}

Figure~\ref{fig:combined_topology_20110906} provides further details on this change in transient brightening behavior by investigating TBs in the context of the magnetic skeleton of the active region, and the magnetic environment for the active epoch (middle column) and two matched quiet epochs (left column: pre-event quiet; right column: post-event quiet). 
Comparing the different epochs, it becomes evident that the largest TB cluster occurs during the active epoch (first row, middle panel) and it occurs in close proximity to the rise time flare ribbons (first row, middle panel, golden contours) within separatrix surfaces in the center of the active region (second row, middle panel). This cluster shows multiple, extended areas of enhanced activity (second row middle panel) and it encloses all of the enhanced high-$\rho_{e}(p)$ area (third row, middle panel) and the majority of high-$r(p)$ areas (fourth row, middle panel).
During the active epoch, in addition to this large cluster, a few more smaller clusters appear. Three of them are associated with enhanced TB activity. All of these clusters are co-spatial with fan traces of isolated nulls and smaller bald patch contours, but occur in regions of low-$\rho_{e}(p)$ and low-$r(p)$. Some small-scale TBs that do not show any correspondence to inferred topological features or regions-of-interest in the magnetic environment also occur.
The pre- and post-event quiet epochs are characterized by only smaller-scale clusters that are all associated with enhanced activity, some are hot spot areas. None of them are co-spatial with locations of enhanced values of the excess magnetic energy density $\rho_{e}(p)$ and the magnetic flux in strong-gradient PIL areas $r(p)$.

In all three epochs, next to various short, intermittent bald patches in weaker magnetic field regions, we find a coherent, long bald patch segment close to the primary strong-field PIL (Fig.~\ref{fig:combined_topology_20110906}, second row). For the active epoch, the longer segment spatially matches the largest cluster of TBs, while for the quiet epochs only a small fraction of the TBs are observed near this longer bald patch segment. We interpret this long, coherent bald patch as signature of a magnetic flux rope, as it is co-spatial with a hot sigmoid observed in 94\AA~and co-spatial with a twisted flux rope found in Non-Force-Free-Field extrapolations by \cite{Prasad:2020}. 
Open magnetic field regions are present in all three epochs but are very small in extent and very localized to the strong negative polarity region in the center of the active region. No transient brightenings are observed in the vicinity of open field, leading us to conclude that no small-scale reconnection between open and closed field lines occurred during these epochs. 

\begin{figure*}
    \centerline{
    \includegraphics[width=0.33\textwidth, clip, trim = 5mm 6mm 15mm 14mm]{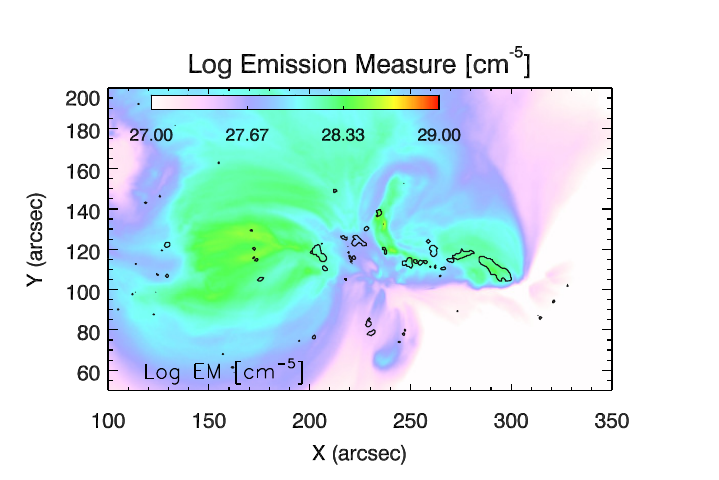}
    \includegraphics[width=0.33\linewidth, clip, trim = 5mm 6mm 15mm 14mm]{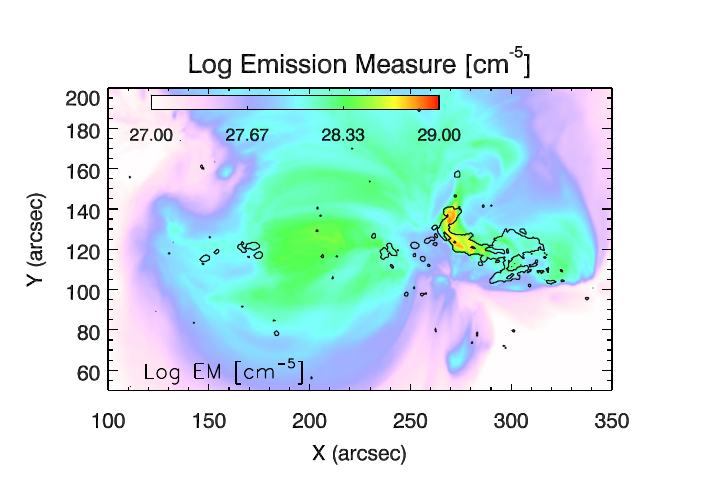}
    \includegraphics[width=0.33\linewidth, clip, trim = 5mm 6mm 15mm 14mm]{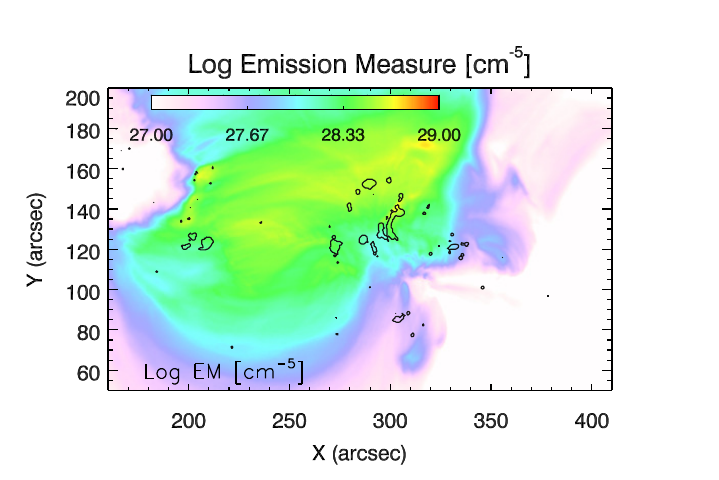}}
    \centerline{
    \includegraphics[width=0.33\textwidth, clip, trim = 5mm 6mm 15mm 14mm]{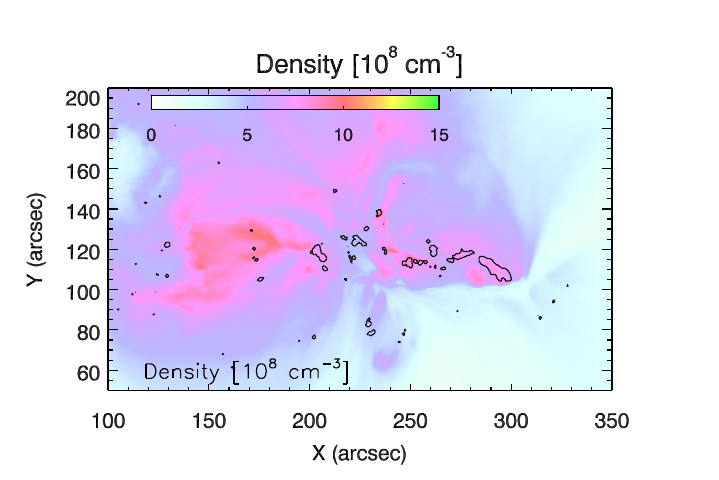}
    \includegraphics[width=0.33\linewidth, clip, trim = 5mm 6mm 15mm 14mm]{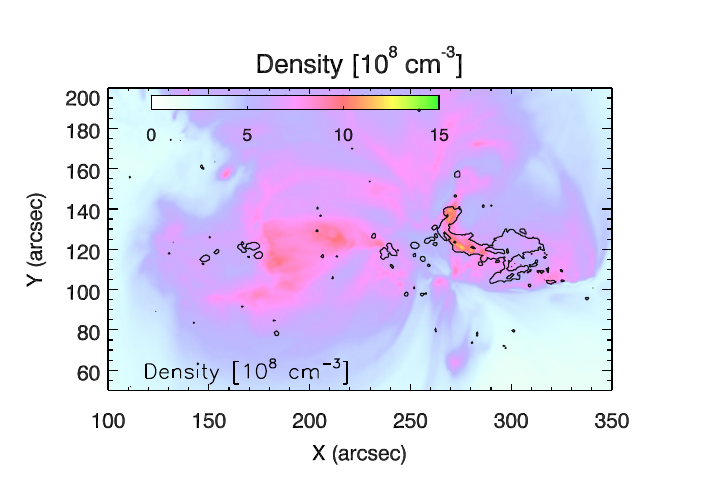}
    \includegraphics[width=0.33\linewidth, clip, trim = 5mm 6mm 15mm 14mm]{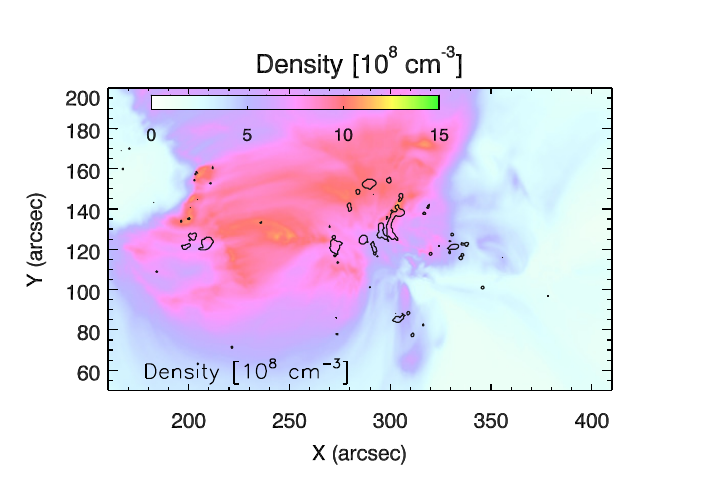}}
        \centerline{
    \includegraphics[width=0.33\textwidth, clip, trim = 5mm 2mm 15mm 14mm]{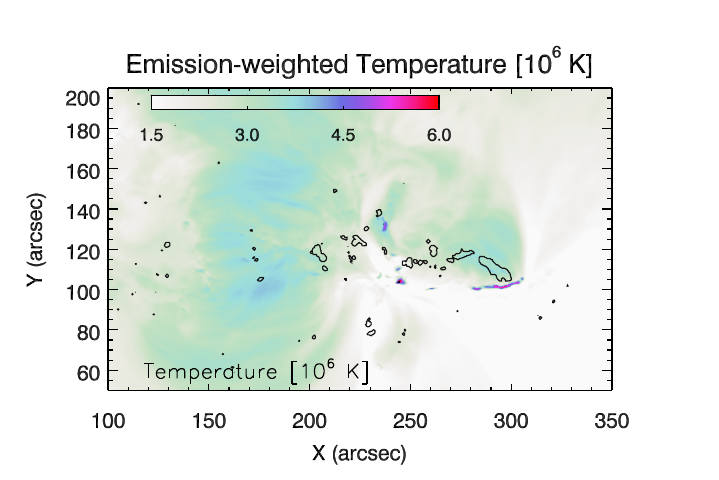}
    \includegraphics[width=0.33\linewidth, clip, trim = 5mm 2mm 15mm 14mm]{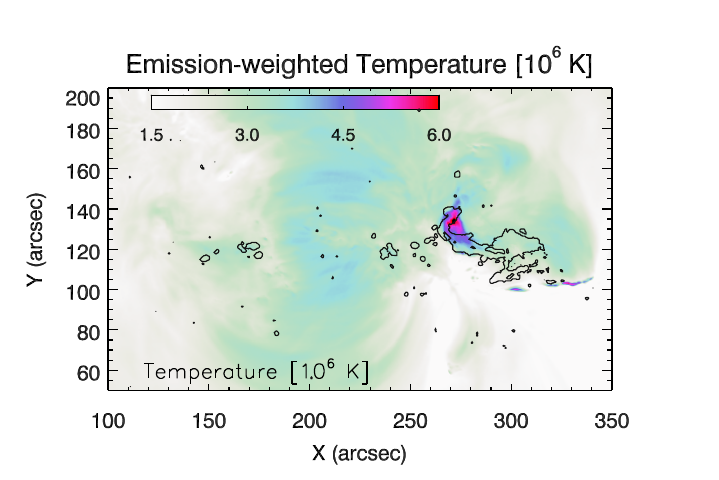}
    \includegraphics[width=0.33\linewidth, clip, trim = 5mm 2mm 15mm 14mm]{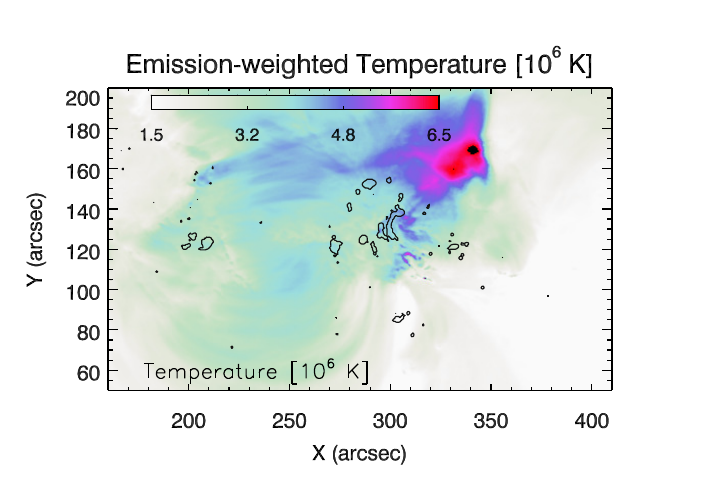}}
    \caption{Same as Fig.~\ref{fig:combined_topology_20101016} but for the active epoch in AR 11283 on September 6, 2011 prior to a X2.1 flare (middle column; 20:12--22:08 UT) and two matched quiet epochs, one prior to the event (left column, pre-event quiet QE\#2, i.e., 16:12--18:08 UT) and one after the event (right column, post-event quiet, i.e., September 7, 2011 00:24--02:20 UT).}
    \label{fig:dem_20110906}
\end{figure*}
Figure~\ref{fig:dem_20110906} shows transient brightenings in the context of the plasma environment. The large TB cluster during the active epoch partially encloses high emission measure areas ($>10^{28}$cm$^{-5}$), indicating in this particular case plasma heating (up to $6$MK). During the pre-event quiet epoch the majority of TBs occur in enhanced emission measure regions ($\sim 5.0\times 10^{28}$cm$^{-5}$), that are moderately dense ($\sim 8\times10^{8}$cm$^{-3}$) but cooler compared to other structures within the active region ($<3.5$MK). A few TB clusters exist in low emission measure areas.
The post-event quiet epoch is dominated by ongoing activity in hot flare loops to the North of the core of the active region after a previous flare. We note that this area is co-spatial with the location of coronal nulls and fan traces of the magnetic skeleton. Transient brightenings randomly appear in this in general emission-measure enhanced post-flare volume, but are not particularly hot.

\begin{figure*}
    \centering
    \includegraphics[width=1.0\linewidth, clip, trim = 0mm 10mm 0mm 20mm]{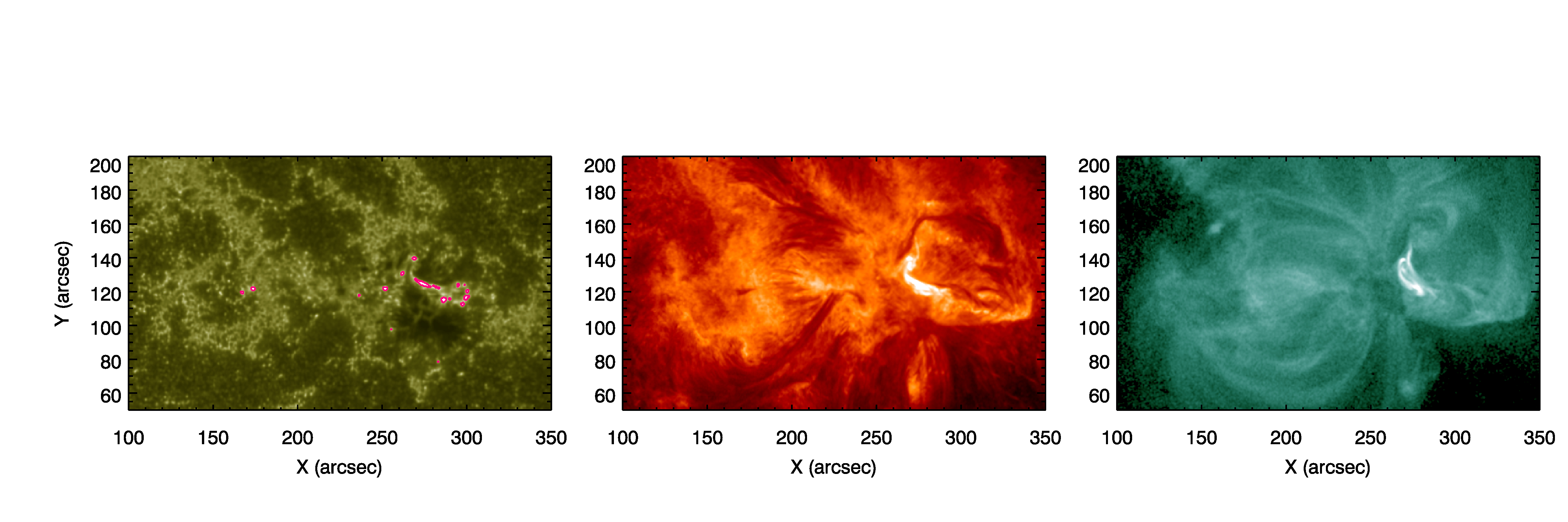}
    \includegraphics[width=1.0\linewidth, clip, trim = 0mm 0mm 0mm 0mm]{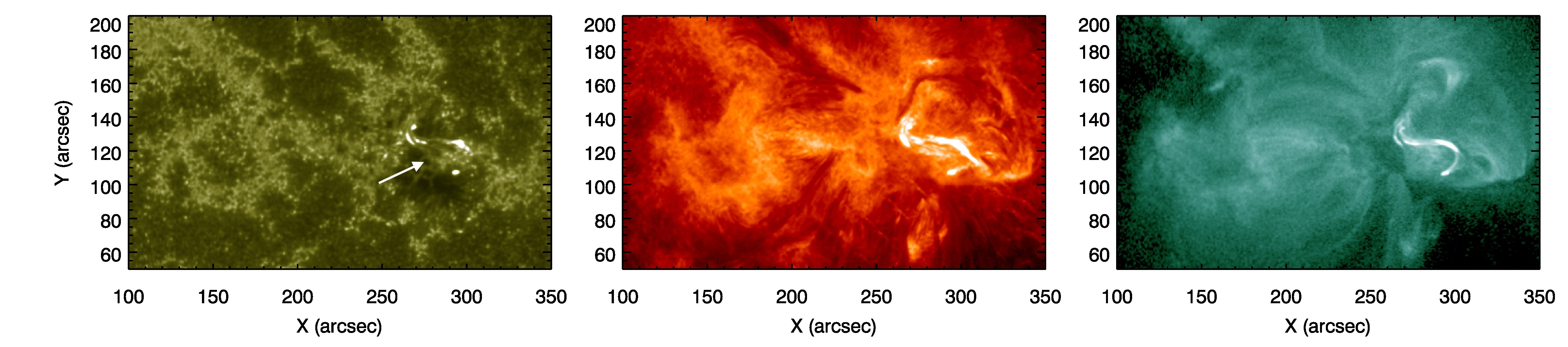}
    \caption{Snapshots of SDO/AIA 1600 (left), 304 (middle) and 94\AA~(right) images at 21:04~UT (top row) and 21:59~UT (bottom row) showing the occurrence of transient brightenings in the upper photosphere/transition region and their corresponding signatures in the chromosphere and corona during the active epoch of AR 11283 on September 6, 2011 prior to a X2.1 flare starting at 22:12~UT. Transient brightenings, indicated by pink contours and a white arrow, are co-spatial with brightenings in 304\AA~and the signature of a hot sigmoid, visible in 94\AA.}
    \label{fig:20110906_tb_combined_channels}
\end{figure*}
Figure~\ref{fig:20110906_tb_combined_channels} provides further observational evidence that the majority of transient brightenings observed during the active epoch are co-spatial with a hot sigmoid that is formed in the core of the active region. Two snapshots of SDO/AIA 1600\AA~data during the active epoch together with instantaneous TB detections and co-temporal 94 and 304\AA~images are shown. A spatial correspondence between the brightenings in 1600\AA, flare ribbon-like signatures in 304\AA, and the hot sigmoid in 94\AA~is observed, indicating that these detected transient brightenings are related to the formation of the flux rope that becomes later part of the eruption \citep{Prasad:2020}. This is in agreement with plasma heating observed at these locations during the active epoch as well (see Fig.~\ref{fig:dem_20110906}).
\subsection{General patterns}
In the following we report on general trends in terms of TB clustering, patterns related to TB occurrence in the context of the magnetic topological skeleton, as well as the magnetic and plasma environment that are observed in all three event cases studied. 
\paragraph{Spatial Clustering during Active Epochs} for the active epoch of each event case under study, there is one dominant, large cluster of transient brightenings, accompanied by smaller-sized clusters. 
The dominant cluster is co-spatial with the location of flare ribbons during the rise time of the associated flare (Figs.~\ref{fig:combined_topology_20101016}, \ref{fig:combined_topology_20101112}, \ref{fig:combined_topology_20110906} first row, middle panels). 
For the quiet epochs (including both pre-event as well as post-event quiet epochs) TB activity is also observed, but these clusters tend to be smaller in size and their appearance does not show a clear spatial pattern (Figs.~\ref{fig:combined_topology_20101016}, \ref{fig:combined_topology_20101112}, \ref{fig:combined_topology_20110906} first row, left and right panels).
\paragraph{Patterns Related to the Topological Skeleton} The majority of transient brightenings, are related to topological features of the magnetic skeleton of the potential field. Most of them are observed at bald patches (Fig.~\ref{fig:combined_topology_20101112}, \ref{fig:combined_topology_20110906}), and near fan traces of coronal nulls (Fig.~\ref{fig:combined_topology_20101016}). We conclude that TBs present the chromospheric signatures of magnetic reconnection episodes.
During active epochs where coherent, longer bald patches segments exist (Fig.~\ref{fig:combined_topology_20101112} \&~\ref{fig:combined_topology_20110906}), the largest cluster of TBs is observed there too, while during quiet epochs some TBs are associated with bald patches that are characterized by intermittent, shorter segments.
Transient brightenings tend to occur at bald patch locations, not next to them as it is traditionally the case for flare ribbons associated with the main flare reconnection.
The large cluster of TBs found in each active epoch occurs within the separatrix surface of a null point, although not always in close proximity to the null fan trace.
We note that although open field footpoints also indicate locations for signatures of reconnection no TBs were observed there.
\paragraph{Patterns related to Magnetic Environment}
during active epochs, transient brightenings tend to occur in regions of increased excess energy density and close to strong-field, high-gradient polarity inversion lines prior to solar flares (Figs.~\ref{fig:combined_topology_20101016}, \ref{fig:combined_topology_20101112}, ~\ref{fig:combined_topology_20110906}, third and fourth rows, middle panels). On the contrary, during quiet epochs the majority of TB clusters appear at locations that do not clearly correlate with areas of enhanced values of these parameters.
\paragraph{Patterns related to Plasma Properties}
TBs during active epochs tend to be associated with the hottest and densest areas within the active region. During quiet epochs, they still appear in enhanced but not areas of interest in regards to their emission measure, density or temperature.
\section{Summary and Discussion}\label{sec:summary_discussion}
This paper examines transient brightenings and their physical relationship to solar energetic events, providing an overview of our planned statistical analysis on their origins and characteristics in different layers of the solar atmosphere. TBs, detected in the upper photosphere and transition region using SDO/AIA 1600\AA~observations, are analyzed via a cumulative approach over two-hour intervals. Heat maps highlight hot spot areas and clustering behavior that would otherwise be difficult to identify. Instantaneous TB masks, which detect TBs at each time step, are also available to study their time evolution and will assist in the parametrization for future statistical analysis.

The origin of TBs and their potential connection to the main flaring event is explored by examining their magnetic environment and the photospheric footprints of the coronal topological skeleton, including the spines and separatrices of coronal null points, open field footpoints, and bald patches. These features are key regions for magnetic reconnection signatures and are potential sources of TBs. We investigate whether specific topologies are more strongly linked to pre-flare TBs compared to quiet epoch TBs. Additionally, we analyze magnetic field parameters, such as locations of strong-gradient polarity inversion lines and areas of free magnetic energy, both known for their importance for the main flare process. To examine connections with higher layers of the solar atmosphere, we use Differential Emission Measure analysis to study plasma properties and heating in the solar corona.

A key open question regarding precursor phenomena is their uniqueness. While precursors have been linked to flaring events in several case studies with known timing and location \citep[][]{Chifor:2007, Joshi:2011, Bamba:2017}, no study has yet examined precursor activity in control data (without flaring) or in a large statistical sample to rule out coincidence. This is a main goal of this paper series. We analyze TB activity during quiet epochs of the same active regions to statistically assess their exclusive role in the pre-flare phase. The methods presented are optimized for large-scale applications, ensuring speed, automation, and objectivity, enabling progress beyond single-case studies.

Transient brightenings prior to three case study events and their quiet epochs are analyzed in this paper for demonstration purposes.  We find that TBs tend to group in one large cluster close to the future flare location and below a null point separatrix surface during active, pre-event epochs, while they appear in smaller clusters that do not exhibit a clear spatial pattern during the matched quiet epochs. This one large cluster is co-spatial with areas of enhanced ``free'' magnetic energy and locations of strong-gradient polarity inversion lines, while the smaller clusters during the quiet epochs are not consistently associated with regions-of-interest of these parameters. The consecutive study of quiet epochs leading up to the X-class flare on September 6, 2011 indicates that the appearance of TBs might even gradually shift towards these flaring-relevant regions-of-interest on a time scale $\sim 4$ hours (Fig.~\ref{fig:evolution_clustering}).

In terms of plasma environment, this one large cluster is characterized by hotter and denser plasma. We note that the magnetic field and plasma environments by themselves do not seem to be unique for active epochs compared to the quiet epochs as high $r(p)$ or $\rho_{e}(p)$ areas are also identified during quiet epochs, as well as hot and dense plasma areas (e.g., compare $\rho_{e}(p)$ in Fig.~\ref{fig:combined_topology_20110906} or temperature in Fig.~\ref{fig:dem_20110906}). The case study results indicate that the \textit{combined analysis} of TBs with their magnetic and plasma environs  might be highly relevant for future flaring, but the significance of this finding needs to be tested through statistical analysis. 

Another important open question regarding precursors is their causal relationship with solar energetic events: does such a relationship exist, and if so, what role do they play in the initiation and triggering of these events? Causal inference cannot be established through single-event case studies; however, previous research has explored the physical role of potential flare precursors in relation to the magnetic configuration of the host active region, laying the groundwork for investigations into causality.

For example, \cite{Masson:2009} reports faint brightenings during the pre-flare phase and suggests low-energetic pre-flare null-point reconnection to be the cause. Null-point reconnection is then reported during the main event in a similar location. This is in agreement with our results for the October 16, 2010 case where the largest cluster of brightenings occurs close to a null point fan trace (see Figure~\ref{fig:combined_topology_20101016}), co-spatial with the flare ribbons for this event.\cite{Mandrini:2014} reports precursor brightenings in SDO/AIA 304 \AA~co-located with quasi-separatrix layers, observed near null point reconnection and the filament, possibly indicating destabilization. The authors could not clearly distinguish between break-out and tether-cutting processes in the pre-event phase, concluding that filament eruption and null point reconnection interacted in a positive feedback loop.

For the September 6, 2011 event case, the largest TB cluster is co-spatial with an extended bald patch segment, indicative of a current channel or magnetic flux rope. Complementary observations in 94\AA~further confirm its formation and we conclude that TBs are highly likely to be small-scale reconnection events associated with its formation. For the November 12, 2010 case, the largest cluster prior to the flare is co-spatial with an extended bald patch contour and occurs near a null point fan trace. It is unclear which topological feature is relevant to the main flare event in this case, but it could be a combination of both. 

Transient brightening clustering during active epochs reported here is observed prior to hot onsets, recently detected in GOES/XRS \citep{Hudson:2021, daSilva:2023} and Solar Orbiter/STIX data \citep{Battaglia:2023}. Our TB analysis ends four minutes prior to the soft X-ray flare onset time and therefore a couple of minutes prior to these observations. Hot onsets are characterized by enhanced isothermal plasma in the temperature range of 10--16~MK and show no signs of gradual increase in temperature, whereas the emission measure steadily increases by about two orders of magnitude, implying a series of incrementally greater energy release \citep{Hudson:2021}. At this point, it is unclear if the TB clustering observed prior to solar flares is related to hot onsets, and while investigating this link is scientifically interesting, it is beyond the scope of this paper.

While some TB clusters are similar in size to UV bursts discovered by IRIS, others are clearly larger, measuring a few arcseconds. We note that the cumulative approach used in this study, to a large degree ``masks" the actual size of individual TBs, as multiple events occurring in close proximity are grouped into the same cluster. We also found that smaller TB clusters, observed during both quiet and active epochs, mostly occur in less conspicuous areas of the corona in terms of plasma environment. They are not associated with plasma heating or accumulation, displaying behavior like that of UV bursts. In addition, most TBs identified in our study occur either at bald patches, near the fan traces of coronal null points, or both -- locations that suggest these are small-scale reconnection events \citep[e.g.][]{Aulanier:1998, Demoulin:1999}. Given the good spatial alignment with the photospheric footprints of these topological features, we therefore conclude that the reconnection site for TBs is lower lying, consistent with the physical interpretation of UV bursts and further reinforcing the connection between the two phenomena. However, the relationship between larger TB clusters during active epochs and UV bursts remains unclear, as UV bursts have not yet been systematically examined during the pre-flare phase. We speculate that there may be two types of TBs: 1) precursors, which cover a larger area and are co-spatial with specific features in the magnetic environment, and 2) signatures of UV bursts.

\section{Conclusions}
The goal of this paper series is to examine whether small-scale activity observed prior to solar energetic events -- commonly referred to as precursors -- are causally linked to the pre-event phase, or if they simply represent random activity in the lower solar atmosphere. This first paper introduces the overarching concept and methodology for the planned statistical analysis, highlighting the following findings based on its application to a few events:
\begin{itemize}
\item Transient brightening activity is observed prior to solar flares but also during quiet epochs of the same active regions when not producing any flares.
\item Prior to flaring, TBs form a large cluster near the future flare ribbon site, co-spatial with reconnection signatures like bald patches and coronal null point fan traces. They are concentrated around strong-gradient polarity inversion lines and regions of increased magnetic energy density.
\item TBs observed during quiet epochs in the same active regions form smaller, irregular clusters, occasionally linked to short, intermittent bald patches and fan traces, but mostly located away from strong-gradient polarity inversion lines and in areas with low energy density.
\end{itemize}

\begin{acknowledgments}
We gratefully acknowledge the support by NASA's Heliophysics Guest Investigator Program (80NSSC21K0738) and the National
Science Foundation under grant no. 2154653. SDO data are courtesy of the NASA/SDO AIA and HMI science teams.
\end{acknowledgments}
\software{Sunpy \citep[version 4.1.7;][]{sunpy_community2020}, Astropy \citep{astropy:2013, astropy:2018, astropy:2022}, hdbscan \citep[][]{McInnes2017}, IDL 8.5, SolarSoft \citep{Freeland:1998, Freeland:2012}, SHTools \citep{Wieczorek:2018}, Basis Pursuit \citep[DEM code;][]{Cheung:2015}}
\facilities{SDO (AIA and HMI)}
\section*{Data Availability}
Replication data for this study can be found on the Harvard Dataverse.
These data include --
Transient Brightenings Data: \dataset[doi:10.7910/DVN/LTHRD4]{https://doi.org/10.7910/DVN/LTHRD4} \citep{tb_data};
Magnetic Environment Data: \dataset[doi:10.7910/DVN/XGZFJL]{https://doi.org/10.7910/DVN/XGZFJL} \citep{mag_data};
Plasma Environment Data: \dataset[doi:10.7910/DVN/CQ2V9Q]{https://doi.org/10.7910/DVN/CQ2V9Q} \citep{dem_data};
Magnetic Skeleton Data: \dataset[doi:10.7910/DVN/DFKU8H]{https://doi.org/10.7910/DVN/DFKU8H} \citep{skeleton_data};
ReadMe files are provided for each deposit documenting the file contents.
%


\begin{thebibliography}{}
\expandafter\ifx\csname natexlab\endcsname\relax\def\natexlab#1{#1}\fi
\providecommand{\url}[1]{\href{#1}{#1}}
\providecommand{\dodoi}[1]{doi:~\href{http://doi.org/#1}{\nolinkurl{#1}}}
\providecommand{\doeprint}[1]{\href{http://ascl.net/#1}{\nolinkurl{http://ascl.net/#1}}}
\providecommand{\doarXiv}[1]{\href{https://arxiv.org/abs/#1}{\nolinkurl{https://arxiv.org/abs/#1}}}

\bibitem[{M.~D. {Altschuler} \& G. {Newkirk}(1969){Altschuler} \& {Newkirk}}]{Altschuler:1969}
{Altschuler}, M.~D., \& {Newkirk}, G. 1969, \bibinfo{title}{{Magnetic Fields and the Structure of the Solar Corona. I: Methods of Calculating Coronal Fields},} \solphys, 9, 131, \dodoi{10.1007/BF00145734}

\bibitem[{M.~J. {Aschwanden}(2004){Aschwanden}}]{Aschwanden:2004}
{Aschwanden}, M.~J. 2004, {Physics of the Solar Corona. An Introduction}

\bibitem[{ {Astropy Collaboration} {et~al.}(2013){Astropy Collaboration}, {Robitaille}, {Tollerud}, {Greenfield}, {Droettboom}, {Bray}, {Aldcroft}, {Davis}, {Ginsburg}, {Price-Whelan}, {Kerzendorf}, {Conley}, {Crighton}, {Barbary}, {Muna}, {Ferguson}, {Grollier}, {Parikh}, {Nair}, {Unther}, {Deil}, {Woillez}, {Conseil}, {Kramer}, {Turner}, {Singer}, {Fox}, {Weaver}, {Zabalza}, {Edwards}, {Azalee Bostroem}, {Burke}, {Casey}, {Crawford}, {Dencheva}, {Ely}, {Jenness}, {Labrie}, {Lim}, {Pierfederici}, {Pontzen}, {Ptak}, {Refsdal}, {Servillat}, \& {Streicher}}]{astropy:2013}
{Astropy Collaboration}, {Robitaille}, T.~P., {Tollerud}, E.~J., {et~al.} 2013, \bibinfo{title}{{Astropy: A community Python package for astronomy},} \aap, 558, A33, \dodoi{10.1051/0004-6361/201322068}

\bibitem[{ {Astropy Collaboration} {et~al.}(2018){Astropy Collaboration}, {Price-Whelan}, {Sip{\H{o}}cz}, {G{\"u}nther}, {Lim}, {Crawford}, {Conseil}, {Shupe}, {Craig}, {Dencheva}, {Ginsburg}, {Vand erPlas}, {Bradley}, {P{\'e}rez-Su{\'a}rez}, {de Val-Borro}, {Aldcroft}, {Cruz}, {Robitaille}, {Tollerud}, {Ardelean}, {Babej}, {Bach}, {Bachetti}, {Bakanov}, {Bamford}, {Barentsen}, {Barmby}, {Baumbach}, {Berry}, {Biscani}, {Boquien}, {Bostroem}, {Bouma}, {Brammer}, {Bray}, {Breytenbach}, {Buddelmeijer}, {Burke}, {Calderone}, {Cano Rodr{\'\i}guez}, {Cara}, {Cardoso}, {Cheedella}, {Copin}, {Corrales}, {Crichton}, {D'Avella}, {Deil}, {Depagne}, {Dietrich}, {Donath}, {Droettboom}, {Earl}, {Erben}, {Fabbro}, {Ferreira}, {Finethy}, {Fox}, {Garrison}, {Gibbons}, {Goldstein}, {Gommers}, {Greco}, {Greenfield}, {Groener}, {Grollier}, {Hagen}, {Hirst}, {Homeier}, {Horton}, {Hosseinzadeh}, {Hu}, {Hunkeler}, {Ivezi{\'c}}, {Jain}, {Jenness}, {Kanarek}, {Kendrew}, {Kern}, {Kerzendorf}, {Khvalko}, {King}, {Kirkby}, {Kulkarni},
  {Kumar}, {Lee}, {Lenz}, {Littlefair}, {Ma}, {Macleod}, {Mastropietro}, {McCully}, {Montagnac}, {Morris}, {Mueller}, {Mumford}, {Muna}, {Murphy}, {Nelson}, {Nguyen}, {Ninan}, {N{\"o}the}, {Ogaz}, {Oh}, {Parejko}, {Parley}, {Pascual}, {Patil}, {Patil}, {Plunkett}, {Prochaska}, {Rastogi}, {Reddy Janga}, {Sabater}, {Sakurikar}, {Seifert}, {Sherbert}, {Sherwood-Taylor}, {Shih}, {Sick}, {Silbiger}, {Singanamalla}, {Singer}, {Sladen}, {Sooley}, {Sornarajah}, {Streicher}, {Teuben}, {Thomas}, {Tremblay}, {Turner}, {Terr{\'o}n}, {van Kerkwijk}, {de la Vega}, {Watkins}, {Weaver}, {Whitmore}, {Woillez}, {Zabalza}, \& {Astropy Contributors}}]{astropy:2018}
{Astropy Collaboration}, {Price-Whelan}, A.~M., {Sip{\H{o}}cz}, B.~M., {et~al.} 2018, \bibinfo{title}{{The Astropy Project: Building an Open-science Project and Status of the v2.0 Core Package},} \aj, 156, 123, \dodoi{10.3847/1538-3881/aabc4f}

\bibitem[{ {Astropy Collaboration} {et~al.}(2022){Astropy Collaboration}, {Price-Whelan}, {Lim}, {Earl}, {Starkman}, {Bradley}, {Shupe}, {Patil}, {Corrales}, {Brasseur}, {N{"o}the}, {Donath}, {Tollerud}, {Morris}, {Ginsburg}, {Vaher}, {Weaver}, {Tocknell}, {Jamieson}, {van Kerkwijk}, {Robitaille}, {Merry}, {Bachetti}, {G{"u}nther}, {Aldcroft}, {Alvarado-Montes}, {Archibald}, {B{'o}di}, {Bapat}, {Barentsen}, {Baz{'a}n}, {Biswas}, {Boquien}, {Burke}, {Cara}, {Cara}, {Conroy}, {Conseil}, {Craig}, {Cross}, {Cruz}, {D'Eugenio}, {Dencheva}, {Devillepoix}, {Dietrich}, {Eigenbrot}, {Erben}, {Ferreira}, {Foreman-Mackey}, {Fox}, {Freij}, {Garg}, {Geda}, {Glattly}, {Gondhalekar}, {Gordon}, {Grant}, {Greenfield}, {Groener}, {Guest}, {Gurovich}, {Handberg}, {Hart}, {Hatfield-Dodds}, {Homeier}, {Hosseinzadeh}, {Jenness}, {Jones}, {Joseph}, {Kalmbach}, {Karamehmetoglu}, {Ka{l}uszy{'n}ski}, {Kelley}, {Kern}, {Kerzendorf}, {Koch}, {Kulumani}, {Lee}, {Ly}, {Ma}, {MacBride}, {Maljaars}, {Muna}, {Murphy}, {Norman}, {O'Steen},
  {Oman}, {Pacifici}, {Pascual}, {Pascual-Granado}, {Patil}, {Perren}, {Pickering}, {Rastogi}, {Roulston}, {Ryan}, {Rykoff}, {Sabater}, {Sakurikar}, {Salgado}, {Sanghi}, {Saunders}, {Savchenko}, {Schwardt}, {Seifert-Eckert}, {Shih}, {Jain}, {Shukla}, {Sick}, {Simpson}, {Singanamalla}, {Singer}, {Singhal}, {Sinha}, {Sip{H{o}}cz}, {Spitler}, {Stansby}, {Streicher}, {{{S}}umak}, {Swinbank}, {Taranu}, {Tewary}, {Tremblay}, {Val-Borro}, {Van Kooten}, {Vasovi{'c}}, {Verma}, {de Miranda Cardoso}, {Williams}, {Wilson}, {Winkel}, {Wood-Vasey}, {Xue}, {Yoachim}, {Zhang}, {Zonca}, \& {Astropy Project Contributors}}]{astropy:2022}
{Astropy Collaboration}, {Price-Whelan}, A.~M., {Lim}, P.~L., {et~al.} 2022, \bibinfo{title}{{The Astropy Project: Sustaining and Growing a Community-oriented Open-source Project and the Latest Major Release (v5.0) of the Core Package},} \apj, 935, 167, \dodoi{10.3847/1538-4357/ac7c74}

\bibitem[{G. {Aulanier} {et~al.}(1998){Aulanier}, {D{\'e}moulin}, {Schmieder}, {Fang}, \& {Tang}}]{Aulanier:1998}
{Aulanier}, G., {D{\'e}moulin}, P., {Schmieder}, B., {Fang}, C., \& {Tang}, Y.~H. 1998, \bibinfo{title}{{Magnetohydrostatic Model of a Bald-Patch Flare},} \solphys, 183, 369, \dodoi{10.1023/A:1005003426798}

\bibitem[{Y. {Bamba} {et~al.}(2014){Bamba}, {Kusano}, {Imada}, \& {Iida}}]{Bamba:2014}
{Bamba}, Y., {Kusano}, K., {Imada}, S., \& {Iida}, Y. 2014, \bibinfo{title}{{Comparison between Hinode/SOT and SDO/HMI, AIA data for the study of the solar flare trigger process},} \pasj, 66, S16, \dodoi{10.1093/pasj/psu091}

\bibitem[{Y. {Bamba} {et~al.}(2017){Bamba}, {Lee}, {Imada}, \& {Kusano}}]{Bamba:2017}
{Bamba}, Y., {Lee}, K.-S., {Imada}, S., \& {Kusano}, K. 2017, \bibinfo{title}{{Study on Precursor Activity of the X1.6 Flare in the Great AR 12192 with SDO, IRIS, and Hinode},} \apj, 840, 116, \dodoi{10.3847/1538-4357/aa6dfe}

\bibitem[{G. {Barnes} {et~al.}(2025){Barnes}, {Dissauer}, {Leka}, \& {Wagner}}]{skeleton_data}
{Barnes}, G., {Dissauer}, K., {Leka}, K.~D., \& {Wagner}, E.~L. 2025, \bibinfo{title}{{Replication Data for ``On the Uniqueness and Causal Relationship of Precursor Activity to Solar Energetic Events: I. Transient Brightenings -- Introduction and Overview'': Magnetic Skeleton Data},}, v1 Harvard Dataverse, \dodoi{provided_when_accepted}

\bibitem[{A.~F. {Battaglia} {et~al.}(2023){Battaglia}, {Hudson}, {Warmuth}, {Collier}, {Jeffrey}, {Caspi}, {Dickson}, {Saqri}, {Purkhart}, {Veronig}, {Harra}, \& {Krucker}}]{Battaglia:2023}
{Battaglia}, A.~F., {Hudson}, H., {Warmuth}, A., {et~al.} 2023, \bibinfo{title}{{The existence of hot X-ray onsets in solar flares},} \aap, 679, A139, \dodoi{10.1051/0004-6361/202347706}

\bibitem[{M. {Battaglia} {et~al.}(2009){Battaglia}, {Fletcher}, \& {Benz}}]{Battaglia:2009}
{Battaglia}, M., {Fletcher}, L., \& {Benz}, A.~O. 2009, \bibinfo{title}{{Observations of conduction driven evaporation in the early rise phase of solar flares},} \aap, 498, 891, \dodoi{10.1051/0004-6361/200811196}

\bibitem[{A.~O. {Benz} {et~al.}(1983){Benz}, {Barrow}, {Dennis}, {Pick}, {Raoult}, \& {Simnett}}]{Benz:1983}
{Benz}, A.~O., {Barrow}, C.~H., {Dennis}, B.~R., {et~al.} 1983, \bibinfo{title}{{X-Ray and Radio Emissions in the Early Stages of Solar Flares},} \solphys, 83, 267, \dodoi{10.1007/BF00148280}

\bibitem[{M.~G. {Bobra} \& S. {Couvidat}(2015){Bobra} \& {Couvidat}}]{BobraCouvidat2015}
{Bobra}, M.~G., \& {Couvidat}, S. 2015, \bibinfo{title}{{Solar Flare Prediction Using SDO/HMI Vector Magnetic Field Data with a Machine-Learning Algorithm},} \apj, 798, 135, \dodoi{10.1088/0004-637X/798/2/135}

\bibitem[{R.~J. G.~B. Campello {et~al.}(2013)Campello, Moulavi, \& Sander}]{Campello:2013}
Campello, R. J. G.~B., Moulavi, D., \& Sander, J. 2013, in Advances in Knowledge Discovery and Data Mining, ed. J.~Pei, V.~S. Tseng, L.~Cao, H.~Motoda, \& G.~Xu (Berlin, Heidelberg: Springer Berlin Heidelberg), 160--172

\bibitem[{R.~J. G.~B. Campello {et~al.}(2015)Campello, Moulavi, Zimek, \& Sander}]{Campello:2015}
Campello, R. J. G.~B., Moulavi, D., Zimek, A., \& Sander, J. 2015, \bibinfo{title}{Hierarchical Density Estimates for Data Clustering, Visualization, and Outlier Detection,} ACM Trans. Knowl. Discov. Data, 10, \dodoi{10.1145/2733381}

\bibitem[{J. {Chen} {et~al.}(2020){Chen}, {Liu}, {Liu}, {Awasthi}, {Zhang}, {Wang}, \& {Kliem}}]{Chen:2020}
{Chen}, J., {Liu}, R., {Liu}, K., {et~al.} 2020, \bibinfo{title}{{Extreme-ultraviolet Late Phase of Solar Flares},} \apj, 890, 158, \dodoi{10.3847/1538-4357/ab6def}

\bibitem[{X. {Cheng} {et~al.}(2012){Cheng}, {Zhang}, {Saar}, \& {Ding}}]{Cheng:2012}
{Cheng}, X., {Zhang}, J., {Saar}, S.~H., \& {Ding}, M.~D. 2012, \bibinfo{title}{{Differential Emission Measure Analysis of Multiple Structural Components of Coronal Mass Ejections in the Inner Corona},} \apj, 761, 62, \dodoi{10.1088/0004-637X/761/1/62}

\bibitem[{M.~C.~M. {Cheung} {et~al.}(2015){Cheung}, {Boerner}, {Schrijver}, {Testa}, {Chen}, {Peter}, \& {Malanushenko}}]{Cheung:2015}
{Cheung}, M. C.~M., {Boerner}, P., {Schrijver}, C.~J., {et~al.} 2015, \bibinfo{title}{{Thermal Diagnostics with the Atmospheric Imaging Assembly on board the Solar Dynamics Observatory: A Validated Method for Differential Emission Measure Inversions},} \apj, 807, 143, \dodoi{10.1088/0004-637X/807/2/143}

\bibitem[{C. {Chifor} {et~al.}(2007){Chifor}, {Tripathi}, {Mason}, \& {Dennis}}]{Chifor:2007}
{Chifor}, C., {Tripathi}, D., {Mason}, H.~E., \& {Dennis}, B.~R. 2007, \bibinfo{title}{{X-ray precursors to flares and filament eruptions},} \aap, 472, 967, \dodoi{10.1051/0004-6361:20077771}

\bibitem[{K. {Cho} {et~al.}(2016){Cho}, {Lee}, {Chae}, {Wang}, {Ahn}, {Yang}, {Lim}, \& {Maurya}}]{Cho:2016}
{Cho}, K., {Lee}, J., {Chae}, J., {et~al.} 2016, \bibinfo{title}{{Strong Blue Asymmetry in H{\ensuremath{\alpha}} Line as a Preflare Activity},} \solphys, 291, 2391, \dodoi{10.1007/s11207-016-0963-5}

\bibitem[{D.~F. {da Silva} {et~al.}(2023){da Silva}, {Hui}, {Sim{\~o}es}, {Valio}, {Costa}, {Hudson}, {Fletcher}, {Hayes}, \& {Hannah}}]{daSilva:2023}
{da Silva}, D.~F., {Hui}, L., {Sim{\~o}es}, P. J.~A., {et~al.} 2023, \bibinfo{title}{{Statistical analysis of the onset temperature of solar flares in 2010-2011},} \mnras, 525, 4143, \dodoi{10.1093/mnras/stad2244}

\bibitem[{P. {D{\'e}moulin} {et~al.}(1999){D{\'e}moulin}, {Aulanier}, \& {Schmieder}}]{Demoulin:1999}
{D{\'e}moulin}, P., {Aulanier}, G., \& {Schmieder}, B. 1999, in Astronomical Society of the Pacific Conference Series, Vol. 184, Third Advances in Solar Physics Euroconference: Magnetic Fields and Oscillations, ed. B.~{Schmieder}, A.~{Hofmann}, \& J.~{Staude}, 65--69

\bibitem[{M.~L. {DeRosa} {et~al.}(2015){DeRosa}, {Wheatland}, {Leka}, {Barnes}, {Amari}, {Canou}, {Gilchrist}, {Thalmann}, {Valori}, {Wiegelmann}, {Schrijver}, {Malanushenko}, {Sun}, \& {R{\'e}gnier}}]{DeRosa:2015}
{DeRosa}, M.~L., {Wheatland}, M.~S., {Leka}, K.~D., {et~al.} 2015, \bibinfo{title}{{The Influence of Spatial resolution on Nonlinear Force-free Modeling},} \apj, 811, 107, \dodoi{10.1088/0004-637X/811/2/107}

\bibitem[{K. {Dissauer} {et~al.}(2025{\natexlab{a}}){Dissauer}, {Barnes}, {Leka}, \& {Wagner}}]{tb_data}
{Dissauer}, K., {Barnes}, G., {Leka}, K.~D., \& {Wagner}, E.~L. 2025{\natexlab{a}}, \bibinfo{title}{{Replication Data for ``On the Uniqueness and Causal Relationship of Precursor Activity to Solar Energetic Events: I. Transient Brightenings -- Introduction and Overview'': Transient Brightenings Data},}, v1 Harvard Dataverse, \dodoi{provided_when_accepted}

\bibitem[{K. {Dissauer} {et~al.}(2025{\natexlab{b}}){Dissauer}, {Barnes}, {Leka}, \& {Wagner}}]{dem_data}
{Dissauer}, K., {Barnes}, G., {Leka}, K.~D., \& {Wagner}, E.~L. 2025{\natexlab{b}}, \bibinfo{title}{{Replication Data for ``On the Uniqueness and Causal Relationship of Precursor Activity to Solar Energetic Events: I. Transient Brightenings -- Introduction and Overview'': Plasma Environment Data},}, v1 Harvard Dataverse, \dodoi{provided_when_accepted}

\bibitem[{K. {Dissauer} {et~al.}(2023){Dissauer}, {Leka}, \& {Wagner}}]{Dissauer:2023}
{Dissauer}, K., {Leka}, K.~D., \& {Wagner}, E.~L. 2023, \bibinfo{title}{{Properties of Flare-imminent versus Flare-quiet Active Regions from the Chromosphere through the Corona. I. Introduction of the AIA Active Region Patches (AARPs)},} \apj, 942, 83, \dodoi{10.3847/1538-4357/ac9c06}

\bibitem[{K. {Dissauer} {et~al.}(2018){Dissauer}, {Veronig}, {Temmer}, {Podladchikova}, \& {Vanninathan}}]{Dissauer:2018a}
{Dissauer}, K., {Veronig}, A.~M., {Temmer}, M., {Podladchikova}, T., \& {Vanninathan}, K. 2018, \bibinfo{title}{{On the Detection of Coronal Dimmings and the Extraction of Their Characteristic Properties},} \apj, 855, 137, \dodoi{10.3847/1538-4357/aaadb5}

\bibitem[{F. {F{\'a}rn{\'\i}k} {et~al.}(1996){F{\'a}rn{\'\i}k}, {Hudson}, \& {Watanabe}}]{Farnik:1996}
{F{\'a}rn{\'\i}k}, F., {Hudson}, H., \& {Watanabe}, T. 1996, \bibinfo{title}{{Spatial Relations between Preflares and Flares},} \solphys, 165, 169, \dodoi{10.1007/BF00149096}

\bibitem[{F. {F{\'a}rn{\'\i}k} {et~al.}(2003){F{\'a}rn{\'\i}k}, {Hudson}, {Karlick{\'y}}, \& {Kosugi}}]{Farnik:2003}
{F{\'a}rn{\'\i}k}, F., {Hudson}, H.~S., {Karlick{\'y}}, M., \& {Kosugi}, T. 2003, \bibinfo{title}{{X-ray and radio observations of the activation stages of an X-class solar flare},} \aap, 399, 1159, \dodoi{10.1051/0004-6361:20021852}

\bibitem[{F. {F{\'a}rn{\'\i}k} \& S.~K. {Savy}(1998){F{\'a}rn{\'\i}k} \& {Savy}}]{Farnik:1998}
{F{\'a}rn{\'\i}k}, F., \& {Savy}, S.~K. 1998, \bibinfo{title}{{Soft X-Ray Pre-Flare Emission Studied in Yohkoh-SXT Images},} \solphys, 183, 339, \dodoi{10.1023/A:1005092927592}

\bibitem[{K. {Florios} {et~al.}(2018){Florios}, {Kontogiannis}, {Park}, {Guerra}, {Benvenuto}, {Bloomfield}, \& {Georgoulis}}]{Florios_etal_2018}
{Florios}, K., {Kontogiannis}, I., {Park}, S.-H., {et~al.} 2018, \bibinfo{title}{{Forecasting Solar Flares Using Magnetogram-based Predictors and Machine Learning},} \solphys, 293, 28, \dodoi{10.1007/s11207-018-1250-4}

\bibitem[{S.~L. {Freeland} \& B.~N. {Handy}(1998){Freeland} \& {Handy}}]{Freeland:1998}
{Freeland}, S.~L., \& {Handy}, B.~N. 1998, \bibinfo{title}{{Data Analysis with the SolarSoft System},} \solphys, 182, 497, \dodoi{10.1023/A:1005038224881}

\bibitem[{S.~L. {Freeland} \& B.~N. {Handy}(2012){Freeland} \& {Handy}}]{Freeland:2012}
{Freeland}, S.~L., \& {Handy}, B.~N. 2012, \bibinfo{title}{{SolarSoft: Programming and data analysis environment for solar physics},}, Astrophysics Source Code Library, record ascl:1208.013

\bibitem[{M.~K. {Georgoulis} {et~al.}(2021){Georgoulis}, {Bloomfield}, {Piana}, {Massone}, {Soldati}, {Gallagher}, {Pariat}, {Vilmer}, {Buchlin}, {Baudin}, {Csillaghy}, {Sathiapal}, {Jackson}, {Alingery}, {Benvenuto}, {Campi}, {Florios}, {Gontikakis}, {Guennou}, {Guerra}, {Kontogiannis}, {Latorre}, {Murray}, {Park}, {von Stachelski}, {Torbica}, {Vischi}, \& {Worsfold}}]{Georgoulis:2021}
{Georgoulis}, M.~K., {Bloomfield}, D.~S., {Piana}, M., {et~al.} 2021, \bibinfo{title}{{The flare likelihood and region eruption forecasting (FLARECAST) project: flare forecasting in the big data \& machine learning era},} Journal of Space Weather and Space Climate, 11, 39, \dodoi{10.1051/swsc/2021023}

\bibitem[{L.~M. {Green} {et~al.}(2018){Green}, {T{\"o}r{\"o}k}, {Vr{\v{s}}nak}, {Manchester}, \& {Veronig}}]{Green:2018}
{Green}, L.~M., {T{\"o}r{\"o}k}, T., {Vr{\v{s}}nak}, B., {Manchester}, W., \& {Veronig}, A. 2018, \bibinfo{title}{{The Origin, Early Evolution and Predictability of Solar Eruptions},} \ssr, 214, 46, \dodoi{10.1007/s11214-017-0462-5}

\bibitem[{N. {Gyenge} {et~al.}(2016){Gyenge}, {Ballai}, \& {Baranyi}}]{Gyenge:2016}
{Gyenge}, N., {Ballai}, I., \& {Baranyi}, T. 2016, \bibinfo{title}{{Statistical study of spatio-temporal distribution of precursor solar flares associated with major flares},} \mnras, 459, 3532, \dodoi{10.1093/mnras/stw859}

\bibitem[{A.~L. {Haynes} \& C.~E. {Parnell}(2007){Haynes} \& {Parnell}}]{Haynes:2007}
{Haynes}, A.~L., \& {Parnell}, C.~E. 2007, \bibinfo{title}{{A trilinear method for finding null points in a three-dimensional vector space},} Physics of Plasmas, 14, 082107, \dodoi{10.1063/1.2756751}

\bibitem[{A.~L. {Haynes} \& C.~E. {Parnell}(2010){Haynes} \& {Parnell}}]{Haynes:2010}
{Haynes}, A.~L., \& {Parnell}, C.~E. 2010, \bibinfo{title}{A method for finding three-dimensional magnetic skeletons,} Physics of Plasmas, 17, 092903, \dodoi{10.1063/1.3467499}

\bibitem[{W. {He} {et~al.}(2023){He}, {Jing}, {Wang}, {Nayak}, \& {Prasad}}]{He:2023}
{He}, W., {Jing}, J., {Wang}, H., {Nayak}, S.~S., \& {Prasad}, A. 2023, \bibinfo{title}{{Coronal Magnetic Field Extrapolation and Topological Analysis of Fine-scale Structures during Solar Flare Precursors},} \apj, 958, 90, \dodoi{10.3847/1538-4357/ad0236}

\bibitem[{A. {Hernandez-Perez} {et~al.}(2019{\natexlab{a}}){Hernandez-Perez}, {Su}, {Thalmann}, {Veronig}, {Dickson}, {Dissauer}, {Joshi}, \& {Chandra}}]{Hernandez-Perez:2019b}
{Hernandez-Perez}, A., {Su}, Y., {Thalmann}, J., {et~al.} 2019{\natexlab{a}}, \bibinfo{title}{{A Hot Cusp-shaped Confined Solar Flare},} \apjl, 887, L28, \dodoi{10.3847/2041-8213/ab5ba1}

\bibitem[{A. {Hernandez-Perez} {et~al.}(2019{\natexlab{b}}){Hernandez-Perez}, {Su}, {Veronig}, {Thalmann}, {G{\"o}m{\"o}ry}, \& {Joshi}}]{Hernandez-Perez:2019a}
{Hernandez-Perez}, A., {Su}, Y., {Veronig}, A.~M., {et~al.} 2019{\natexlab{b}}, \bibinfo{title}{{Pre-eruption Processes: Heating, Particle Acceleration, and the Formation of a Hot Channel before the 2012 October 20 M9.0 Limb Flare},} \apj, 874, 122, \dodoi{10.3847/1538-4357/ab09ed}

\bibitem[{J.~T. {Hoeksema} {et~al.}(2014){Hoeksema}, {Liu}, {Hayashi}, {Sun}, {Schou}, {Couvidat}, {Norton}, {Bobra}, {Centeno}, {Leka}, {Barnes}, \& {Turmon}}]{Hoeksema:2014}
{Hoeksema}, J.~T., {Liu}, Y., {Hayashi}, K., {et~al.} 2014, \bibinfo{title}{{The Helioseismic and Magnetic Imager (HMI) Vector Magnetic Field Pipeline: Overview and Performance},} \solphys, 289, 3483, \dodoi{10.1007/s11207-014-0516-8}

\bibitem[{H.~S. {Hudson} {et~al.}(2021){Hudson}, {Sim{\~o}es}, {Fletcher}, {Hayes}, \& {Hannah}}]{Hudson:2021}
{Hudson}, H.~S., {Sim{\~o}es}, P. J.~A., {Fletcher}, L., {Hayes}, L.~A., \& {Hannah}, I.~G. 2021, \bibinfo{title}{{Hot X-ray onsets of solar flares},} \mnras, 501, 1273, \dodoi{10.1093/mnras/staa3664}

\bibitem[{H.~S. {Hudson} {et~al.}(2020){Hudson}, {Simoes}, {Fletcher}, {Hayes}, \& {Hannah}}]{Hudson:2020}
{Hudson}, H.~S., {Simoes}, P. J.~A., {Fletcher}, L., {Hayes}, L.~A., \& {Hannah}, I.~G. 2020, \bibinfo{title}{{Hot X-ray Onsets of Solar Flares},} arXiv e-prints, arXiv:2007.05310.
\newblock \doarXiv{2007.05310}

\bibitem[{N. {Ishiguro} \& K. {Kusano}(2017){Ishiguro} \& {Kusano}}]{IshiguroKusano2017}
{Ishiguro}, N., \& {Kusano}, K. 2017, \bibinfo{title}{{Double Arc Instability in the Solar Corona},} \apj, 843, 101, \dodoi{10.3847/1538-4357/aa799b}

\bibitem[{M. {Janvier} {et~al.}(2016){Janvier}, {Savcheva}, {Pariat}, {Tassev}, {Millholland}, {Bommier}, {McCauley}, {McKillop}, \& {Dougan}}]{Janvier:2016}
{Janvier}, M., {Savcheva}, A., {Pariat}, E., {et~al.} 2016, \bibinfo{title}{{Evolution of flare ribbons, electric currents, and quasi-separatrix layers during an X-class flare},} \aap, 591, A141, \dodoi{10.1051/0004-6361/201628406}

\bibitem[{C. {Jiang} {et~al.}(2018){Jiang}, {Feng}, \& {Hu}}]{Jiang:2018}
{Jiang}, C., {Feng}, X., \& {Hu}, Q. 2018, \bibinfo{title}{{Formation and Eruption of an Active Region Sigmoid. II. Magnetohydrodynamic Simulation of a Multistage Eruption},} \apj, 866, 96, \dodoi{10.3847/1538-4357/aadd08}

\bibitem[{C. {Jiang} {et~al.}(2013){Jiang}, {Feng}, {Wu}, \& {Hu}}]{Jiang:2013}
{Jiang}, C., {Feng}, X., {Wu}, S.~T., \& {Hu}, Q. 2013, \bibinfo{title}{{Magnetohydrodynamic Simulation of a Sigmoid Eruption of Active Region 11283},} \apjl, 771, L30, \dodoi{10.1088/2041-8205/771/2/L30}

\bibitem[{C. {Jiang} {et~al.}(2014){Jiang}, {Wu}, {Feng}, \& {Hu}}]{Jiang:2014}
{Jiang}, C., {Wu}, S.~T., {Feng}, X., \& {Hu}, Q. 2014, \bibinfo{title}{{Formation and Eruption of an Active Region Sigmoid. I. A Study by Nonlinear Force-free Field Modeling},} \apj, 780, 55, \dodoi{10.1088/0004-637X/780/1/55}

\bibitem[{B. {Joshi} {et~al.}(2011){Joshi}, {Veronig}, {Lee}, {Bong}, {Tiwari}, \& {Cho}}]{Joshi:2011}
{Joshi}, B., {Veronig}, A.~M., {Lee}, J., {et~al.} 2011, \bibinfo{title}{{Pre-flare Activity and Magnetic Reconnection during the Evolutionary Stages of Energy Release in a Solar Eruptive Flare},} \apj, 743, 195, \dodoi{10.1088/0004-637X/743/2/195}

\bibitem[{M.~D. {Kazachenko} {et~al.}(2012){Kazachenko}, {Canfield}, {Longcope}, \& {Qiu}}]{Kazachenko:2012}
{Kazachenko}, M.~D., {Canfield}, R.~C., {Longcope}, D.~W., \& {Qiu}, J. 2012, \bibinfo{title}{{Predictions of Energy and Helicity in Four Major Eruptive Solar Flares},} \solphys, 277, 165, \dodoi{10.1007/s11207-011-9786-6}

\bibitem[{M.~D. {Kazachenko} {et~al.}(2017){Kazachenko}, {Lynch}, {Welsch}, \& {Sun}}]{Kazachenko:2017}
{Kazachenko}, M.~D., {Lynch}, B.~J., {Welsch}, B.~T., \& {Sun}, X. 2017, \bibinfo{title}{{A Database of Flare Ribbon Properties from the Solar Dynamics Observatory. I. Reconnection Flux},} \apj, 845, 49, \dodoi{10.3847/1538-4357/aa7ed6}

\bibitem[{K. {Kusano} {et~al.}(2012){Kusano}, {Bamba}, {Yamamoto}, {Iida}, {Toriumi}, \& {Asai}}]{Kusano_etal_2012}
{Kusano}, K., {Bamba}, Y., {Yamamoto}, T.~T., {et~al.} 2012, \bibinfo{title}{{Magnetic Field Structures Triggering Solar Flares and Coronal Mass Ejections},} \apj, 760, 31, \dodoi{10.1088/0004-637X/760/1/31}

\bibitem[{K.~D. {Leka} \& G. {Barnes}(2003){Leka} \& {Barnes}}]{params}
{Leka}, K.~D., \& {Barnes}, G. 2003, \bibinfo{title}{{Photospheric Magnetic Field Properties of Flaring vs. Flare-Quiet Active Regions I: Data, General Analysis Approach, and Sample Results},} \apj, 595, 1277

\bibitem[{K.~D. {Leka} \& G. {Barnes}(2007){Leka} \& {Barnes}}]{dfa3}
{Leka}, K.~D., \& {Barnes}, G. 2007, \bibinfo{title}{{Photospheric Magnetic Field Properties of Flaring vs. Flare-Quiet Active Regions. IV: A Statistically Significant Sample},} \apj, 656, 1173, \dodoi{10.1086/510282}

\bibitem[{K.~D. {Leka} {et~al.}(2018){Leka}, {Barnes}, \& {Wagner}}]{Leka:2018}
{Leka}, K.~D., {Barnes}, G., \& {Wagner}, E. 2018, \bibinfo{title}{{The NWRA Classification Infrastructure: description and extension to the Discriminant Analysis Flare Forecasting System (DAFFS)},} Journal of Space Weather and Space Climate, 8, A25, \dodoi{10.1051/swsc/2018004}

\bibitem[{K.~D. {Leka} {et~al.}(2023){Leka}, {Dissauer}, {Barnes}, \& {Wagner}}]{Leka:2023}
{Leka}, K.~D., {Dissauer}, K., {Barnes}, G., \& {Wagner}, E.~L. 2023, \bibinfo{title}{{Properties of Flare-imminent versus Flare-quiet Active Regions from the Chromosphere through the Corona. II. Nonparametric Discriminant Analysis Results from the NWRA Classification Infrastructure (NCI)},} \apj, 942, 84, \dodoi{10.3847/1538-4357/ac9c04}

\bibitem[{K.~D. {Leka} {et~al.}(2025){Leka}, {Dissauer}, {Barnes}, \& {Wagner}}]{mag_data}
{Leka}, K.~D., {Dissauer}, K., {Barnes}, G., \& {Wagner}, E.~L. 2025, \bibinfo{title}{{Replication Data for ``On the Uniqueness and Causal Relationship of Precursor Activity to Solar Energetic Events: I. Transient Brightenings -- Introduction and Overview'': Magnetic Environment Data},}, v1 Harvard Dataverse, \dodoi{provided_when_accepted}

\bibitem[{J.~R. {Lemen} {et~al.}(2012){Lemen}, {Title}, {Akin}, {Boerner}, {Chou}, {Drake}, {Duncan}, {Edwards}, {Friedlaender}, {Heyman}, {Hurlburt}, {Katz}, {Kushner}, {Levay}, {Lindgren}, {Mathur}, {McFeaters}, {Mitchell}, {Rehse}, {Schrijver}, {Springer}, {Stern}, {Tarbell}, {Wuelser}, {Wolfson}, {Yanari}, {Bookbinder}, {Cheimets}, {Caldwell}, {Deluca}, {Gates}, {Golub}, {Park}, {Podgorski}, {Bush}, {Scherrer}, {Gummin}, {Smith}, {Auker}, {Jerram}, {Pool}, {Soufli}, {Windt}, {Beardsley}, {Clapp}, {Lang}, \& {Waltham}}]{Lemen:2012}
{Lemen}, J.~R., {Title}, A.~M., {Akin}, D.~J., {et~al.} 2012, \bibinfo{title}{{The Atmospheric Imaging Assembly (AIA) on the Solar Dynamics Observatory (SDO)},} \solphys, 275, 17, \dodoi{10.1007/s11207-011-9776-8}

\bibitem[{D.~W. {Longcope}(2005){Longcope}}]{Longcope:2005}
{Longcope}, D.~W. 2005, \bibinfo{title}{Topological Methods for the Analysis of Solar Magnetic Fields,} Living Reviews in Solar Physics, 2, 7, \dodoi{10.12942/lrsp-2005-7}

\bibitem[{C.~H. {Mandrini} {et~al.}(2014){Mandrini}, {Schmieder}, {D{\'e}moulin}, {Guo}, \& {Cristiani}}]{Mandrini:2014}
{Mandrini}, C.~H., {Schmieder}, B., {D{\'e}moulin}, P., {Guo}, Y., \& {Cristiani}, G.~D. 2014, \bibinfo{title}{{Topological Analysis of Emerging Bipole Clusters Producing Violent Solar Events},} \solphys, 289, 2041, \dodoi{10.1007/s11207-013-0458-6}

\bibitem[{S. {Masson} {et~al.}(2009){Masson}, {Pariat}, {Aulanier}, \& {Schrijver}}]{Masson:2009}
{Masson}, S., {Pariat}, E., {Aulanier}, G., \& {Schrijver}, C.~J. 2009, \bibinfo{title}{{The Nature of Flare Ribbons in Coronal Null-Point Topology},} \apj, 700, 559, \dodoi{10.1088/0004-637X/700/1/559}

\bibitem[{L. McInnes {et~al.}(2017)McInnes, Healy, \& Astels}]{McInnes2017}
McInnes, L., Healy, J., \& Astels, S. 2017, \bibinfo{title}{hdbscan: Hierarchical density based clustering,} The Journal of Open Source Software, 2, \dodoi{10.21105/joss.00205}

\bibitem[{W.~D. {Pesnell} {et~al.}(2012){Pesnell}, {Thompson}, \& {Chamberlin}}]{Pesnell:2012}
{Pesnell}, W.~D., {Thompson}, B.~J., \& {Chamberlin}, P.~C. 2012, \bibinfo{title}{{The Solar Dynamics Observatory (SDO)},} \solphys, 275, 3, \dodoi{10.1007/s11207-011-9841-3}

\bibitem[{H. {Peter} {et~al.}(2014){Peter}, {Tian}, {Curdt}, {Schmit}, {Innes}, {De Pontieu}, {Lemen}, {Title}, {Boerner}, {Hurlburt}, {Tarbell}, {Wuelser}, {Mart{\'\i}nez-Sykora}, {Kleint}, {Golub}, {McKillop}, {Reeves}, {Saar}, {Testa}, {Kankelborg}, {Jaeggli}, {Carlsson}, \& {Hansteen}}]{Peter:2014}
{Peter}, H., {Tian}, H., {Curdt}, W., {et~al.} 2014, \bibinfo{title}{{Hot explosions in the cool atmosphere of the Sun},} Science, 346, 1255726, \dodoi{10.1126/science.1255726}

\bibitem[{D.~I. {Pontin} \& E.~R. {Priest}(2022){Pontin} \& {Priest}}]{Pontin:2022}
{Pontin}, D.~I., \& {Priest}, E.~R. 2022, \bibinfo{title}{Magnetic reconnection: MHD theory and modelling,} Living Reviews in Solar Physics, 19, 1, \dodoi{10.1007/s41116-022-00032-9}

\bibitem[{A. {Prasad} {et~al.}(2020){Prasad}, {Dissauer}, {Hu}, {Bhattacharyya}, {Veronig}, {Kumar}, \& {Joshi}}]{Prasad:2020}
{Prasad}, A., {Dissauer}, K., {Hu}, Q., {et~al.} 2020, \bibinfo{title}{{Magnetohydrodynamic Simulation of Magnetic Null-point Reconnections and Coronal Dimmings during the X2.1 Flare in NOAA AR 11283},} \apj, 903, 129, \dodoi{10.3847/1538-4357/abb8d2}

\bibitem[{J. {Qiu} \& J. {Cheng}(2017){Qiu} \& {Cheng}}]{Qiu:2017}
{Qiu}, J., \& {Cheng}, J. 2017, \bibinfo{title}{{Gradual Solar Coronal Dimming and Evolution of Coronal Mass Ejection in the Early Phase},} \apjl, 838, L6, \dodoi{10.3847/2041-8213/aa6798}

\bibitem[{J. {Qiu} {et~al.}(2007){Qiu}, {Hu}, {Howard}, \& {Yurchyshyn}}]{Qiu:2007}
{Qiu}, J., {Hu}, Q., {Howard}, T.~A., \& {Yurchyshyn}, V.~B. 2007, \bibinfo{title}{{On the Magnetic Flux Budget in Low-Corona Magnetic Reconnection and Interplanetary Coronal Mass Ejections},} \apj, 659, 758, \dodoi{10.1086/512060}

\bibitem[{J. {Qiu} {et~al.}(2002){Qiu}, {Lee}, {Gary}, \& {Wang}}]{Qiu:2002}
{Qiu}, J., {Lee}, J., {Gary}, D.~E., \& {Wang}, H. 2002, \bibinfo{title}{{Motion of Flare Footpoint Emission and Inferred Electric Field in Reconnecting Current Sheets},} \apj, 565, 1335, \dodoi{10.1086/324706}

\bibitem[{J. {Qiu} {et~al.}(2004){Qiu}, {Wang}, {Cheng}, \& {Gary}}]{Qiu:2004}
{Qiu}, J., {Wang}, H., {Cheng}, C.~Z., \& {Gary}, D.~E. 2004, \bibinfo{title}{{Magnetic Reconnection and Mass Acceleration in Flare-Coronal Mass Ejection Events},} \apj, 604, 900, \dodoi{10.1086/382122}

\bibitem[{S. {R{\'e}gnier}(2012){R{\'e}gnier}}]{Regnier:2012}
{R{\'e}gnier}, S. 2012, \bibinfo{title}{{Magnetic Energy Storage and Current Density Distributions for Different Force-Free Models},} \solphys, 277, 131, \dodoi{10.1007/s11207-011-9830-6}

\bibitem[{J. {Saqri} {et~al.}(2020){Saqri}, {Veronig}, {Heinemann}, {Hofmeister}, {Temmer}, {Dissauer}, \& {Su}}]{Saqri:2020}
{Saqri}, J., {Veronig}, A.~M., {Heinemann}, S.~G., {et~al.} 2020, \bibinfo{title}{{Differential Emission Measure Plasma Diagnostics of a Long-Lived Coronal Hole},} \solphys, 295, 6, \dodoi{10.1007/s11207-019-1570-z}

\bibitem[{K.~H. {Schatten} {et~al.}(1969){Schatten}, {Wilcox}, \& {Ness}}]{Schatten:1969}
{Schatten}, K.~H., {Wilcox}, J.~M., \& {Ness}, N.~F. 1969, \bibinfo{title}{A model of interplanetary and coronal magnetic fields,} \solphys, 6, 442, \dodoi{10.1007/BF00146478}

\bibitem[{P.~H. {Scherrer} {et~al.}(2012){Scherrer}, {Schou}, {Bush}, {Kosovichev}, {Bogart}, {Hoeksema}, {Liu}, {Duvall}, {Zhao}, {Title}, {Schrijver}, {Tarbell}, \& {Tomczyk}}]{Scherrer:2012}
{Scherrer}, P.~H., {Schou}, J., {Bush}, R.~I., {et~al.} 2012, \bibinfo{title}{{The Helioseismic and Magnetic Imager (HMI) Investigation for the Solar Dynamics Observatory (SDO)},} \solphys, 275, 207, \dodoi{10.1007/s11207-011-9834-2}

\bibitem[{C.~J. {Schrijver}(2007){Schrijver}}]{Schrijver2007}
{Schrijver}, C.~J. 2007, \bibinfo{title}{{A Characteristic Magnetic Field Pattern Associated with All Major Solar Flares and Its Use in Flare Forecasting},} \apjl, 655, L117, \dodoi{10.1086/511857}

\bibitem[{N. {Seehafer}(1986){Seehafer}}]{Seehafer:1986}
{Seehafer}, N. 1986, \bibinfo{title}{{On the Magnetic Field Line Topology in Solar Active Regions},} \solphys, 105, 223, \dodoi{10.1007/BF00172044}

\bibitem[{A.~C. {Sterling} \& R.~L. {Moore}(2001){Sterling} \& {Moore}}]{Sterling:2001}
{Sterling}, A.~C., \& {Moore}, R.~L. 2001, \bibinfo{title}{{EIT Crinkles as Evidence for the Breakout Model of Solar Eruptions},} \apj, 560, 1045, \dodoi{10.1086/322241}

\bibitem[{A.~C. {Sterling} \& R.~L. {Moore}(2004){Sterling} \& {Moore}}]{Sterling:2004}
{Sterling}, A.~C., \& {Moore}, R.~L. 2004, \bibinfo{title}{{Evidence for Gradual External Reconnection before Explosive Eruption of a Solar Filament},} \apj, 602, 1024, \dodoi{10.1086/379763}

\bibitem[{A.~C. {Sterling} \& R.~L. {Moore}(2005){Sterling} \& {Moore}}]{Sterling:2005}
{Sterling}, A.~C., \& {Moore}, R.~L. 2005, \bibinfo{title}{{Slow-Rise and Fast-Rise Phases of an Erupting Solar Filament, and Flare Emission Onset},} \apj, 630, 1148, \dodoi{10.1086/432044}

\bibitem[{S.~J. {Tappin}(1991){Tappin}}]{Tappin:1991}
{Tappin}, S.~J. 1991, \bibinfo{title}{{Do all solar flares have X-ray precursors?},} \aaps, 87, 277

\bibitem[{ {The SunPy Community} {et~al.}(2020){The SunPy Community}, Barnes, Bobra, Christe, Freij, Hayes, Ireland, Mumford, Perez-Suarez, Ryan, Shih, Chanda, Glogowski, Hewett, Hughitt, Hill, Hiware, Inglis, Kirk, Konge, Mason, Maloney, Murray, Panda, Park, Pereira, Reardon, Savage, Sipőcz, Stansby, Jain, Taylor, Yadav, Rajul, \& Dang}]{sunpy_community2020}
{The SunPy Community}, Barnes, W.~T., Bobra, M.~G., {et~al.} 2020, \bibinfo{title}{The SunPy Project: Open Source Development and Status of the Version 1.0 Core Package,} The Astrophysical Journal, 890, 68, \dodoi{10.3847/1538-4357/ab4f7a}

\bibitem[{V.~S. {Titov} \& P. {D{\'e}moulin}(1999){Titov} \& {D{\'e}moulin}}]{Titov:1999}
{Titov}, V.~S., \& {D{\'e}moulin}, P. 1999, \bibinfo{title}{{Basic topology of twisted magnetic configurations in solar flares},} \aap, 351, 707

\bibitem[{V.~S. {Titov} {et~al.}(1993){Titov}, {Priest}, \& {D\'emoulin}}]{Titov:1993}
{Titov}, V.~S., {Priest}, E.~R., \& {D\'emoulin}, P. 1993, \bibinfo{title}{{Conditions for the appearance of ``bald patches'' at the solar surface},} \aap, 276, 564

\bibitem[{K. {Vanninathan} {et~al.}(2015){Vanninathan}, {Veronig}, {Dissauer}, {Madjarska}, {Hannah}, \& {Kontar}}]{Vanninathan:2015}
{Vanninathan}, K., {Veronig}, A.~M., {Dissauer}, K., {et~al.} 2015, \bibinfo{title}{{Coronal Response to an EUV Wave from DEM Analysis},} \apj, 812, 173, \dodoi{10.1088/0004-637X/812/2/173}

\bibitem[{H. {Wang} {et~al.}(2017){Wang}, {Liu}, {Ahn}, {Xu}, {Jing}, {Deng}, {Huang}, {Liu}, {Kusano}, {Fleishman}, {Gary}, \& {Cao}}]{Wang:2017}
{Wang}, H., {Liu}, C., {Ahn}, K., {et~al.} 2017, \bibinfo{title}{{High-resolution observations of flare precursors in the low solar atmosphere},} Nature Astronomy, 1, 0085, \dodoi{10.1038/s41550-017-0085}

\bibitem[{B.~T. {Welsch} {et~al.}(2009){Welsch}, {Li}, {Schuck}, \& {Fisher}}]{Welsch_etal_2009}
{Welsch}, B.~T., {Li}, Y., {Schuck}, P.~W., \& {Fisher}, G.~H. 2009, \bibinfo{title}{{What is the Relationship Between Photospheric Flow Fields and Solar Flares?},} \apj, 705, 821, \dodoi{10.1088/0004-637X/705/1/821}

\bibitem[{M.~A. {Wieczorek} \& M. {Meschede}(2018){Wieczorek} \& {Meschede}}]{Wieczorek:2018}
{Wieczorek}, M.~A., \& {Meschede}, M. 2018, \bibinfo{title}{SHTools -- Tools for working with spherical harmonics,} Geochemistry, Geophysics, Geosystems, 19, 2574, \dodoi{10.1029/2018GC007529}

\bibitem[{T. {Wiegelmann} \& T. {Sakurai}(2021){Wiegelmann} \& {Sakurai}}]{Wiegelmann2021}
{Wiegelmann}, T., \& {Sakurai}, T. 2021, \bibinfo{title}{{Solar force-free magnetic fields},} Living Reviews in Solar Physics, 18, 1, \dodoi{10.1007/s41116-020-00027-4}

\bibitem[{T.~N. {Woods} {et~al.}(2011){Woods}, {Hock}, {Eparvier}, {Jones}, {Chamberlin}, {Klimchuk}, {Didkovsky}, {Judge}, {Mariska}, {Warren}, {Schrijver}, {Webb}, {Bailey}, \& {Tobiska}}]{Woods:2011}
{Woods}, T.~N., {Hock}, R., {Eparvier}, F., {et~al.} 2011, \bibinfo{title}{{New Solar Extreme-ultraviolet Irradiance Observations during Flares},} \apj, 739, 59, \dodoi{10.1088/0004-637X/739/2/59}

\bibitem[{P.~R. {Young} {et~al.}(2018){Young}, {Tian}, {Peter}, {Rutten}, {Nelson}, {Huang}, {Schmieder}, {Vissers}, {Toriumi}, {Rouppe van der Voort}, {Madjarska}, {Danilovic}, {Berlicki}, {Chitta}, {Cheung}, {Madsen}, {Reardon}, {Katsukawa}, \& {Heinzel}}]{Young:2018}
{Young}, P.~R., {Tian}, H., {Peter}, H., {et~al.} 2018, \bibinfo{title}{{Solar Ultraviolet Bursts},} \ssr, 214, 120, \dodoi{10.1007/s11214-018-0551-0}

\bibitem[{Q.~M. {Zhang} {et~al.}(2017){Zhang}, {Su}, \& {Ji}}]{Zhang:2017}
{Zhang}, Q.~M., {Su}, Y.~N., \& {Ji}, H.~S. 2017, \bibinfo{title}{{Pre-flare coronal dimmings},} \aap, 598, A3, \dodoi{10.1051/0004-6361/201629477}

\bibitem[{C. {Zhu} {et~al.}(2024){Zhu}, {DeVore}, {Dahlin}, {Qiu}, {Kazachenko}, {Uritsky}, \& {Vandervelde}}]{Zhu:2024}
{Zhu}, C., {DeVore}, C.~R., {Dahlin}, J.~T., {et~al.} 2024, \bibinfo{title}{{Large-scale Coronal Dimming Foreshadowing a Solar Eruption on 2011 October 1},} \apj, 961, 218, \dodoi{10.3847/1538-4357/ad1603}

\end{thebibliography}

%
\end{document}